\pdfoutput=1
\documentclass[11pt,twoside,a4paper,cmspaper,final,collab]{cms-tdr}
\def\svnVersion{8354cdb}\def\svnDate{2025/10/14}\def\cmsCernNoTag{CERN-EP-2024-313}\def\cmsCernDate{\today}\def\cmsMessage{Published in the Journal of High Energy Physics as \href{http://dx.doi.org/10.1007/JHEP10(2025)074}{\doi{10.1007/JHEP10(2025)074}.}}
\begin{document}\cmsNoteHeader{HIG-22-004}

\newcommand{\PA}{\ensuremath{\mathrm{A}}\xspace}
\newcommand{\mtt}{\ensuremath{m_{\PGt\PGt}}\xspace}
\newcommand{\mlltt}{\ensuremath{m_{\Pell\Pell\PGt\PGt}}\xspace}
\newcommand{\mh}{\ensuremath{m_{\Ph}}\xspace}
\newcommand{\mA}{\ensuremath{m_{\PA}}\xspace}
\newcommand{\ggA}{\ensuremath{\Pg\Pg\to\PA}\xspace}
\newcommand{\bbA}{\ensuremath{\bbbar\PA}\xspace}
\newcommand{\order}{\ensuremath{\mathcal{O}}}
\newcommand{\ptem}{\ensuremath{\pt^{\Pe(\PGm)}}\xspace}
\newcommand{\dR}{\ensuremath{{\Delta}R}\xspace}
\newcommand{\Irelem}{\ensuremath{I_{\text{rel}}^{\Pe(\PGm)}}\xspace}
\newcommand{\Irele}{\ensuremath{I_{\text{rel}}^{\Pe}}\xspace}
\newcommand{\Irelm}{\ensuremath{I_{\text{rel}}^{\PGm}}\xspace}
\newcommand{\yDT}{\ensuremath{y^{\text{DT}}}\xspace}
\newcommand{\Dj}{\ensuremath{D_{\text{jet}}}\xspace}
\newcommand{\De}{\ensuremath{D_{\Pe}}\xspace}
\newcommand{\Dm}{\ensuremath{D_{\PGm}}\xspace}
\newcommand{\etau}{\ensuremath{\Pe\tauh}\xspace}
\newcommand{\mutau}{\ensuremath{\PGm\tauh}\xspace}
\newcommand{\tautau}{\ensuremath{\tauh\tauh}\xspace}
\newcommand{\tanbeta}{\ensuremath{\tan\beta}\xspace}
\newcommand{\BSMEFT}{\ensuremath{\text{M}_{\text{h,EFT}}^{\text{125}}}\xspace}
\newcommand{\alphas}{\ensuremath{\alpha_{\text{S}}}\xspace}
\newcommand{\muR}{\ensuremath{\mu_{\text{R}}}\xspace}
\newcommand{\muF}{\ensuremath{\mu_{\text{F}}}\xspace}
\newcommand{\zerobtag}{\ensuremath{\text{{\textit{no b-tag}}}}\xspace}
\newcommand{\btag}{\ensuremath{\text{{\textit{b-tag}}}}\xspace}
\newcommand{\DNLL}{\ensuremath{-2\Delta(\log\mathrm{L})}\xspace}
\newcommand{\NLL}{\ensuremath{-2\log\mathrm{L}}\xspace}
\newlength\cmsTabSkip\setlength{\cmsTabSkip}{1ex}
\providecommand{\cmsTable}[1]{\resizebox{\textwidth}{!}{#1}}
\providecommand{\cmsLeft}{left\xspace}
\providecommand{\cmsRight}{right\xspace}

\cmsNoteHeader{HIG-22-004}
\title{Search for a heavy pseudoscalar Higgs boson decaying to a \texorpdfstring{125\GeV}{125 GeV} Higgs boson and a \texorpdfstring{\PZ}{Z} boson in final states with two tau and two light leptons in proton-proton collisions at \texorpdfstring{$\sqrt{s} = 13\TeV$}{sqrt(s) = 13 TeV}}

\date{\today}
\abstract{
A search for a heavy pseudoscalar Higgs boson, \PA, decaying to a 125\GeV Higgs boson \Ph and a \PZ boson is presented. The \Ph boson is identified via its decay to a pair of tau leptons, while the \PZ boson is identified via its decay to a pair of electrons or muons. The search targets the production of the \PA boson via the gluon-gluon fusion process, $\Pg\Pg\to\PA$, and in association with bottom quarks, $\bbbar\PA$.  The analysis uses a data sample corresponding to an integrated luminosity of 138\fbinv collected with the CMS detector at the CERN  LHC in proton-proton collisions at a centre-of-mass energy of $\sqrt{s} = 13\TeV$. Constraints are set on the product of the cross sections of the \PA production mechanisms and the $\PA\to\PZ\Ph$ decay branching fraction. The observed (expected) upper limit at 95\% confidence level ranges from 0.049 (0.060)\unit{pb} to 1.02 (0.79)\unit{pb} for the $\Pg\Pg\to\PA$ process and from 0.053 (0.059)\unit{pb} to 0.79 (0.61)\unit{pb} for the $\bbbar\PA$ process in the probed range of the \PA boson mass, $m_{\PA}$, from 225\GeV to 1\TeV. The results of the search are used to constrain parameters within the ${\text{M}_{\text{h,EFT}}^{\text{125}}}$ benchmark scenario of the minimal supersymmetric extension of the standard model. Values of $\tan\beta$ below 2.2 are excluded in this scenario at 95\% confidence level for all $m_{\PA}$ values in the range from 225 to 350\GeV.}

\hypersetup{%
pdfauthor={CMS Collaboration},%
pdftitle={Search for a heavy pseudoscalar Higgs boson decaying to a 125 GeV Higgs boson and a Z boson in final states with two tau and two light leptons in proton-proton collisions at sqrt(s) = 13  TeV},%
pdfsubject={CMS},%
pdfkeywords={CMS, Higgs, 2HDM, MSSM}} %

\maketitle 
\section{Introduction}
\label{sec:introduction}
The 2012 observation of a Higgs-like  boson with a mass of approximately 125\GeV at
the CERN LHC~\cite{Aad:2012tfa,Chatrchyan:2012xdj,Chatrchyan:2013lba} completed the set of 
particles predicted by the standard model (SM).
In the years since, LHC  collaborations have measured the properties of this Higgs boson across various
production modes and decay channels, including refined measurements of its mass, coupling strengths to
fermions and gauge bosons, and spin-parity quantum numbers~\cite{ATLAS:2022vkf, CMS:2022dwd}.
To date, these refined measurements of the observed boson are compatible with SM expectations.
In the SM, the Higgs field is introduced as a complex doublet in the electroweak sector,
and the Higgs boson emerges as a massive scalar state with couplings to the massive
fermions and gauge bosons from the spontaneous breaking of the electroweak gauge symmetry.
These couplings are found to be in good agreement, within the currently attained 
experimental precision of 5--20\%~\cite{Khachatryan:2016vau,Sirunyan:2018koj,Aad:2019mbh,Sirunyan:2019twz}, 
with the expectation for a SM Higgs boson with mass $125.38 \pm 0.14\GeV$~\cite{CMS:2020xrn}. 
The SM still leaves several fundamental questions open, including the presence
of dark matter and the observed baryon asymmetry in our universe.
Beyond-the-SM (BSM) scenarios seek to extend our understanding to explain such phenomena,
for example, by adding structure to its Higgs sector.  
Two Higgs Doublet Models (2HDMs) are BSM theories introducing a second Higgs doublet, which, following
electroweak symmetry breaking, yields five mass eigenstates~\cite{PhysRevD.8.1226, Branco:2011iw}.
Two are charged ($\PSHpm$), two are neutral scalars (\Ph, \PH), and one is a massive pseudoscalar boson \PA, the subject of this search. 
Given current experimental constraints, most of the 2HDMs associate the lighter scalar boson \Ph with the observed 125\GeV Higgs boson and we follow this convention throughout the paper.
The 2HDMs are motivated because they contain, or allow, for additional sources of $CP$
violation that could explain the observed baryon asymmetry in our universe~\cite{Fromme:2006cm}.
The presence of two Higgs doublets is also a requirement in the minimal supersymmetric
extension of the SM (MSSM)~\cite{Fayet:1974pd,Fayet:1977yc}, which offers a dark matter candidate,
protects the Higgs boson mass from receiving large radiative corrections, and provides conditions
for the unification of gauge interactions at the $10^{16}\GeV$ scale~\cite{Haber:1984rc}.

The MSSM Higgs sector is a Type-II 2HDM, which at tree-level is charaterized by two parameters, usually taken to be $\tanbeta=v_2/v_1$, the ratio of the vacuum expectation values of the two Higgs doublets, and \mA, the mass of the \PA  boson. 
The MSSM Higgs boson masses receive large contributions from radiative loop corrections related to
the supersymmetric (SUSY) partners of the SM particles.
In many MSSM scenarios, the mass scale of these supersymmetric partners, $M_{\text{SUSY}}$,
is assumed to be ${\mathcal{O}}(1\TeV)$. With this assumption, the predicted value for \mh falls below 125\GeV at $\tanbeta\lesssim 5-7$. 
Thus, higher scales for $M_{\text{SUSY}}$ are required to support the mass of the observed Higgs boson,
$\mh\approx 125\GeV$, in the whole parameter space, including values of \tanbeta as low as $\approx$1. 
This paper addresses a specific MSSM benchmark scenario, called \BSMEFT~\cite{Bahl:2019ago}. This scenario uses an effective field theory (EFT) approach, where the growing logarithmic corrections associated with the large 
values of $M_{\text{SUSY}}$ are resummed, and $M_{\text{SUSY}}$ can reach $10^{16}\GeV$ and is adjusted to values 
that are compatible with $\mh\approx125\GeV$ at each point in the $\mA$-$\tanbeta$
parameter space individually. The adjusted value of $M_{\text{SUSY}}$  varies in the range from $10^4$\GeV for $\tanbeta \sim 10$ to  $10^8$\GeV at $\tanbeta \sim 1$.

\begin{figure}[htb!]
\centering
  \includegraphics[width=0.49\textwidth]{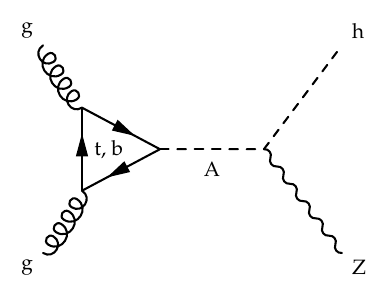}
  \includegraphics[width=0.49\textwidth]{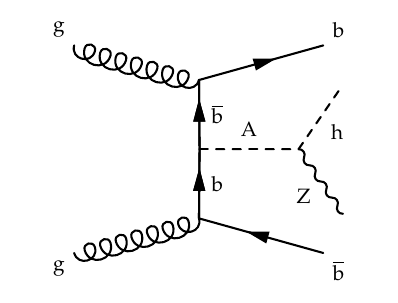}
 \caption{Feynman diagrams representing the production of the pseudoscalar A boson via gluon-gluon $\mA$-$\tanbeta$
 fusion (\cmsLeft) and associated production with a bottom quark-antiquark pair (\cmsRight). In each case, the \PA  boson decays to an SM-like \Ph boson and a \PZ  boson.}
 \label{fig:Feynman}
\end{figure}

This paper reports a search for a pseudoscalar \PA  boson decaying to a 125\GeV Higgs 
boson \Ph and a \PZ boson in proton-proton ($\Pp\Pp$) collisions at $\sqrt{s}=13\TeV$. 
The search uses data collected at the LHC in the period from the year 2016 to 2018, by the CMS experiment, 
corresponding to an integrated luminosity of 138\fbinv.
The analysis targets \PA  boson production via both gluon-gluon fusion, \ggA, and in association
with \PQb quarks, \bbA. The Feynman diagrams for both production processes are shown in Fig.~\ref{fig:Feynman}. Signal mass hypotheses are tested in the range from 225\GeV, near the kinematic threshold for decays to $\PZ\Ph$, up to 1\TeV. The analysis strategy is optimized for a resolved topology and relies on conventional techniques to identify isolated leptons. This strategy sustains high sensitivity to the signal for \mA up to 1\TeV. At higher \PA masses, the \PZ and \Ph bosons are produced with large Lorentz boost, causing the decay products of each boson to be collimated and thus overlapping. As a consequence, the sensitivity of the analysis, targeting resolved topology, rapidly degrades at $\mA>1\TeV$. It should also be emphasized that most BSM scenarios predict rapidly decreasing branching fractions of bosonic decays of the \PA boson with increasing \mA. 

Previous searches by the ATLAS and CMS Collaborations for $\PA\to\PZ\Ph$ targeted final states with two light leptons from the \PZ  boson decay plus two tau leptons from the \Ph boson decay in $\Pp\Pp$ collision dataset at 
$\sqrt{s}=8\TeV$~\cite{Aad:2015wra, Khachatryan:2015tha} and $\sqrt{s}=13\TeV$~\cite{CMS:2019kca}.
The ATLAS and CMS Collaborations have also used $\Pp\Pp$ collisions at $\sqrt{s}=13\TeV$ 
to search for the $CP$-odd \PA  
boson decaying to the same intermediate $\PZ\Ph$ state, but with the Higgs boson \Ph decaying 
to a pair of bottom quarks~\cite{Aaboud:2017cxo, Sirunyan:2019xls,ATLAS:2022enb}.
These analyses set model-independent and model-dependent limits in the context of 2HDMs,
including supersymmetric models.

In this analysis, the \Ph boson is identified by its decay to a pair of tau leptons.
Three possible $\PGt\PGt$ decay channels are considered: $\Pe\tauh$, $\PGm\tauh$,
and $\tauh\tauh$, where \tauh denotes hadronic \PGt lepton decays. Throughout the paper, 
neutrinos are omitted from the notation of the final states. These three decay channels are combined 
with \PZ boson decays into two light leptons, $\PZ\to\Pellp\Pellm$ ($\Pell=\Pe,\PGm$), 
resulting in six distinct final states of the \PA  boson decay. To account for the 
missing transverse momentum that results from the neutrinos in the final states, we use 
a modified version of the {SVFit} algorithm~\cite{Bianchini:2016yrt}, {FastMTT}~\cite{Matyszkiewicz:2025gal}, 
to reconstruct the four-vector of \Ph, while constraining its mass to 125\GeV. The signal is 
extracted from the distributions of the reconstructed four-lepton mass obtained in these individual 
search channels. Furthermore, to increase the sensitivity to different production modes of the \PA  boson, events are split 
into two categories depending on the presence of a \PQb quark in the event.

Relative to the CMS search performed using the 2016 data~\cite{CMS:2019kca}, 
this analysis benefits from the increased integrated luminosity and novel machine learning 
based identification of hadronic decays of $\tau$ leptons~\cite{CMS:2022prd} and of jets originating from
bottom quarks~\cite{Bols:2020bkb}.
It also includes the production of an \PA  boson in association with \PQb quarks, a process not considered previously, and extends the range
of probed masses of the pseudoscalar boson up to 1\TeV.
This analysis provides complementary results to the recent ATLAS search in the $(\PZ\to\PGn\PAGn/\Pell\Pell)(\Ph\to\bbbar)$ channels performed on 139\fbinv of data collected at the same centre-of-mass energy~\cite{ATLAS:2022enb}.

A complete set of tabulated results of the current analysis for all tested mass hypotheses is available in the {HEPData} record~\citep{hepdata}. 

\section{The CMS detector}
\label{sec:detector}

The central feature of the CMS apparatus is a superconducting solenoid of 6\unit{m}
internal diameter, providing a magnetic field of 3.8\unit{T}. Within the
solenoid volume are a silicon pixel and strip tracker, a lead tungstate crystal
electromagnetic calorimeter (ECAL), and a brass and scintillator hadron calorimeter,
each composed of a barrel and two endcap sections. Forward calorimeters extend
the pseudorapidity ($\eta$) coverage provided by the barrel and endcap detectors.
Muons are reconstructed using gas-ionization detectors embedded in the steel flux-return yoke outside the solenoid.
Events of interest are selected using a two-tiered 
trigger system. The first level (L1), composed of custom hardware processors, uses 
information from the calorimeters and muon detectors to select events at a rate of 
around 100\unit{kHz} within a fixed latency of about 4\mus~\cite{Sirunyan:2020zal}. 
The second level, known as the high-level trigger, consists of a farm of 
processors running a version of the full event reconstruction software optimized 
for fast processing and reduces the event rate to around 1\unit{kHz} before data 
storage~\cite{Khachatryan:2016bia}. A more detailed description of the CMS detector,
together with a definition of the coordinate system used and the relevant kinematic
variables, can be found in Ref.~\cite{CMS:2008xjf}.

\section{Event reconstruction}\label{sec:reconstruction}

The reconstruction of the $\Pp\Pp$ collision products is based on the particle-flow 
(PF) algorithm~\cite{Sirunyan:2017ulk}, which combines information from all 
CMS subdetectors to reconstruct a set of particle candidates (PF candidates), 
identified as charged and neutral hadrons, electrons, photons, and muons.  
The primary vertex (PV) is taken to be the vertex corresponding to 
the hardest scattering in the event, evaluated using tracking information alone, 
as described in Ref.~\cite{CMS-TDR-15-02}. Secondary vertices, which are displaced 
from the PV, might be associated with decays of long-lived particles emerging 
from the PV. Any other collision vertices in the event are associated with 
additional, mostly soft, inelastic $\Pp\Pp$ collisions, referred to as pileup (PU). In the 
2016 (2017--2018) datasets,  the average number of PU $\Pp\Pp$ collisions was 23 (32). 

Electrons are reconstructed using tracks from hits in the tracking system and 
calorimeter deposits in the ECAL~\cite{CMS:2020uim}. To 
increase their purity, reconstructed electrons are required to pass a multivariate 
electron identification discriminant, which combines information on track quality, 
shower shape, and kinematic quantities. For this analysis, a working point with 
an identification efficiency of 90\% is used, for a rate of jets misidentified as 
electrons of $\approx$1\%. Muons in the event are reconstructed by combining 
the information from the tracker and the muon detectors~\cite{Sirunyan:2018fpa}.
The mere presence of hits in the muon detectors leads to a strong suppression 
of particles misidentified as muons. Additional identification requirements on 
the track fit quality and the compatibility of individual track segments with the 
fitted track can reduce the misidentification rate further. For this analysis, 
muon identification requirements with an efficiency of $\approx$99\% are chosen, 
with a misidentification rate below 0.2\% for hadrons. 

The contributions from backgrounds to the electron and muon selections are further 
reduced by requiring the corresponding lepton to be isolated from any hadronic 
activity in the detector. This property is quantified by an isolation variable
  \begin{equation}
    \Irelem=\frac{1}{\ptem}\left(\sum\pt^{\text{charged}} + \max\left(0,\sum\et^{\text{neutral}}+\sum\et^{\gamma}-\pt^{\text{PU}}\right)\right),
    \label{isoeq}
  \end{equation}
where \ptem corresponds to the electron (muon) \pt and $\sum\pt^{\text{charged}}$, 
$\sum\et^{\text{neutral}}$, and $\sum\et^{\gamma}$ to the \pt (or transverse energy 
\et) sum of all charged particles, neutral hadrons, and photons, in a predefined 
cone of radius $\dR = \sqrt{\smash[b]{\left(\Delta\eta\right)^{2}+\left(\Delta
\varphi\right)^{2}}}$ around the lepton direction at the PV, where $\Delta\eta$ 
and $\Delta\varphi$ (measured in radians) correspond to the angular distances of 
the particle to the lepton in the $\eta$ and azimuthal angle $\varphi$ directions. The 
chosen cone size is $\dR=0.3$ (0.4) for electrons (muons). The lepton itself is 
excluded from the calculation. To mitigate any distortions from PU, only those 
charged particles whose tracks are associated with the PV are included. 
Since an unambiguous association with the PV is not possible for neutral hadrons 
and photons, an estimate of the contribution from PU ($\pt^{\text{PU}}$) is 
subtracted from the sum of $\sum\et^{\text{neutral}}$ and $\sum\et^{\gamma}$. 
This estimate is obtained from the mean energy flow in the case of \Irele
and from tracks not associated with the PV in the case of 
\Irelm. For negative values, the neutral part of $I_{\mathrm{rel}}$ is set to zero.

For each event, hadronic jets are clustered from the PF candidates using the infrared and collinear safe 
anti-\kt algorithm~\cite{Cacciari:2008gp, Cacciari:2011ma} with a distance parameter of 0.4. 
Jet momentum is determined as the vector sum of all particle momenta in the jet, and is found from simulation to be, on average, 
within 5 to 10\% of the true momentum over the whole \pt spectrum and detector acceptance. 
Pileup can contribute extraneous tracks and calorimetric energy depositions to the jet momentum measurement. 
To mitigate this effect, charged particles identified to be originating from pileup vertices are discarded and an offset correction is applied to account for the remaining contributions. 
Jet energy corrections are derived from simulation to bring the measured response of jets to that of particle level jets on average. 
In situ measurements of the momentum balance in dijet, $\text{photon} + \text{jet}$, $\PZ + \text{jet}$, and multijet events are used 
to account for any residual differences in the jet energy scale between data and simulation~\cite{CMS:2016lmd}. 
The jet energy resolution amounts typically to 15--20\% at 30\GeV, 10\% at 100\GeV, and 5\% at 1\TeV~\cite{CMS:2016lmd}. 
Additional selection criteria are applied to each jet to remove jets potentially dominated by anomalous contributions from various subdetector components or reconstruction failures. 

To identify jets resulting from the hadronization of \PQb 
quarks (\PQb jets) the {DeepJet} algorithm is used, as described in 
Refs.~\cite{Sirunyan:2017ezt,Bols:2020bkb}. In this analysis, a working point 
of this algorithm is chosen that corresponds to a \PQb jet identification 
efficiency of $\approx$80\% for a misidentification rate for jets originating
from light-flavour quarks or gluons of \order(1\%)~\cite{CMS-DP-2018-058}. Jets 
with $\pt>30\GeV$ and $\abs{\eta}<4.7$, and \PQb jets with $\pt>20\GeV$ and 
$\abs{\eta}<2.4$ are used in the analysis of the 2016 data. 
From 2017 onwards, after the upgrade of the 
silicon pixel detector, the \PQb jet $\eta$ range is extended to $\abs{\eta}<2.5$. 

Jets are also used as seeds for the reconstruction of \tauh candidates. This 
is done by utilizing the features of the PF candidates within the distance parameter 
of jets using the ``hadrons-plus-strips'' 
algorithm, as described in Refs.~\cite{Sirunyan:2018pgf,CMS:2022prd}. Decays to 
one or three charged hadrons with up to two neutral pions with $\pt>2.5\GeV$ are 
used (referred to as \tauh decay mode thereafter). Neutral pions are reconstructed as strips with dynamic size in $\eta$-$\phi$ 
from reconstructed photons and electrons contained in the seeding jet, where the 
electrons originate from photon conversions. The strip size varies as a function of 
the \pt of the electron or photon candidates. The \tauh decay mode is then obtained 
by combining the charged hadrons with the strips. To distinguish \tauh candidates 
from jets originating from the hadronization of quarks or gluons, and from electrons 
or muons, the DeepTau~\cite{CMS:2022prd} (DT) algorithm is used. This algorithm uses information from the event, such as tracking, impact parameter, calorimeter cluster composition, and the kinematic and object identification properties of the PF candidates in the vicinity of the \tauh candidate, as well as quantities that estimate the PU density of the event. 
This process results in a multiclassification output $\yDT_{\alpha}$ ($\alpha=\PGt$, \Pe, \PGm, jet) that quantifies compatibility 
of \tauh candidate with the hypothesis of a genuine \PGt lepton, an electron, 
a muon, or the hadronization of a quark or gluon.
From this output, three discriminants are built according to 
  \begin{equation}
    D_{\alpha} = \frac{\yDT_{\PGt}}{\yDT_{\PGt}+\yDT_{\alpha}}, \quad
    \alpha=\Pe,\,\PGm,\,\text{jet}.
    \label{Deq}
  \end{equation}

For the analysis presented here, predefined thresholds on the \De, \Dm and \Dj discriminants corresponding to the medium working point of the DeepTau algorithm~\cite{CMS:2022prd} are chosen depending on the $\Ph\to\PGt\PGt$ final state, for which the \tauh selection efficiencies and misidentification rates are given in 
Table~\ref{tab:dt-working-points}. The $\PZ\to\PGt\PGt$ decay is used to measure the \tauh 
identification efficiency. Samples of $\PZ\to\Pe\Pe$ and $\PZ\to\PGm\PGm$ decays are employed to 
measure the $\Pe\to\tauh$ and $\PGm\to\tauh$ misidentification rates, respectively. Samples 
of $\PW(\to\Pell\nu)+{\text{jets}}$ events and top quark-antiquark pairs are used to measure 
the $\text{jet}\to\tauh$ misidentification rate.

While neutrinos cannot be detected directly, they contribute to the missing transverse momentum.
The missing transverse momentum vector \ptvecmiss is computed as the negative
vector sum of the transverse momenta of all the PF candidates in an event~\cite{CMS:2019ctu}.
The \ptvecmiss is modified to account for corrections to the energy scale of the reconstructed jets in the event. With \ptmiss we refer to the magnitude of this quantity.

\begin{table}[htbp]
  \topcaption{
    Efficiencies for the identification of \tauh decays and corresponding 
    misidentification rates (given in parentheses) for the working points of \De, \Dm, 
    and \Dj, chosen for the $\Ph\to\PGt\PGt$ selection, depending on the $\PGt\PGt$ final state. The numbers are given as percentages. Efficiencies and misidentification rates are determined from dedicated studies~\cite{CMS:2022prd}.}
  \label{tab:dt-working-points}
  \centering
  \begin{tabular}{lccc}
    $\PGt\PGt$ channel & \De (\%) & \Dm (\%) & \Dj (\%) \\
    \hline
     \etau   & $>$80 ($<$0.5) & $>$99 ($<$0.5) & $>$65 (1--3) \\
     \mutau  & $>$95 (1--2)    & $>$97 ($<$0.1) & $>$65 (1--3) \\
     \tautau & $>$95 (1--2)    & $>$99 ($<$0.5) & $>$65 (1--3) \\
  \end{tabular}
\end{table}

The mass of the \PA  boson candidate is reconstructed using the {FastMTT} algorithm~\cite{Matyszkiewicz:2025gal},
which uses a simplified mass likelihood function to reduce the
computation time.
This algorithm makes use of \ptvecmiss and its uncertainty, and the four-vectors of the
reconstructed visible \PGt lepton decay products to calculate an estimate of the mass of the
parent boson and the full four-momenta of the \Ph decay products. Compared to the
{SVFit} algorithm, the {FastMTT} algorithm
removes the contributions of the leptonic and hadronic \PGt lepton decay matrix elements to the likelihood function, and assumes that visible \PGt lepton decay products move collinearly with the original \PGt lepton momentum.
This gives a mass resolution that is similar to that of the {SVFit} algorithm, but the computation time
is reduced by two orders of magnitude. Further improvement in the four-lepton mass resolution
for $\Ph \to \PGt\PGt$ decays is achieved by imposing the mass constraint $\mtt=\mh=125\GeV$. 

In summary, three mass reconstruction techniques have been studied in the course of this analysis: 
in the first, the four-lepton mass is computed using the leptons from the \PZ boson decay and 
only visible decay products of the \PGt leptons, denoted $\mlltt^{\text{vis}}$. This method 
yields a mass resolution of 20--30\%. In the second, the four-lepton mass is computed using 
the \PZ boson decay leptons and the {FastMTT}-corrected \PGt lepton four-vectors with no mass constraint, 
denoted $\mlltt^{\text{corr}}$. This method yields a resolution on \mA of 10--15\%. In the third, the 
four-lepton mass is computed with the \PZ decay leptons and the {FastMTT}-corrected \PGt four 
vectors with a mass constraint of $\mtt = 125\GeV$ imposed, denoted $\mlltt^{\text{cons}}$. This method yields the best experimental resolution on \mA of 5--7\% and eliminates the bias in the mean value of the reconstructed mass observed with the other two methods.

\begin{figure}[h!]
\centering
  \includegraphics[width=0.49\textwidth]{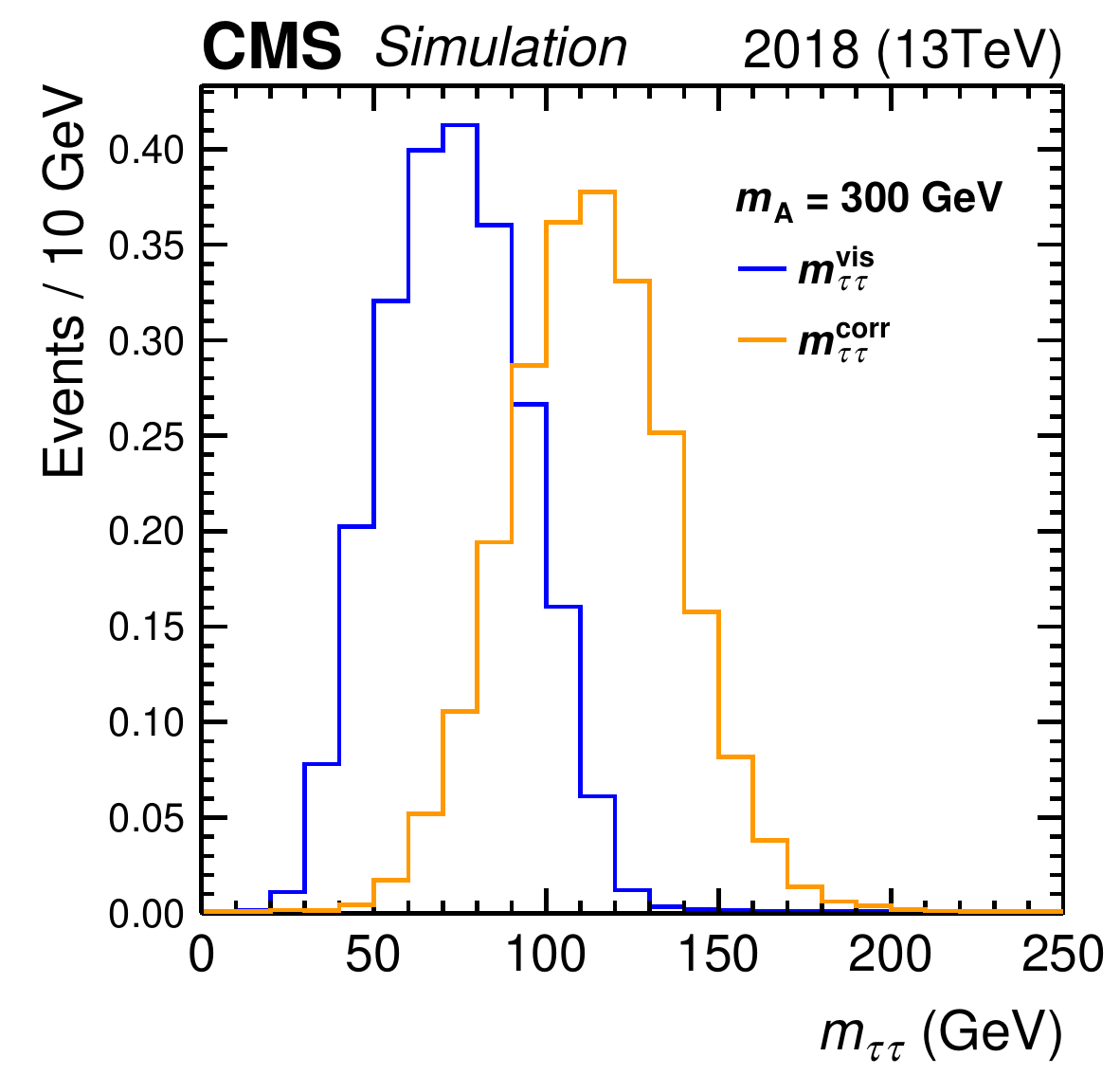}
  \includegraphics[width=0.49\textwidth]{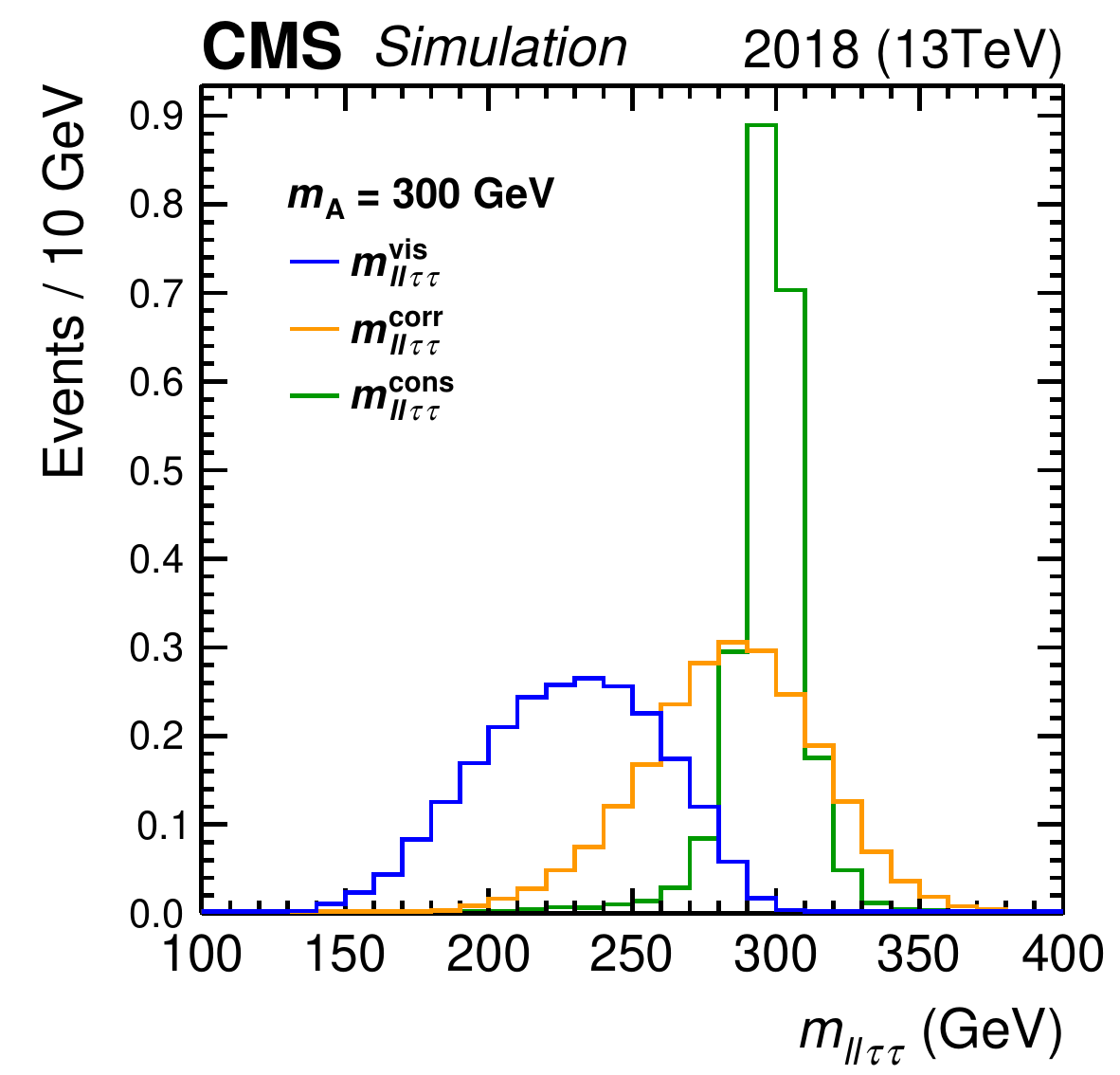}
\caption{
The distribution of the reconstructed mass of the $\Ph\to\PGt\PGt$ candidate (\cmsLeft plot) 
and of the $\PA\to\PZ\Ph\to(\Pell\Pell)(\PGt\PGt)$ candidate (\cmsRight plot) in a 2018 simulated 
sample of \ggA events with $\mA=300\GeV$. Several methods of mass reconstruction are compared: 
1) using only the visible decay products of \PGt lepton ($\mtt^{\text{vis}}$ in the \cmsLeft plot and 
$\mlltt^{\text{vis}}$ in the \cmsRight plot, blue histograms), 2) using the {FastMTT} 
algorithm to correct for missing momentum carried away by neutrinos in the \PGt lepton decays 
($\mtt^{\text{corr}}$ in the \cmsLeft plot and $\mlltt^{\text{corr}}$ in the \cmsRight plot, orange 
histograms), and 3) using the {FastMTT} algorithm with a mass constraint of 125\GeV for 
the $\Ph\to\PGt\PGt$ candidate ($\mlltt^{\text{cons}}$ in the \cmsRight plot, green histogram).
}
 \label{fig:mllttc}
\end{figure}

These strategies are compared in Fig.~\ref{fig:mllttc}, where all final states of the \PA  boson decay are combined and the distributions are obtained after applying a selection of $\PZ\to\Pellp\Pellm$ and $\Ph\to\PGt\PGt$ candidates, as described in Section~\ref{sec:selection}. The analysis employs the second method to 
obtain the best estimate for the mass of the \Ph candidate from the {{FastMTT}}-corrected four 
vectors of \PGt candidates, $\mtt^{\text{corr}}$. This variable is used in the event selection, as described in 
Section~\ref{sec:selection}. The third method is used to reconstruct the mass of \PA  candidate, 
$\mlltt^{\text{cons}}$. This observable serves as a final discriminant in the statistical inference.  

\section{Data and simulated samples}\label{sec:samples}

The data sample analyzed in this search corresponds to an integrated luminosity of 138\fbinv from $\Pp\Pp$ 
collisions at a centre-of-mass energy at 13\TeV, collected with the CMS detector at the LHC. 

Simulated signal events with a $CP$-odd Higgs boson \PA  produced in 
gluon-gluon fusion and associated production with \PQb quarks, decaying to a 125\GeV 
Higgs boson and a \PZ
boson are generated at leading order (LO) precision in the strong coupling constant \alpS 
using \MGvATNLO v2.6.5~\cite{Alwall:2014hca}, assuming a narrow signal width. The generated \PA  boson mass points lie in the range 225\GeV 
to 1\TeV.
The $\PA\to\PZ\Ph$ decaying to $\Pell\Pell\PGt\PGt$ is simulated 
with \textsc{MadSpin}~\cite{Artoisenet:2012st}.  In the gluon-gluon fusion production mode, up to 
one additional jet is included in the matrix element calculations, following the MLM matching scheme with parton showers~\cite{Alwall:2007fs}.

In the signal event simulation, the value of $m_{\Ph}$ is set to 125 GeV and the additional 
$CP$-even \PH boson, $CP$-odd \PA  boson, and charged Higgs boson masses are assumed to be at least 225\GeV. The 
discrete $\mathbb{Z}_2$ symmetry is broken as in the minimal supersymmetric standard model~\cite{Martin:1997ns}, and $CP$ is enforced to be conserved at tree level in the 2HDM Higgs $\mathbb{Z}_2$
sector~\cite{Branco:2011iw}. 

The background samples consist of all SM processes with non-negligible yield in the signal region,
including those with a 125\GeV Higgs boson present.
The $\PZ\Ph$, $\PW\Ph$, and $\ttbar\Ph$ processes, in which an \Ph decaying to two \PGt leptons is generated
in association with a \PZ boson, \PW boson, or top quark pair, respectively,
are generated at next-to-LO (NLO) precision in \alpS with 
\POWHEG~v2.0~\cite{Nason:2004rx,Frixione:2007vw,Alioli:2010xd,Luisoni:2013cuh,Granata:2017iod,
Hartanto:2015uka}. The contribution from processes with a \Ph boson decaying to two \PW bosons, 
produced via $\PW\Ph$ or $\PZ\Ph$, are generated at NLO precision in \alpS using \POWHEG 
and \textsc{JHUGen} v7.7.2~\cite{Bolognesi:2012mm} programs. Contributions from \Ph produced via gluon-gluon fusion or vector boson fusion, where the \Ph decays to two \PGt leptons or two \PW bosons, is negligible.
The \Ph boson samples are normalized to their inclusive cross sections, and 
the branching fractions used are those recommended by the LHC Higgs Working 
Group~\cite{LHCHiggsCrossSectionWorkingGroup:2016ypw}, assuming an $m_{\Ph}$ mass of 125.38 GeV~\cite{CMS:2020xrn}.

The $\Pg\Pg\to\PZ\PZ$ process is generated at LO precision in \alpS with
\textsc{mcfm}~v7.0.1~\cite{Campbell:2011bn}. The $\qqbar\to\PV\PV$ processes, where \PV is either a \PW or a 
\PZ boson, are generated with NLO precision in \alpS using \POWHEG~\cite{Nason:2013ydw} 
or with \MGvATNLO with the FxFx jet matching and merging scheme~\cite{Frederix:2012ps}.  

Triboson, $\PZ+\text{jets}$, $\ttbar\PW$, and $\ttbar\PZ$ production are generated via \MGvATNLO,
with the scheme applied either at NLO precision in \alpS exploiting the FxFx jet matching and merging scheme, or at LO precision in \alpS with the MLM jet matching and 
merging scheme. For $\PZ+\text{jets}$, supplementary samples are generated with up 
to four outgoing partons in the hard interaction to increase the number of simulated events in 
regions of high signal purity. The $\ttbar$ background processes is generated at NLO 
precision in \alpS with \POWHEG~\cite{Alioli:2011as}.  The 
$\Pg\Pg\to\PZ\PZ$, $\qqbar\to\PV\PV$, triboson, $\PZ+\text{jets}$, \ttbar, $\ttbar\PW$, and $\ttbar\PZ$ 
are normalized to their cross sections at NLO precision in \alpS or 
higher~\cite{Caola:2016trd,Campbell:2011bn,Gehrmann:2014fva, Melnikov:2006kv, Czakon:2011xx, 
Campbell:2012dh, Garzelli:2012bn}. 

For all simulated samples, the NNPDF3.1~\cite{Ball:2017nwa} parton 
distribution functions (PDFs) are used for the simulation. Parton showering and hadronization, as well as the \PGt lepton decays, are 
modeled using version 8.230 of the \PYTHIA event generator~\cite{Sjostrand:2014zea}. 
The description of the underlying event is parameterized according to the CP5~\cite{Sirunyan:2019dfx} 
tunes for the simulation. Additional inclusive inelastic $\Pp\Pp$ collisions generated with \PYTHIA 
are added according to the expected pileup profile in data. All generated events are 
passed through a \GEANTfour-based~\cite{Agostinelli:2002hh} simulation of the CMS 
detector and reconstructed using the same version of the CMS event reconstruction 
software used for the data. The details of the various generator and simulation programs used are summarized in Table~\ref{tab:mc-generators}.

\begin{table*}[htbp]
\centering
\topcaption{
Summary of Monte Carlo programs and their purposes.
}
\label{tab:mc-generators}
\cmsTable{
\begin{tabular}{l l l}
   {Program} & {Version / Scheme} & {Role / Process Simulated} \\
   \hline
    \MGvATNLO      & v2.6.5 (LO, MLM)     & Signal production ($\ggA, \bbA$), narrow-width approximation \\
    \textsc{MadSpin}                 & \NA                    & Decay of \PGt leptons with proper spin correlations \\
    \POWHEG                  & v2.0 (NLO)           & $\PZ\Ph$, $\PW\Ph$, $\ttbar\Ph$ and $\ttbar$ backgrounds at NLO in \alpS \\
    \textsc{JHUGen}                  & v7.7.2 (NLO)         & $\PW\Ph/\PZ\Ph$ with $\Ph\to \PW\PW^{(*)}$ at NLO, including spin effects \\
    \textsc{mcfm}                    & v7.0.1 (LO)          & $\Pg\Pg\to \PZ\PZ$ continuum background at LO \\
    \MGvATNLO      & FxFx (NLO)           & $\qqbar\to\PV\PV$ diboson backgrounds at NLO with jet matching \\
    \MGvATNLO     & MLM (LO)             & Triboson, $\PZ+\text{jets}$, $\ttbar\PW$, $\ttbar\PZ$ ($\PZ+\text{jets}$ also with $\leq$ 4 partons) \\
    \PYTHIA                  & v8.230 (CP5 tune)    & Parton shower, hadronization, \PGt decays \\
    NNPDF3.1                & \NA                   & Parton distribution functions for all samples \\
   \GEANTfour                  & \NA                    & Full CMS detector simulation and event reconstruction \\

\end{tabular}}
\end{table*}

\section{Event selection}\label{sec:selection}

Events are selected online using single-lepton triggers targeting leptons resulting from \PZ boson decays. The nominal online \pt thresholds for the single-electron (single-muon) trigger are 25--35\GeV (24--27\GeV), depending on the data-taking period.
Events are selected if either of the two leptons assigned to a
$\PZ\to\Pe\Pe$ ($\PZ\to\PGm\PGm$) decay satisfies the single-electron (single-muon) trigger.
The offline \pt selection of the triggering lepton is required to be 1\GeV higher than the nominal \pt threshold
of the corresponding trigger.
Leptons selected by the trigger are required to geometrically match with a selected offline lepton.
Corrections are applied to account for small differences in the trigger selection efficiencies
measured in simulation and data.

The light leptons and the \tauh candidates that do not pass the online triggers selections are required to have $\pt>10\GeV$
and $\pt>20\GeV$, respectively.
Constraints on $\abs{\eta}$ arising from detector geometry are
$\abs{\eta^{\Pe}}<2.5$ for electrons, $\abs{\eta^{\PGm}}<2.4$ for muons, and
$\abs{\eta^{\tauh}}<2.3$ for \tauh candidates.
These constraints are applied to all selected electrons and muons of the event whether or not they pass the 
online trigger criteria.  The light leptons in an event are required to be separated from each
other by $\Delta R > 0.3$, while the \tauh candidates must be separated from each other and
from any other lepton by $\Delta R > 0.5$. 
The resulting selected events are made mutually exclusive by discarding events that
have additional identified and isolated electrons or muons.

The \PZ boson candidates are reconstructed from pairs of same-flavour and opposite-charge light leptons
satisfying $60<m_{\Pellp\Pellm} < 120\GeV$.
In events with multiple \PZ boson candidates, we choose the one with the mass closest to the \PZ boson mass.
The leptons associated with the $\Ph\to\PGt\PGt$ decay $(\Pe\tauh$, $\PGm\tauh, \tauh\tauh)$ are required to have opposite charge and an isolation 
requirement of $\Irelem<0.15$, whereas other identification requirements are described in Section~\ref{sec:reconstruction}. The \tauh candidates associated with the \Ph boson must satisfy the 
\tauh identification with efficiencies detailed in Table~\ref{tab:dt-working-points}. In the following, the light lepton (\Pe or \PGm) and \tauh identification criteria described above are referred to as ‘nominal’ lepton identification criteria.

Events with at least one identified \PQb jet, according to the criteria 
given in Section~\ref{sec:reconstruction}, fall into the \btag category, used 
to target \PQb quark associated \PA  boson production. All other events are placed
into the \zerobtag category, which is used to target gluon-gluon fusion production of the \PA  boson.  

To further improve search sensitivity, the {FastMTT} algorithm is employed to account for unmeasured momentum carried away by \PGt decay neutrinos. The FastMTT-corrected mass of the $\Ph\to\PGt\PGt$ candidate, $\mtt^{\text{corr}}$, is required to be within the 90--160\GeV mass range.

\section{Modeling of signal and background}\label{sec:background_estimation}

Backgrounds with prompt lepton decays ($\PZ\PZ$, $\ttbar\PZ$, triboson, and
SM processes producing a \Ph boson), and the acceptance of
signal processes (\ggA and \bbA) are estimated from simulation.
Background processes are scaled by their theoretical cross sections calculated at the
highest order available, as described in Section~\ref{sec:samples}, whereas normalizations
of signal processes are extracted from fits to data as described in Section~\ref{sec:results}.

Reducible background, arising from the misidentification of one or both \PGt candidates, is estimated from data. The dominant contributions come
from the \ttbar, $\PZ+\text{jets}$, and $\PW\PZ+\text{jets}$ processes, where 
at least one \PGt candidate is mimicked by a hadronic jet. 
This  background is evaluated using a "misidentification factor" method, that involves measuring the probabilities
of misidentifying a hadronic jet as a prompt light lepton or \tauh candidate.
The misidentification factors for electrons, muons, and \tauh are measured in a control region dominated
by nonprompt or misidentified leptons, and are defined as the fraction of objects passing the nominal selections in the sample of objects passing loose selections.
When selecting signal candidate events, any events with \PGt candidates passing loose 
criteria but failing the nominal criteria are assigned to an application region (AR). They are used along with 
the misidentification factors to estimate the contribution from the reducible background in the signal region, \ie, the region that contains events with leptons or \tauh passing identification and isolation.  The nominal selection uses the standard criteria defined in Table~\ref{tab:dt-working-points} for identifying events in the signal region, while the loose selection applies only basic requirements and more relaxed criteria.

Loose identification criteria imposed on leptons are summarized in the following.
\begin{itemize}

\item Muon candidates: loosened quality criteria on the muon track and relaxed cut on the relative isolation variable (see Eq.~\ref{isoeq}), $I_{\mathrm{rel}(\mu)} < 0.5$. 
\item  Electron candidates: relaxed cut on the relative isolation variable, $I_{\mathrm{rel}(e)} < 0.5$, and loosened cut on the electron multivariate discriminant. 
\item \tauh candidates : loosened cut on $D_{\text{jet}}$ discriminant (see Eq.~\ref{Deq}).
\end{itemize}

Loose identification criteria lead to an increase in the lepton misidentification rate by one to three orders of magnitude, depending on the lepton flavor (\Pe, \PGm, or \tauh), lepton \pt and $\eta$, and for \tauh also on the decay mode.

An estimate of the background from misidentified leptons and \tauh in the signal region is obtained 
by applying suitably chosen weights to the events selected in the AR. Applied weights are computed according to the following relations:
  \begin{equation}
  \begin{aligned}  
    w_{1} &= \frac{f_1}{1-f_1}, \\
    w_{2} &= \frac{f_2}{1-f_2}, \\
    w_{12} &= -\frac{f_1f_2}{(1-f_1)(1-f_2)},
  \end{aligned}
  \label{eq:reducible_weight}
  \end{equation}
where: weight $w_1$ ($w_2$) is applied to events where the first (second) \PGt candidate fails the 
nominal identification criteria, whereas the second (first) \PGt candidate passes the nominal 
identification criteria; weight $w_{12}$ is applied to events where both \PGt candidates fail 
the nominal identification criteria;  and $f_1$ and $f_2$ are the misidentification factors of the 
first and second \PGt candidates. The \PGt candidates are sorted by descending visible transverse momentum, 
with the leading candidate defined as the `first' and the subleading as the `second'.  The negative sign of the weight $w_{12}$ accounts for double 
counting of events from \ttbar and $\PZ+\text{jets}$ processes with both \PGt candidates failing 
the nominal identification criteria. 

The misidentification factors $f_{i}$ are measured in event samples that have no 
overlap with the signal region. A validation region, defined to be orthogonal to the signal and 
measurement regions, is used to test the robustness of the reducible background estimate. Systematic 
uncertainties corresponding to the possible differences between the true and estimated reducible 
background rates in the signal region are estimated from these closure tests.  

The reducible background estimate involves four regions: the control region enriched in Drell--Yan (DY) events (used to measure misidentification factors), the application region (events failing one or two \PGt ID criteria and used for applying fake-rate weights), the validation region (same-sign \PGt candidates used to test the method), and the signal region (events with two \PGt candidates passing the nominal ID). A summary of the definitions and purposes of all control regions discussed in this Section, along with the definition of the signal region, is provided in Table~\ref{tab:analysis_regions}. It should be noted that the selection criteria defining the regions listed in Table 3 ensure their mutual orthogonality.

The dedicated DY control region requires an additional hadronic 
jet that is misidentified as a light lepton or a \tauh candidate. This control region is used to measure the 
misidentification (fake) rates for jets misidentified as \tauh candidates, electrons, or muons. The estimation of misidentification factors relies on reconstructing 
an opposite-charge, same-flavour lepton pair compatible with a \PZ boson, and requiring one additional loosely selected lepton or \tauh candidate (i.e., passing an identification working point looser than the one used in the signal region). 

The requirements on the leptons originating from the \PZ 
boson are the same as those defined in Section~\ref{sec:selection}, but they must fulfil a more 
stringent dilepton mass cut with $81 < m_{\Pellp\Pellm} < 120\GeV$. The lower threshold on the 
dilepton mass is tightened to suppress the contribution of DY events affected by final-state radiation. These events typically have
 lower dilepton mass compared to DY events without final-state radiation.

\begin{table*}[htbp]
\centering
\topcaption{Summary of regions used in the misidentification factor method.}
\label{tab:analysis_regions}
\begin{tabular}{lcl}
\multicolumn{3}{c}{Common for all regions} \\
\hline
\multicolumn{3}{c}{$\Pep\Pem$ or $\PGmp\PGmm$ pair consistent with \PZ decay} \\
 & & \\
 & & \\
Definition   &  &  Purpose \\
\multicolumn{3}{c}{Determination region (DR)} \\
\hline
one and only one & &  determination of lepton \\
 $\PGt$ candidate & &  misidentification factors \\
passing loose lepton id. & & ($f_{1,2}$ in Eq.~\ref{eq:reducible_weight}) \\
 & &  \\
\multicolumn{3}{c}{Application region (AR)} \\
\hline  
$\PGtp\PGtm$ pair where both &  & construction of the reducible \\
\PGt candidates pass loose lepton &  &  background model by applying \\
id. but at least one fails nominal id. &  &  misidentification factors \\
& &  \\
\multicolumn{3}{c}{Validation region (VR)} \\
\hline
same sign $\PGtpm\PGtpm$ pair & & validation of the reducible \\
where both \PGt canidates &  & background model, assessment of \\
pass nominal lepton id. & & related systematic uncertainties \\             
& & \\
\multicolumn{3}{c}{Signal region (SR)} \\
\hline
$\PGtp\PGtm$ pair where both \PGt &  & selection of events into final sample, \\
candidates pass nominal lepton id. & & where the signal is extracted \\
\end{tabular}
\end{table*}

After reconstructing the $\PZ\to\Pellp\Pellm$ candidate, the misidentification factor is estimated by applying the 
lepton identification algorithm to the additional loosely identified light lepton or \tauh candidate 
in the event. Orthogonality to the signal region is achieved by rejecting events with extra \PGt 
candidates (either light lepton or \tauh), passing loose identification criteria. The misidentification factors are measured in different bins of lepton \pt, and are further split between reconstructed decay 
modes for the \tauh candidate, and for muons and electrons in bins of lepton $\eta$, based on the barrel 
and endcap regions.  The events where the \PGt candidates are genuine \tauh, electrons, or muons,
are estimated from simulation and subtracted from data so that the misidentification factors are measured 
for genuine hadronic jets only. The misidentification factors obtained for electrons (muons) are 
$<$2 (5)\% in the barrel and endcap regions for lepton $\pt>10\GeV$, whereas for \tauh candidates with $\pt>20\GeV$ the misidentification factors vary between 2 and 20\%, depending 
on the decay mode and discriminator working point.

The measured misidentification factors are validated in another region that consists of events with a \PZ boson candidate and two additional $\tau$ candidates. To ensure that the validation region is not contaminated with signal events or contributions from backgrounds with genuine prompt leptons, the two additional \PGt candidates are required to have the same charge. All other selection criteria are identical to those defining the signal region. The reducible background in the validation region is constructed in the same way as in the nominal analysis. The purity of the reducible background in the validation region amounts to more than 95\% in all analyzed channels.

Given the limited statistical power of the data sample in the validation region, the observed data yields and shapes of the $\mlltt^{\text{cons}}$ distribution are compared with the predicted yields and shapes of the reducible background by combining all three data-taking periods. The comparison revealed only modest differences between the observed data and the model of the reducible background. The saturated goodness of fit (GoF) test~\cite{Cousins:2018tiz}, quantifying consistency of the observed data with the background model, yields $p$-values of 0.31, 0.48, and 0.82 for the \tautau, \etau, and \mutau channels, respectively. This test is performed using the CMS statistical toolkit {\textsc{Combine}}~\cite{CMS:2024onh}. The differences between the observed data and the model, together with the statistical uncertainties due to the limited data sample in the validation region, are accounted for by assigning a systematic uncertainty in the yield of the reducible background, estimated to be 20--30\% depending on the \Ph decay channel. Given that the comparison is performed for the combined dataset, these uncertainties are 
correlated among data-taking periods, but uncorrelated among \Ph decay channels. 

The limited statistical power of the data sample in the AR, along with the contribution of events with 
negative weights (Eq.~\ref{eq:reducible_weight}), hampers the construction of smooth templates to model the shape of the $\mlltt^{\text{cons}}$ distribution of the reducible background. The shape of the distribution is therefore taken from a data region with same-sign \PGt candidates that pass loose identification and isolation requirements.
This region has higher statistical power than the AR region, resulting in a smoother shape for the $\mlltt^{\text{cons}}$ distribution, which is normalized 
to the estimated yield of the reducible background in the signal region. Consistency of shapes obtained from the statistically limited AR and the region with same-sign \PGt candidates, passing loose identification criteria, is verified with a Kolmogorov-Smirnov test, yielding $p$-values between 0.56 and 0.97 depending on the di-\PGt decay mode.  

\section{Systematic uncertainties}
\label{sec:systematics}
The dominant systematic uncertainties considered in the analysis are summarized in Table~\ref{tab:uncertainties}.

The uncertainties in the \tauh identification efficiency are estimated in $\PZ\to\PGt\PGt$ and $\PW\to\PGt\PGn$ 
decays from control samples. These uncertainties are partially correlated across data-taking years. The size of uncertainties varies in the range of 2--10\% per \tauh, depending on \pt and the decay mode of the \tauh candidate~\cite{CMS:2022prd}. 
The uncertainties in the \tauh energy scale amount 
to 0.5--1.1\%, depending on the \tauh decay mode. They are predominantly of statistical origin and 
uncorrelated between data-taking years. Uncertainties in the \tauh identification efficiency and momentum scale affect both the normalization of simulated processes and the shape of the $\mlltt^{\text{cons}}$ distribution. 

Uncertainties in the identification and isolation efficiencies of electrons and muons are 1.5\% and are correlated across all years. These uncertainties result in normalization variations between 1.5\% and 4.5\%, depending on the analyzed final state.  
Uncertainties in the muon momentum scale amount to less than 0.3\% and have a negligible impact on the analysis. 
The uncertainty in the electron energy scale, which is derived from the calibration of the ECAL crystals and 
applied on an event-by-event basis, is less than 2\%. The uncertainty in the single-lepton trigger efficiency results in a normalization uncertainty of about 2\% for both single-electron and single-muon triggers.

Uncertainties in both the identification efficiency for \PQb jets and in the misidentification rates for 
light-flavour or \PQc quarks or gluon jets range from the subpercent level to $\mathcal{O}(10\%)$, depending on jet flavour and $\pt$. The uncertainty in the identification efficiency of \PQb jets causes variation of 1--4\% (0.3--1\%) in the normalization of the \bbA signal and background processes with genuine \PQb jets in the \btag (\zerobtag) category. The 
uncertainty in the misidentification efficiency of the light-flavour, or \PQc quarks, or gluon jets modifies normalization of the \ggA signal and background processes with no genuine \PQb jets by 5--10\% in the \btag category and has an impact of $\mathcal{O}(0.1\%)$ on the normalization of processes without genuine \PQb jets in the \zerobtag category.

An uncertainty related to the energy carried by unclustered particle candidates, which are not contained in jets in the event~\cite{CMS:2019ctu}, is propagated on an event-by-event basis to \ptmiss, resulting in a variation of up to 10\% in the shape of the $\mlltt^{\text{cons}}$ distribution. A normalization-altering effect of 1--3\% is introduced by this uncertainty by requiring $90<\mtt^{\text{corr}}<160\GeV$. The jet energy scale and resolution affect both the selection efficiencies and shapes of the $\mlltt^{\text{cons}}$ distributions. The jet energy scale is responsible for a 1--3\% variation in the number of selected background and signal events; the jet energy resolution contributes an additional 0.5--1\%. The calibration accuracy of the unclustered and jet energy scales and jet energy resolution is mainly affected by statistical limitations of the measurements, the time-dependence of the data-taking conditions, and the aging of the detector. For this reason, the respective uncertainties are uncorrelated between data-taking years. 

Theoretical uncertainties related to the choice of PDFs, and the renormalization (\muR) and factorization (\muF) scales, affecting both the acceptance and cross section of the dominant background processes, are estimated from simulation separately for each process. Uncertainties due to the choice of \muR and \muF in the calculation of the matrix elements are obtained from an independent variation of these scales by factors of 0.5 and 2, omitting the variations where one scale is multiplied by 2 and the corresponding other scale by 0.5. The uncertainties are then obtained from an envelope of these variations. The uncertainties due to PDF variations and the uncertainty in \alphas are obtained following the PDF4LHC recommendations~\cite{Butterworth:2015oua}, taking the root mean square of the variation of the results when using different replicas of the default NNPDF3.1 set.

Combining \muR and \muF scale uncertainties with the PDF set uncertainty for the $\qqbar\to \PZ\PZ$ process leads to a normalization uncertainty of 5\%. For the $\Pg\Pg\to\PZ\PZ$ process, a normalization uncertainty of 15\% is obtained. It covers variations of \muR and \muF, PDF set and \alphas uncertainties, and uncertainty that accounts for effects of interference with the process mediated by the off-shell Higgs boson, $\mathrm{gg}\to\Ph^{*}\to\PZ\PZ$~\cite{Caola:2016trd,Sirunyan:2019twz}. The uncertainties in cross sections of the $\ttbar\PZ$ and triboson production amount to 25\%~\cite{Sirunyan:2017uzs} and dominate normalization uncertainty for these processes. 

The uncertainty in the theoretical calculations of the SM $\Ph\to\PGt\PGt$ branching fraction, amounting 
to 2\%~\cite{LHCHiggsCrossSectionWorkingGroup:2016ypw} is applied to both the signal samples as well 
as all backgrounds that include the $\Ph\to\PGt\PGt$ process. The inclusive uncertainty for $\PZ\Ph$ production related to the PDFs amounts to 1.3\%, whereas the uncertainty for the variation of \muR and \muF is 0.9\%~\cite{LHCHiggsCrossSectionWorkingGroup:2016ypw}. For the subleading \Ph boson processes $\Pg\Pg\to\Ph\to\PZ\PZ$, and $\ttbar\Ph$ the inclusive uncertainties related to the PDFs amount to 3.2 and 3.6\% and the uncertainties for the variation of \muR and \muF are 3.9 and 8\%, respectively~\cite{LHCHiggsCrossSectionWorkingGroup:2016ypw}.

For the MSSM parameter scan, theoretical uncertainties in the \ggA and \bbA cross sections are accounted for as described in Ref.~\cite{Bagnaschi:2021jaj}. This includes uncertainties in \muR, \muF, PDFs, and \alphas. Uncertainties are evaluated separately for each $\mA$-$\tanbeta$ point under consideration. They are typically 5--20\% (10--25\%)) for \ggA (\bbA) production.

Uncertainties in the estimated yield of background with misidentified leptons comprise two components.
Statistical uncertainties arise from the limited statistical power of the data sample in the AR. These are the dominant uncertainties in the analysis and they range between 10--20\% (20--40\%) in the \zerobtag (\btag) category and are uncorrelated between channels, event categories, and data-taking periods. Uncertainties related to the misidentification factor method are estimated in the sideband region with same-sign \PGt pairs, as described in Section~\ref{sec:background_estimation}. Normalization uncertainties of 20--30\% are assigned to the estimated yield of this background. They are correlated between data-taking periods and event categories but uncorrelated across di-\PGt final states.

The integrated luminosities of the 2016, 2017, and 2018 data-taking periods are  known 
with uncertainties in the 1.2--2.5\% range~\cite{CMS-LUM-17-003,CMS-PAS-LUM-17-004,CMS-PAS-LUM-18-002}, while the total integrated luminosity for the years 2016--2018 has an uncertainty of 1.6\%.  Uncertainties related to the finite number of simulated events, referred to in Table~\ref{tab:uncertainties} as bin-by-bin statistical uncertainties, are taken into account using the Barlow-Beeston "lite" method~\cite{Barlow:1993dm, CMS:2024onh, Conway:2011in}. They are considered for all bins of the distributions used to extract the results. They are uncorrelated across different samples and across bins of a single distribution.

The uncertainty in the predicted yield of the reducible background and theoretical uncertainties in cross sections of the $\qqbar\to \PZ\PZ$, $\Pg\Pg\to \PZ\PZ$, and $\ttbar\PZ$ processes are the dominant factors limiting the sensitivity of the search. Their combined effect reduces sensitivity of the analysis by 15--20\% (20--25\%) in terms of expected upper limits on the rate of the \ggA (\bbA) process.

\begin{table*}[htbp]
\centering
\topcaption{
Dominant sources of systematic uncertainty are considered in this analysis. 
The symbol $\dagger$ indicates uncertainties that affect both the shape and normalization
of the final $\mlltt^\mathrm{cons}$ distributions.
Uncertainties without $\dagger$ affect only normalization. 
The magnitude column indicates an approximation of the associated change in normalization. 
The uncertainties in each group are listed in descending order of their impact on the analysis sensitivity. 
}
\label{tab:uncertainties}
\cmsTable{
\begin{tabular}{lll}
Source of uncertainty & Magnitude & Process\\
\hline
\multicolumn{3}{c}{Experimental uncertainties}\\
 \tauh id.$^{\dagger}$                                 & 2--10\%     & all simulations  \\
 \PGm trigger                              & 2\%          & all simulations  \\
 \PGm id.\ \& isolation                    & 1.5--4.5\% & all simulations  \\
 \Pe trigger                               & 2\%          & all simulations  \\
 \Pe id.\ \& isolation                     & 1.5--4.5\% & all simulations  \\
 limited MC event count                   & bin-by-bin uncertainties & all simulations  \\
 \tauh energy scale$^{\dagger}$            & 0.5--1.5\% & all simulations  \\
 integrated luminosity                    & $<$2\%          & all simulations  \\
 \Pe energy scale$^{\dagger}$           & 1--2\%     & all simulations  \\
 \PQb jet identification efficiency       & 1--4\%  & all simulations \\
 \PQb jet misidentification rate          & 5--10\% & all simulations \\
 jet energy scale$^{\dagger}$             & 1--3\%  & all simulations \\
 \ptvecmiss unclustered energy scale$^{\dagger}$      & 1--3\% & all simulations \\
 jet energy resolution$^{\dagger}$        & $<$1\%      & all simulations \\[\cmsTabSkip]
 
 \multicolumn{3}{c}{Uncertainties in reducible background estimate} \\          
                               &        & misidentified $\tau$ leptons \\
   normalization uncertainty   &  30\%                        &  \etau channel \\
                               &  20\%                        &  \mutau channel \\
                               &  20\%                        &  \tautau channel \\
   event count in AR           &  20--40\%                  & (\btag category) \\
                               &  10--20\%                  & (\zerobtag category) \\ [\cmsTabSkip]
                               
 \multicolumn{3}{c}{Theoretical uncertainties in background estimate}  \\
 $\qqbar\to \PZ\PZ$ normalization                  & 5\%  & $\qqbar\to \PZ\PZ$\\
 $\Pg\Pg\to \PZ\PZ$ normalization                  & 15\% & $\Pg\Pg \to \PZ\PZ$\\
 $\ttbar\PZ$ normalization         & 25\% & $\ttbar\PZ$\\
 triboson    normalization         & 25\% & triboson\\
 \muF and \muR scales    & 1--8\% & Higgs bkg. \\
 theoretical uncertainty in $\mathcal{B}(\Ph\to\PGt\PGt)$  & $<$2\%  & \ggA, \bbA, Higgs bkg. \\
 PDFs     & 1.3--3.6\% & Higgs bkg. \\ [\cmsTabSkip]
 
 \multicolumn{3}{c}{Theoretical uncertainties in the signal estimate (applied in the MSSM interpretation)}  \\
 signal cross section & &   \\
 (\muF, \muR scale, PDFs, \alphas) & 5--20\% (10--25\%) & \ggA (\bbA) \\ 
\end{tabular}}
\end{table*}

\section{Results}
\label{sec:results}

The discriminating variable for this analysis is $\mlltt^{\text{cons}}$, 
the mass of the four-lepton final state where four-vector of the $\Ph\to\PGt\PGt$ candidate is
reconstructed by the {FastMTT} algorithm~\cite{Matyszkiewicz:2025gal} with the mass constraint 
$\mtt=125\GeV$. Six final states corresponding to each combination 
of $\PZ\to \Pe\Pe,\PGm\PGm$ and $\Ph\to\Pe\tauh, \PGm\tauh,\tauh\tauh$ 
decays are considered. Each final state comprises two event categories, \zerobtag and \btag, 
as described in Section~\ref{sec:selection}. Data are examined for the presence of signal using 
the CMS statistical analysis toolkit \textsc{Combine}~\cite{CMS:2024onh}, which is based on 
the \textsc{RooFit}~\cite{Verkerke:2003ir} and \textsc{RooStats}~\cite{Moneta:2010pm} frameworks. 

The signal extraction is performed with a simultaneous binned likelihood fit, combining 36 
$\mlltt^{\text{cons}}$ distributions, which correspond to 6 analyzed final states times 2 event categories times 3 data-taking periods.
The likelihood used to infer the signal has the following form:
  \begin{equation}
    \mathcal{L}\left(\{k_{i}\},\{\mu_{s}\},\{\theta_{j}\}\right) =\prod
    \limits_{i}\mathcal{P}\Bigl(k_{i}|\sum\limits_{s}\mu_{s}\,S_{si}(\{\theta_{j}
    \})+\sum\limits_{b}B_{bi}(\{\theta_{j}\})\Bigr)\,
    \prod\limits_{j}\mathcal{C}(\widetilde{\theta}_{j}|\theta_{j}),
    \label{eq:likelihood}
  \end{equation}

where $\mu_{s}$ denotes the value of the signal strength modifier associated with the signal $s$ (either \ggA or \bbA) that maximizes the likelihood function 
(\ie, the best-fit signal strength), with the signal strength modifier defined as the ratio between the measured signal rate and the rate assumed for the signal model. 
Furthermore,  $i$ labels the bins of the discriminating distributions,
split by the final state, signal category (\zerobtag or \btag),
and data-taking year. The function $\mathcal{P}(k_{i}
|\sum\mu_{s}\,S_{si}(\{\theta_{j}\})+\sum B_{bi}(\{\theta_{j}\}))$ corresponds to 
the Poisson probability to observe $k_{i}$ events in bin $i$ for a prediction of 
$\sum \mu_{s}\,S_{si}$ signal and $\sum B_{bi}$ background events. The predictions 
for $S_{si}$ and $B_{bi}$ are obtained from the signal and background models. 
The parameters $\mu_{s}$ act as linear 
scaling factors of the corresponding signal $s$. Systematic uncertainties are 
incorporated in the form of constraint terms for additional nuisance parameters 
$\{\theta_{j}\}$ in the likelihood, appearing as a product with predefined 
probability density functions $\mathcal{C}(\widetilde{\theta}_{j}|\theta_{j})$, where 
$\widetilde{\theta}_{j}$ corresponds to the nominal value for $\theta_{j}$. The 
predefined uncertainties in the $\widetilde{\theta}_{j}$, as discussed in 
Section~\ref{sec:systematics}, may be constrained by the fit to the 
data. 

The test statistic used for the inference of the signal is the profile likelihood 
ratio as discussed in Ref.~\cite{CMS:2012zhx}: 
  \begin{equation}
    q_{\mu_{s}}=-2\ln
    \left(
    \frac{\mathcal{L}(\left.\{k_{i}\}\right|\sum\limits_{s}\mu_{s}\,S_{si}
    (\{\hat{\theta}_{j,\mu_{s}}\})+\sum\limits_{b}B_{bi}(\{\hat{\theta}_{j,\mu_{s}}\}))}
    {\mathcal{L}(\left.\{k_{i}\}\right|\sum\limits_{s}\hat{\mu}_{s}\,S_{si}(\{\hat{
    \theta}_{j,\hat{\mu}_{s}}\})+\sum\limits_{b}B_{bi}(\{\hat{\theta}_{j,\hat{\mu}_{
    s}}\}))}\right) , \;\; 0\leq \hat{\mu}_{s} \leq \mu_{s},
    \label{eq:profile-likelihood-ratio}
  \end{equation}
where $\mu_{s}$ are tested values of the parameters of interest (POIs); $\hat{\theta}_{j,\mu_{s}}$ are the values of the nuisance parameters that maximize the likelihood function for specified values of $\mu_{s}$, i.e. these are conditional maximum likelihood estimators of $\theta_{j}$ and thus are functions of $\mu_{s}$; 
 $\hat{\mu}_{s}$ and $\hat{\theta}_{j,\hat{\mu}_{s}}$ are unconditional maximum likelihood estimators of POIs and nuisance parameters. 
The index of $q_{\mu_{s}}$ indicates that the test statistic is evaluated for specified values of $\mu_{s}$.

Compared to the unbinned approach, the binned likelihood fit does not require the introduction of analytical probability density functions for description of the background and signal models, thereby simplifying statistical inference. The likelihood function is rigorously defined and properly accounts for the analysis bins with zero observed yields. The background and signal models are represented as binned templates derived from simulated samples or the sideband region in data as described in Section~\ref{sec:background_estimation}. The binning has been optimised to maintain high sensitivity in the entire probed range of \mA while ensuring reasonable statistical population across all bins of the final discriminant in the combined background template, with the statistical uncertainty ranging between 0.5\% and 30\%.

The statistical model includes two POIs: the rate of \ggA production and the rate of \bbA production.
Constraints on the production rates of the signal processes are derived using
the modified frequentist \CLs method~\cite{Junk:1999kv,Read:2002hq}. 

\begin{figure}[hbtp]
\centering
  \includegraphics[width=0.49\textwidth]{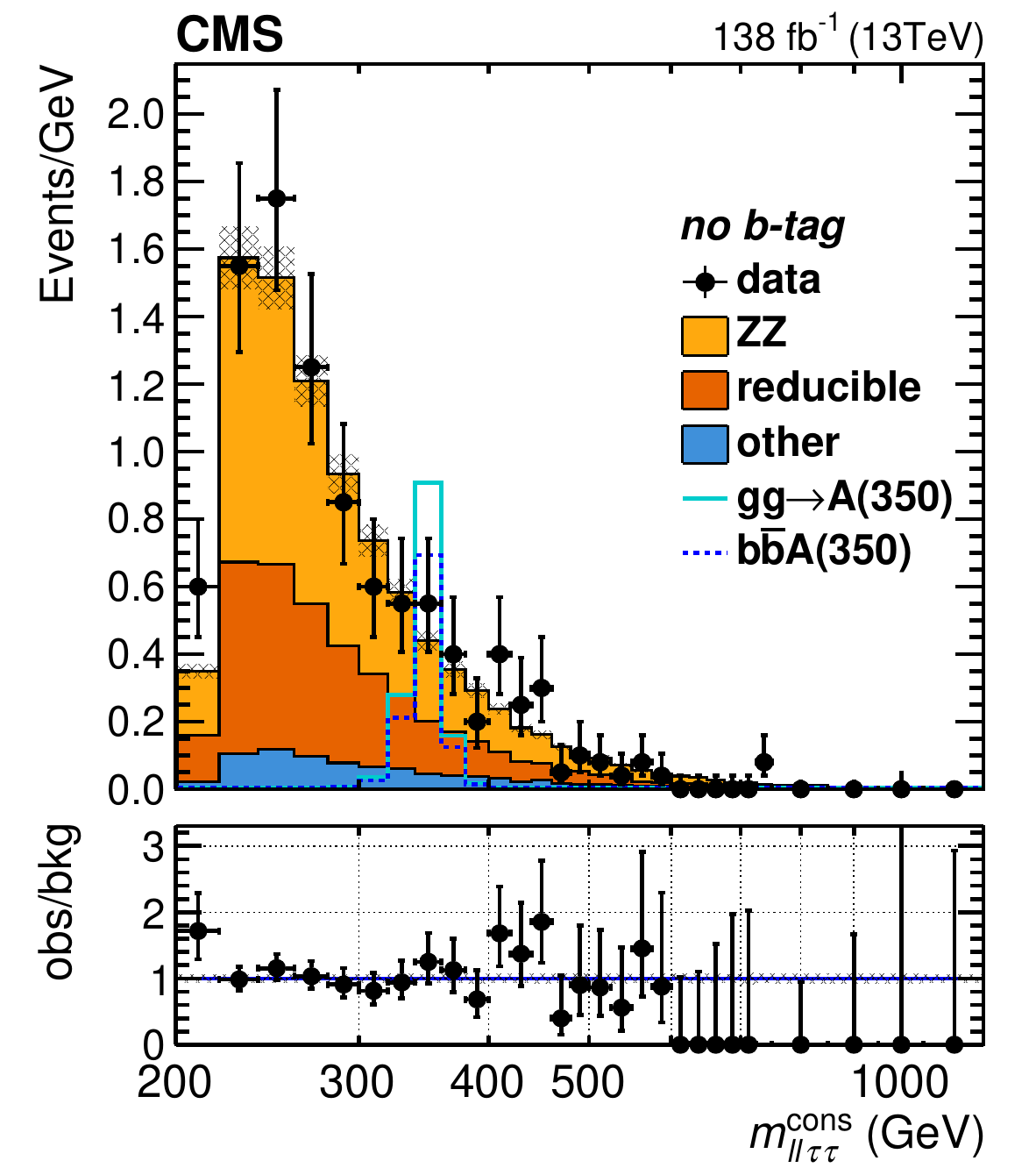}
  \includegraphics[width=0.49\textwidth]{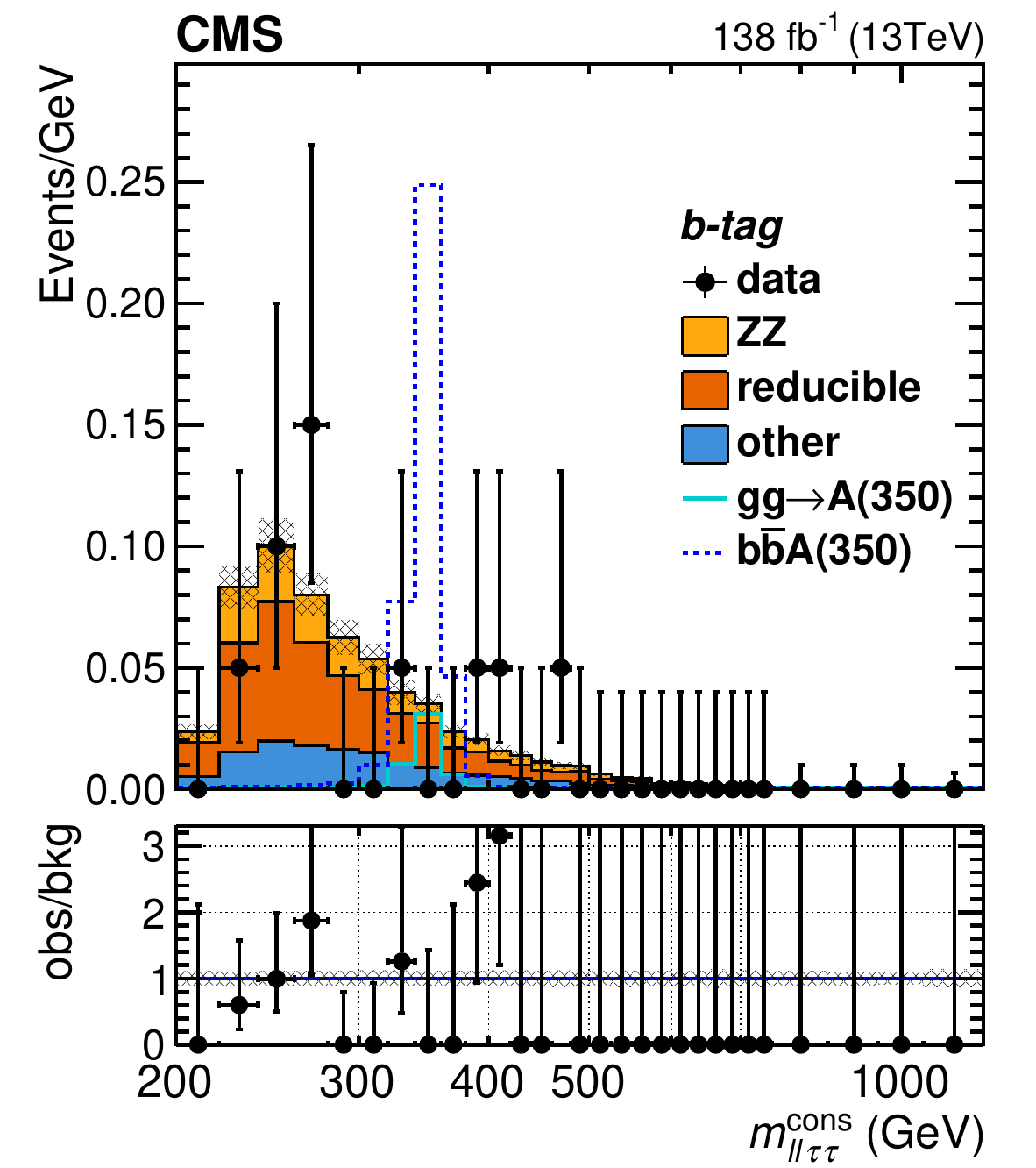}
 \caption{
The reconstructed four-lepton mass, $\mlltt^{\text{cons}}$, in the \zerobtag (\cmsLeft plot) and 
\btag (\cmsRight plot) categories. Background distributions are shown after performing 
a maximum likelihood fit to the data under a background-only hypothesis. Signal samples 
corresponding to the \ggA and \bbA production modes of a pseudoscalar Higgs boson with a 
mass of \mA = 350\GeV, are overlaid to illustrate the expected signal contribution. 
Signal yields are computed by setting $\sigma{\mathcal{B}}(\PA\to\PZ\Ph)$ to a benchmark 
value of 1\unit{pb} for both \ggA and \bbA processes. Hatched bands indicate uncertainties in the 
total background. Contents of each bin, along with the corresponding uncertainties, are divided by the bin width.
}
 \label{fig:mllttcFinal}
\end{figure}

The $\mlltt^{\text{cons}}$ distributions, 
combined across all search channels, 
are presented in Fig.~\ref{fig:mllttcFinal}. The distributions are shown 
separately for \zerobtag and \btag categories. 

Expected and observed event yields in the final selected sample are reported in Table~\ref{tab:contributions}.

\begin{table}[h!]
\begin{center}
\topcaption{Expected and observed yields in the final selected sample. The $\PZ\to\Pe\Pe$ and $\PZ\to\PGm\PGm$ samples and all three data-taking periods are combined for the final results. Numbers are reported individually for \zerobtag and \btag categories and three analyzed di-\PGt decay modes: \etau, \mutau, and \tautau, combining $\PZ\to \mathrm{ee},\mu\mu$ channels and three data-taking years. Background yields and related uncertainties are obtained after 
performing a maximum likelihood fit to the data under a background-only hypothesis. Signal yields are 
computed for representative chosen mass hypotheses of \mA = 250, 350, 500, and 800\GeV, by 
setting $\sigma{\mathcal{B}}(\PA\to\PZ\Ph)$ to a benchmark value of 1\unit{pb} for both the \ggA and \bbA processes.}
\cmsTable{
\begin{tabular}{lcccccc}
\multirow{2}{*}{Process} &\multicolumn{3}{c}{\zerobtag}&\multicolumn{3}{c}{\btag} \\
                         & \etau & \mutau & \tautau & \etau & \mutau & \tautau \\
\hline

$\PZ\PZ$  &  $25.6\pm3.4$ &  $36.4\pm4.1$ &  $41.9\pm5.2$ &  $0.75\pm0.13$ &  $1.09\pm0.18$ &  $1.28\pm0.21$ \\
reducible &  $18.7\pm5.2$ &  $14.8\pm3.4$ &  $33.5\pm8.4$ &  $3.41\pm1.09$ &  $1.69\pm0.49$ &  $1.21\pm0.35$ \\
other     &  $\phantom{2}4.4\pm1.1$ &  $\phantom{3}6.6\pm1.5$ &   $\phantom{4}7.0\pm1.5$ &  $0.99\pm0.26$ &  $1.43\pm0.37$ &  $0.68\pm0.17$ \\
total bkg. &  $48.7\pm6.4$ &  $57.8\pm5.6$ &  $82.4\pm9.9$ &  $5.15\pm1.14$ &  $4.21\pm0.68$ &  $3.17\pm0.47$ \\ [\cmsTabSkip]

observed        & 58               &  57              &  81              & 4   &  2    & 4      \\ [\cmsTabSkip]

\ggA (250\GeV) & 4.70  &  6.84  &  8.66   & 0.16  &  0.24  & 0.30    \\
\bbA (250\GeV) & 3.70  &  5.44  &  7.04   & 1.24  &  1.72  & 2.26    \\
[\cmsTabSkip]
\ggA (350\GeV) & 6.66  &  9.54  &  12.36  & 0.24  &  0.34  & 0.44    \\
\bbA (350\GeV) & 5.10  &  7.34  &  9.40   & 1.88  &  2.64  & 3.40    \\ [\cmsTabSkip]
\ggA (500\GeV) & 8.94  &  12.68 &  16.44  & 0.38  &  0.52  & 0.68    \\
\bbA (500\GeV) & 6.74  &  9.38  &  12.38  & 2.72  &  3.78  & 4.98    \\ [\cmsTabSkip]

\ggA (800\GeV) & 11.90 &  16.76 &  22.20  & 0.56  &  0.76  & 1.06    \\
\bbA (800\GeV) & 8.72  &  12.28 &  16.38  & 3.82  &  5.26  & 7.14    \\ [\cmsTabSkip]
\end{tabular}}
\label{tab:contributions}
\end{center}
\end{table}

It should be noted that \PQb jets in the \bbA process have a relatively soft $\pt$ spectrum, and about 
two-thirds of \bbA events have \PQb jets outside the acceptance of \PQb tagging algorithm. As a consequence, 
about 75\% of selected \bbA events contribute to the \zerobtag category, with the remaining 25\% assigned 
to the \btag category.

\begin{figure}[hbtp]
\centering
  \includegraphics[width=0.49\textwidth]{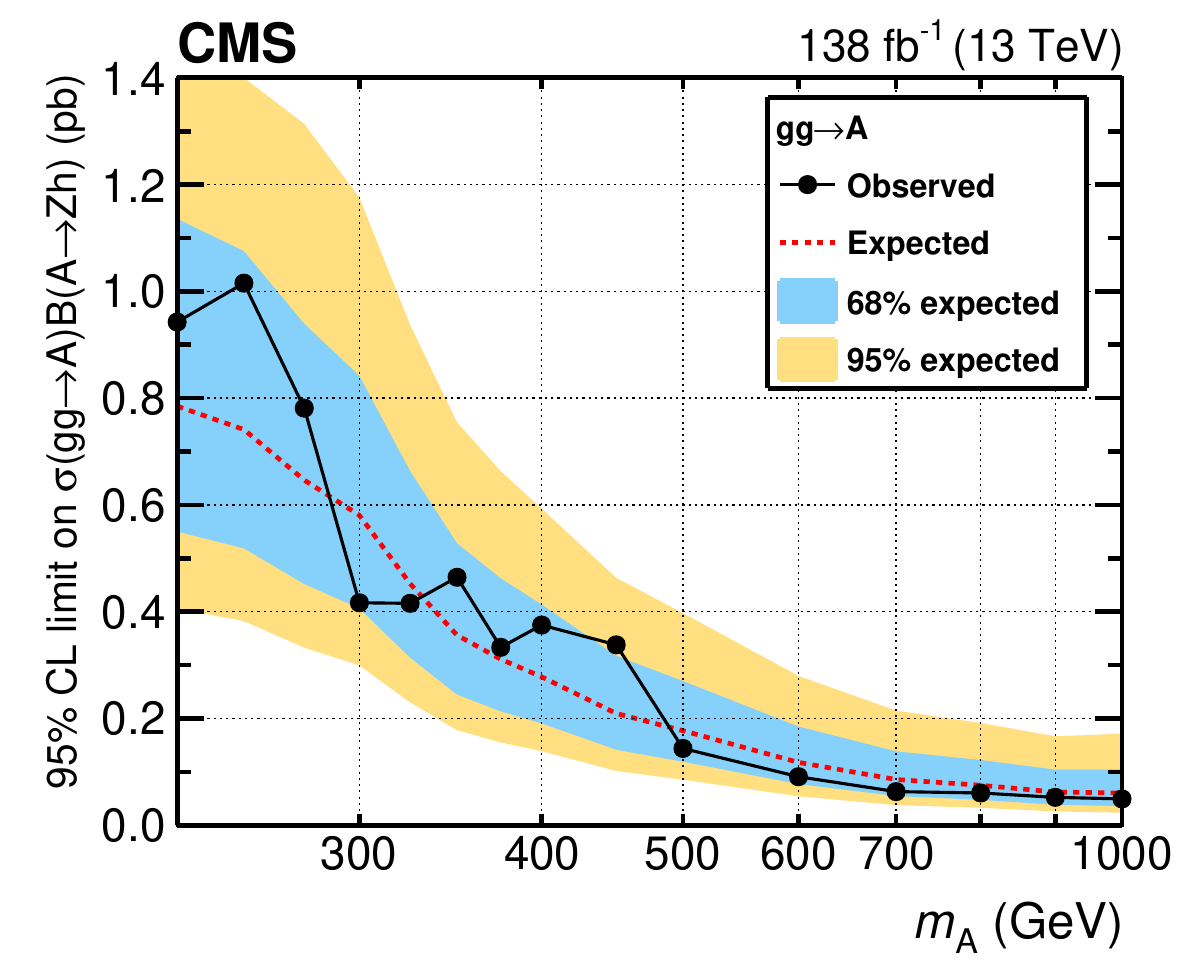}
  \includegraphics[width=0.49\textwidth]{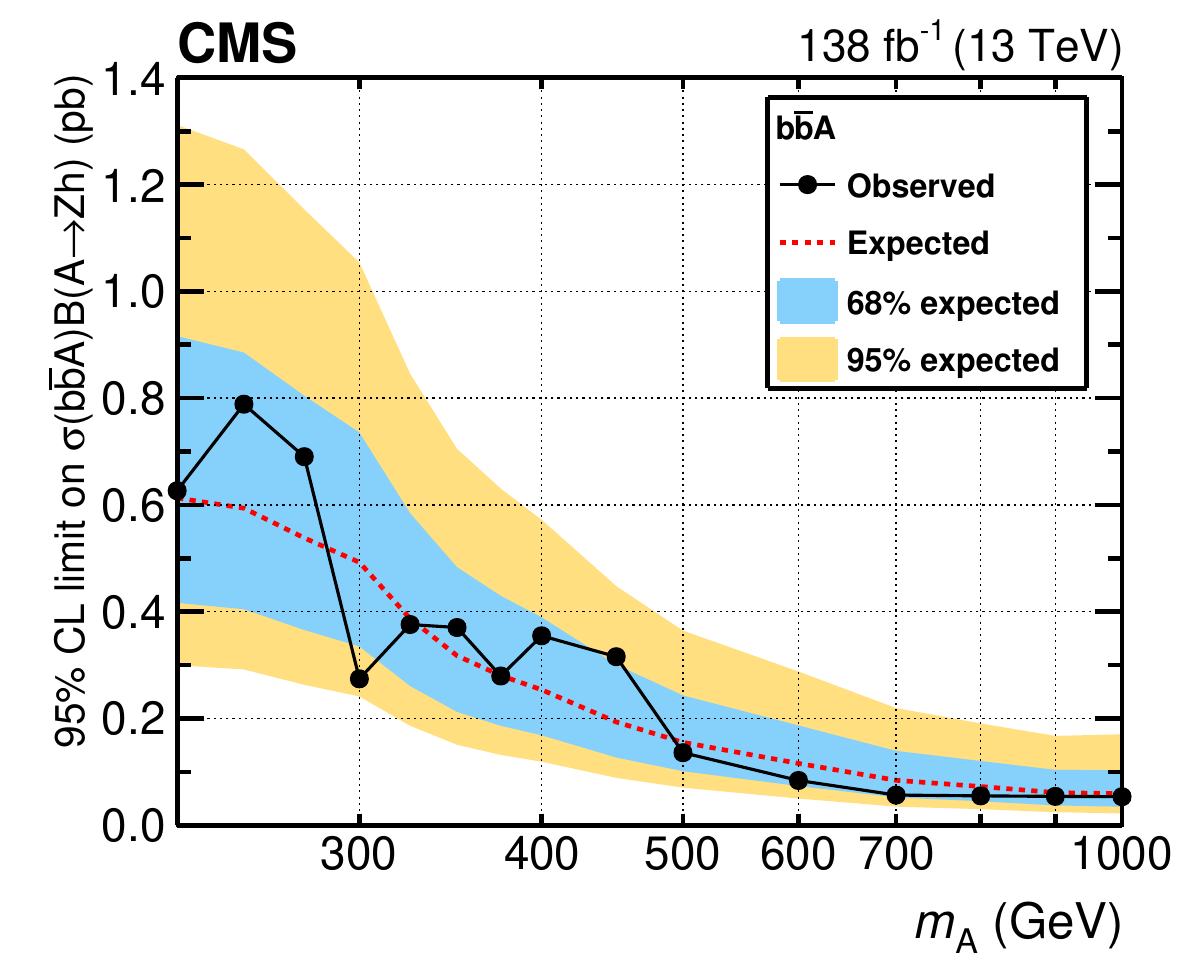}
 \caption{
The expected and observed upper limits at 95\% \CL on the production cross section times branching fraction 
of the $\PA\to\PZ\Ph$ decay for \ggA (\cmsLeft plot) and \bbA (\cmsRight plot) processes as functions of \mA. 
The limits for the \ggA (\bbA) process are derived with the rate of other process fixed to zero. The 
branching fraction of the $\Ph\to\PGt\PGt$ decay is set to the value predicted in the SM, ${\mathcal{B}}(\Ph\to\PGt\PGt)=0.062$~\cite{LHCHiggsCrossSectionWorkingGroup:2016ypw}.
 }
\label{fig:expLimits}
\end{figure}

The statistical inference did not reveal any evidence for the presence of a signal. The compatibility 
of the observed distributions of the  $\mlltt^{\text{cons}}$ discriminant with a null hypothesis is evaluated with a GoF test, 
confronting data against a background-only expectation. The $p$-value returned by a GoF test based on the 
saturated model for the test statistics~\cite{Cousins:2018tiz}, is 0.89. The results of the analysis are used 
to constrain the rate of the signal processes. The upper limits at 95\% confidence level (\CL) on the \PA  
production cross section times branching fraction of $\PA\to\PZ\Ph$ are shown in Fig.~\ref{fig:expLimits} 
for all channels combined. The statistical inference is performed with one POI,
corresponding to the rate of the probed process, whereas the rate of the other process is fixed to zero. This interpretation targets BSM scenarios, where one of the two processes, either \ggA or \bbA, prevails, whereas the contribution of the other process to the signal is negligible. The example of such scenarios is given by Type-II 2HDM, where at low \tanbeta values the \PA boson production is dominated by the \ggA process, while the \bbA process has vanishing cross section. The branching fraction of the $\Ph\to\PGt\PGt$ 
decay is set to the value predicted in the SM, ${\mathcal{B}}(\Ph\to\PGt\PGt)=0.062$~\cite{LHCHiggsCrossSectionWorkingGroup:2016ypw}. The observed (expected) 
limits range from 0.049 (0.060)\unit{pb} at $\mA=1\TeV$ to 1.02 (0.79)\unit{pb} at $\mA=250$~(225)\GeV 
for the \ggA process. For the \bbA process, the observed (expected) limits range from 0.053 (0.059)\unit{pb} 
at $\mA=1\TeV$ to 0.79 (0.61)\unit{pb} at $\mA=250$~(225)\GeV.

\begin{figure*}[hbtp]
  \centering
    \includegraphics[width=0.49\textwidth]{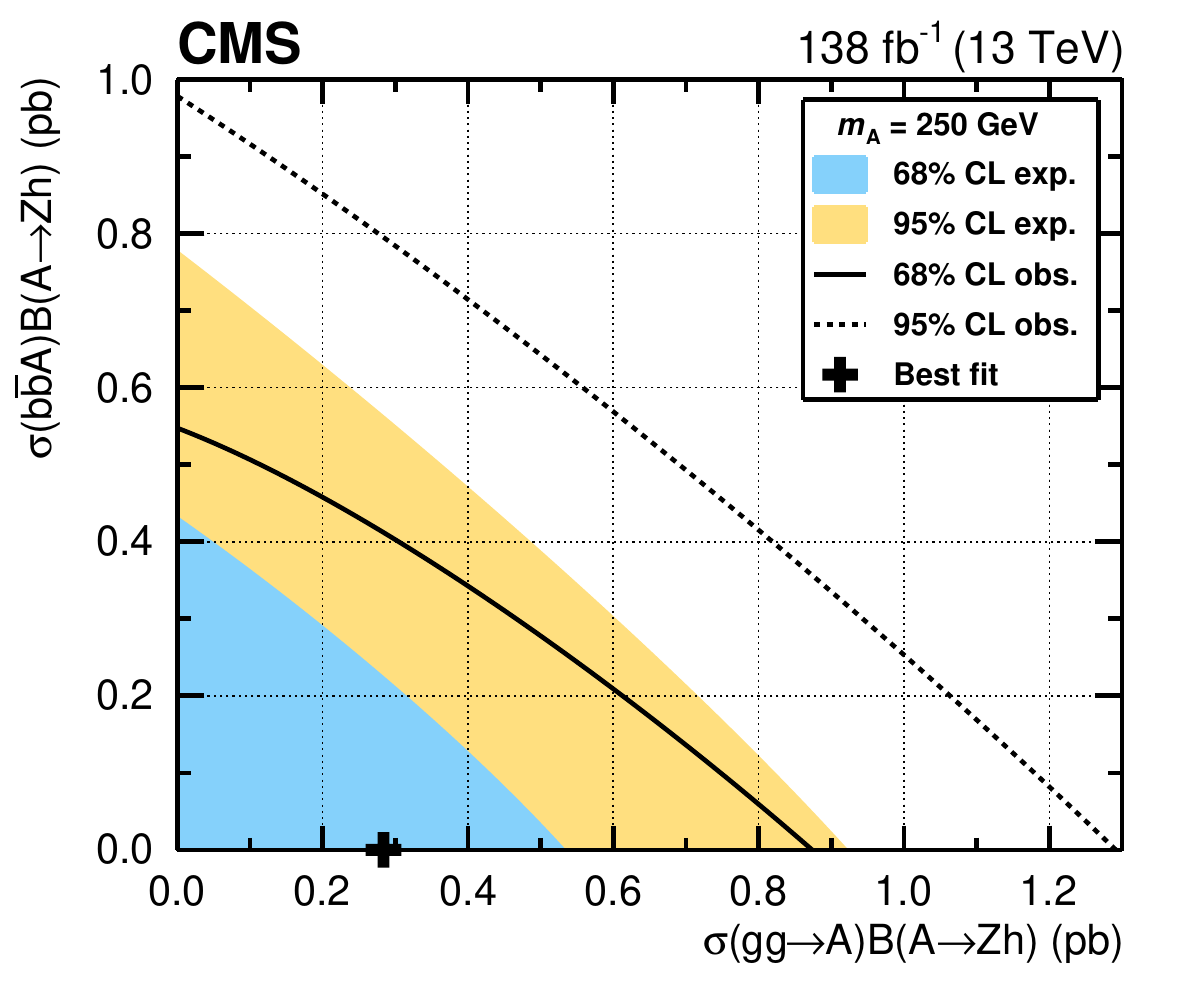}
    \includegraphics[width=0.49\textwidth]{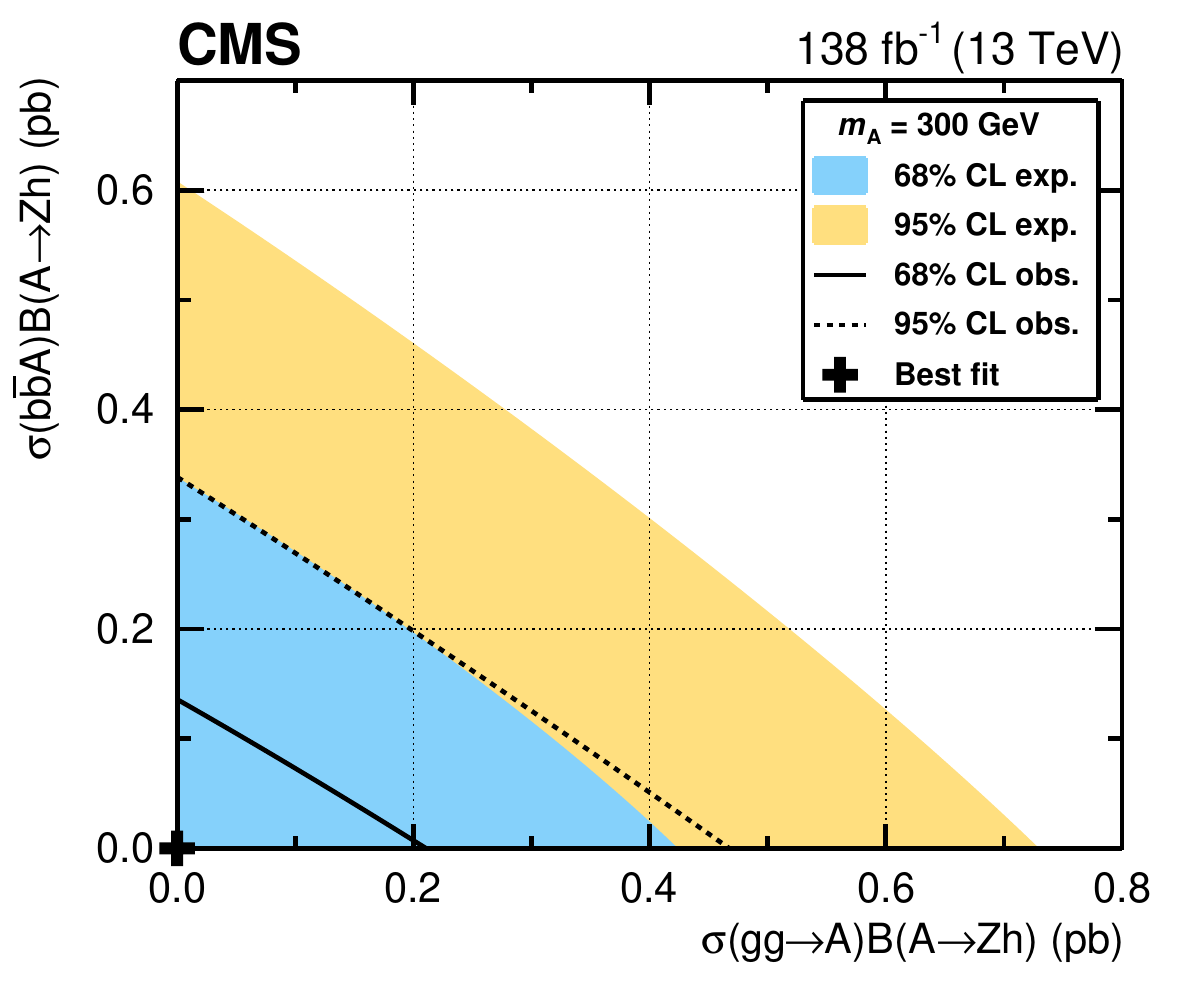}\\
    \includegraphics[width=0.49\textwidth]{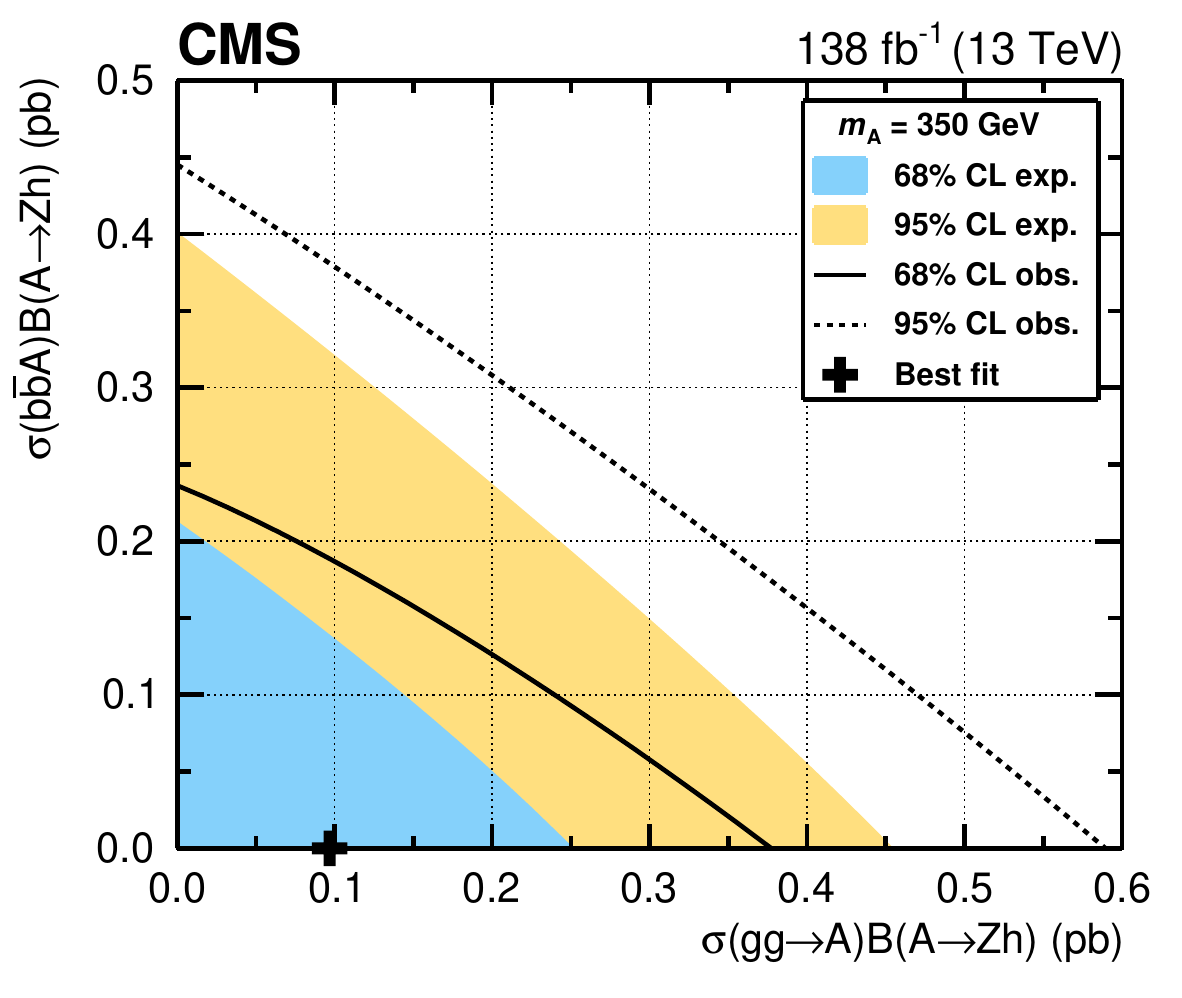}
    \includegraphics[width=0.49\textwidth]{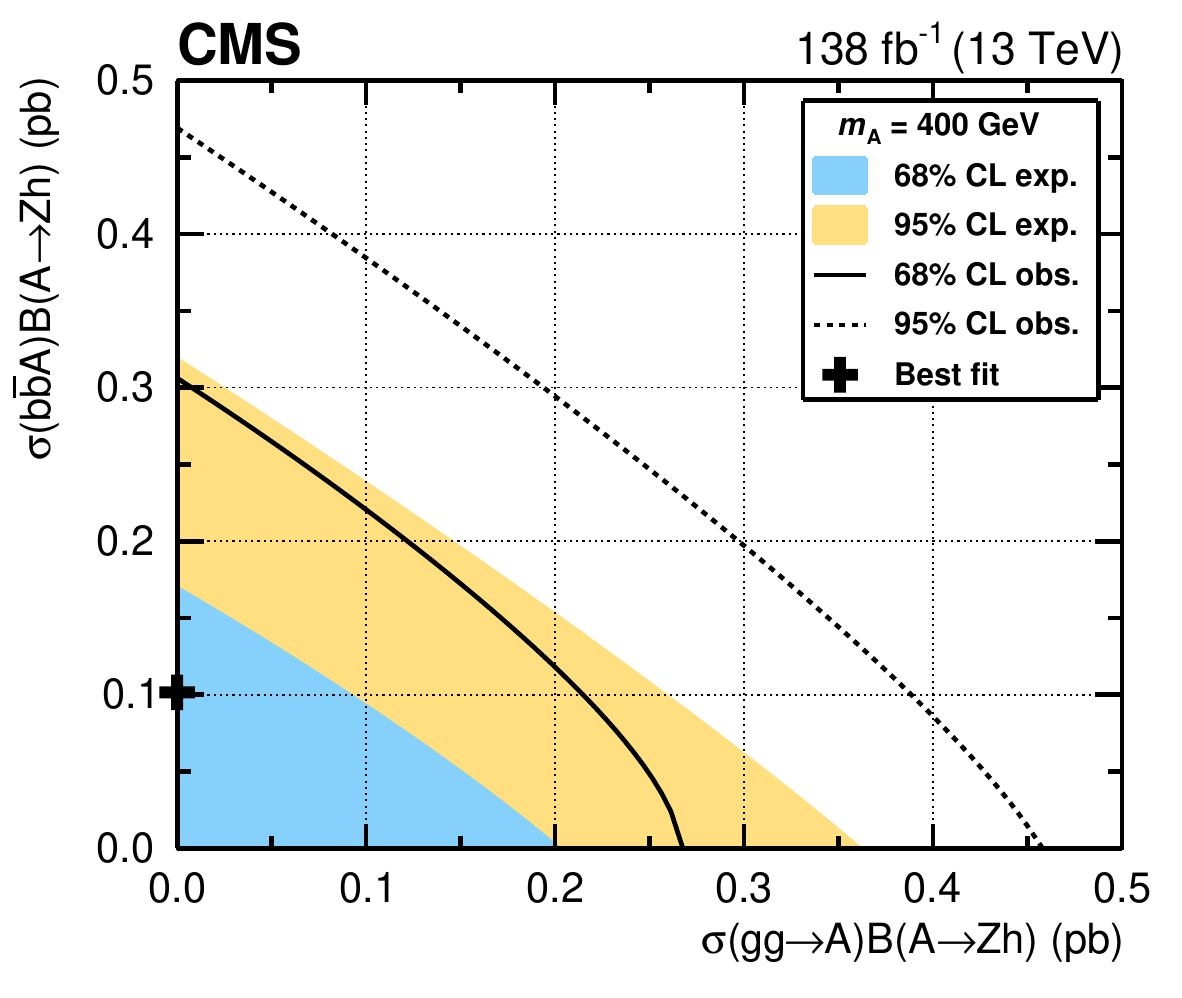}
  \caption{Two-dimensional constraints on the cross section times branching fraction for the two production mechanisms. The confidence level intervals are derived for mass hypotheses of $m_\mathrm{A}=250$ (upper \cmsLeft plot), 300 (upper \cmsRight plot), 350 (lower \cmsLeft plot), and 400\GeV (lower \cmsRight plot). The branching fraction of the $\Ph\to\PGt\PGt$ decay is set to the value predicted in the SM, ${\mathcal{B}}(\Ph\to\PGt\PGt)=0.062$~\cite{LHCHiggsCrossSectionWorkingGroup:2016ypw}. Computation of the best fit point and determination of the observed and expected 68\% and 95\% \CL contours are described in the text.}
  \label{fig:2D_limits_1}
\end{figure*}

\begin{figure*}[hbtp]
\centering
    \includegraphics[width=0.49\textwidth]{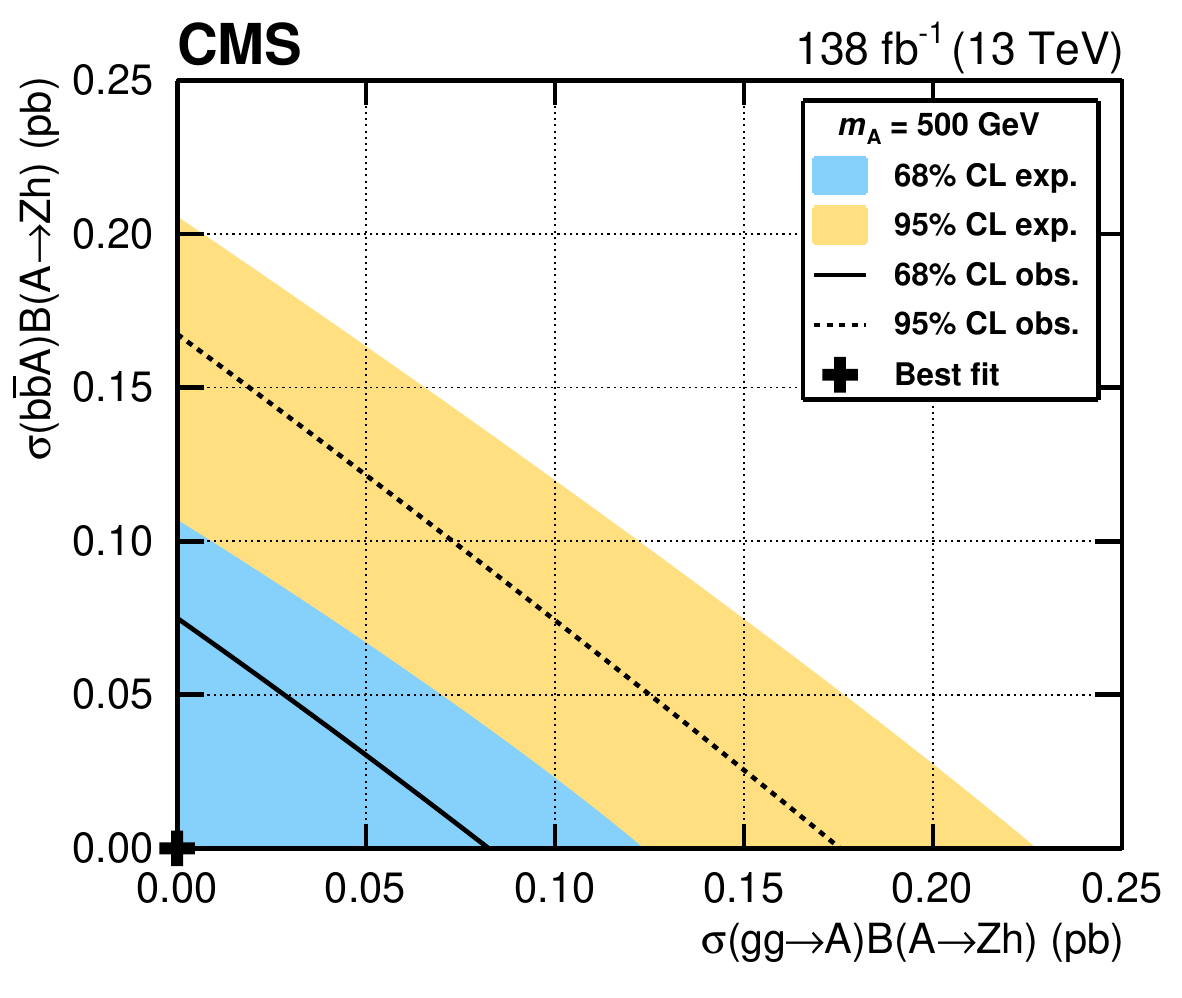}
    \includegraphics[width=0.49\textwidth]{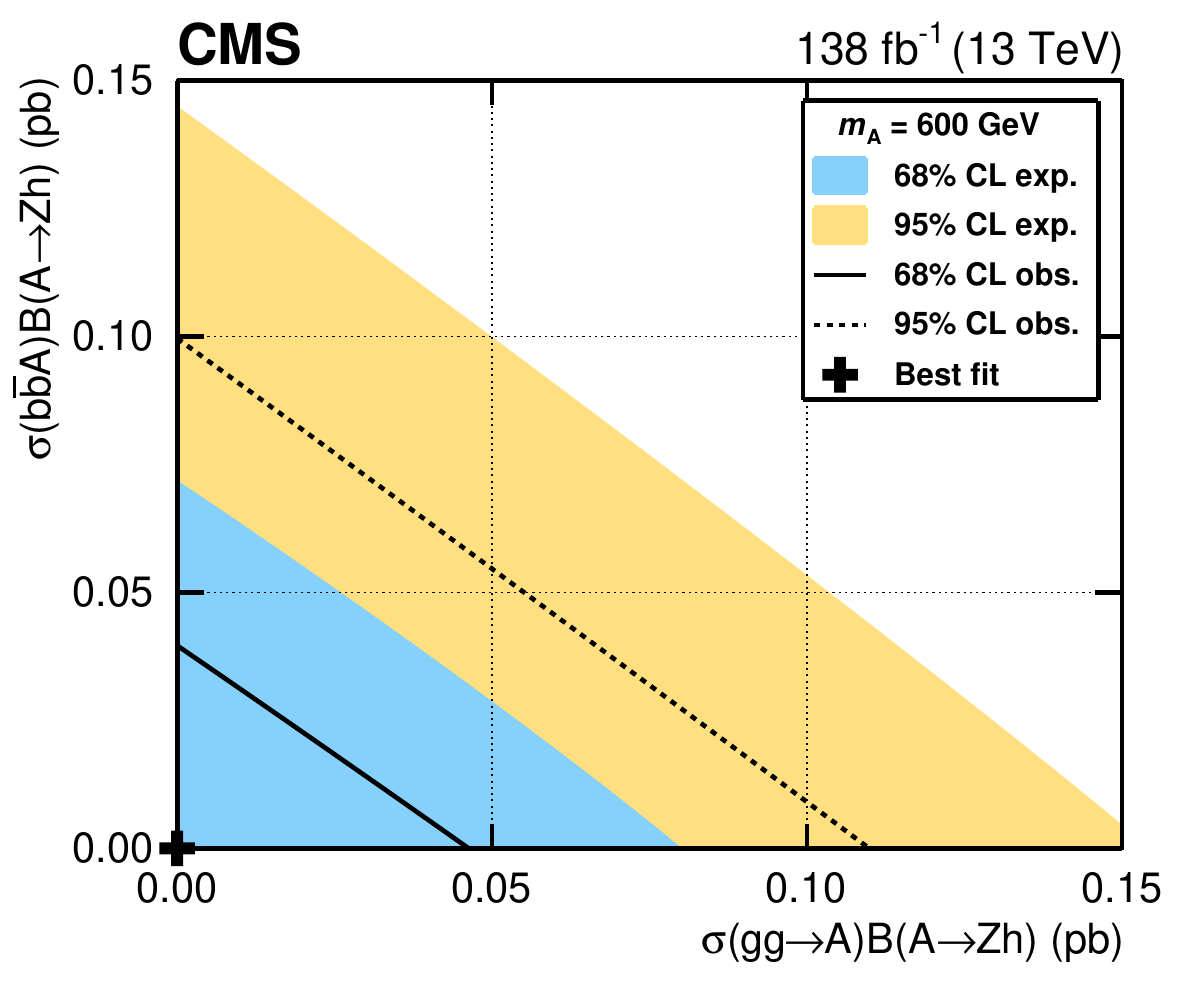}\\
    \includegraphics[width=0.49\textwidth]{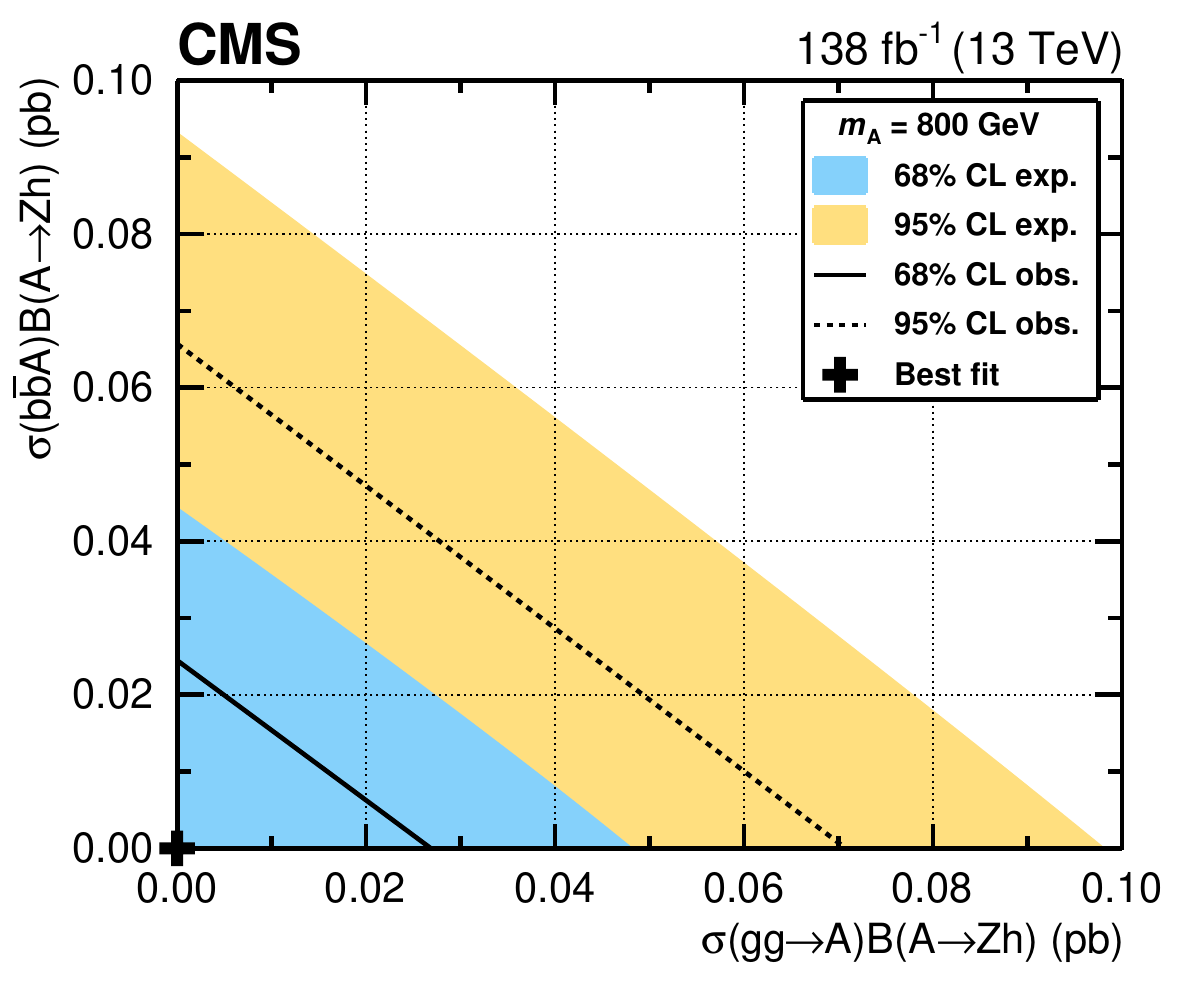}
    \includegraphics[width=0.49\textwidth]{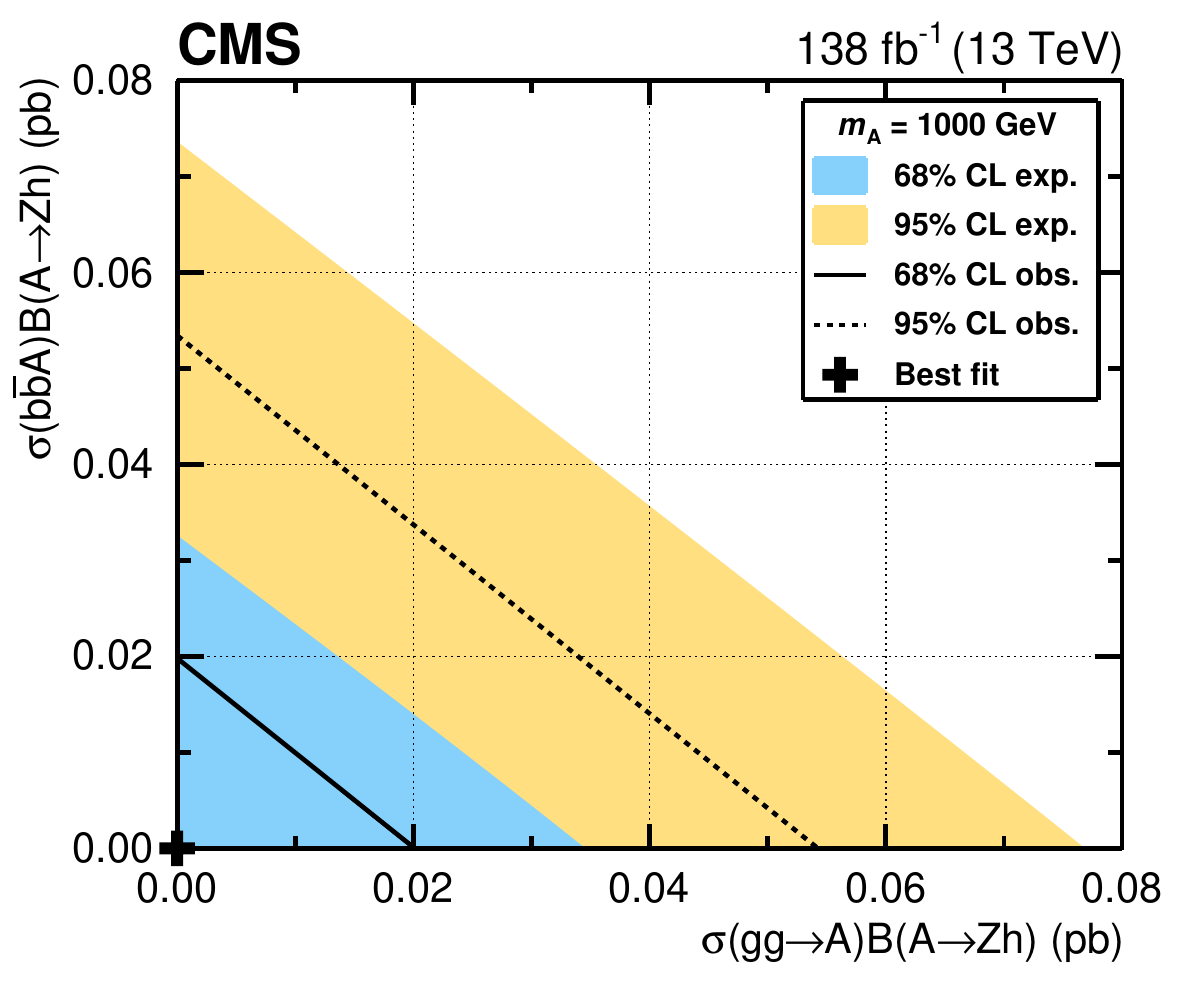}\\

  \caption{Same as Fig.~\ref{fig:2D_limits_1} but for mass hypotheses of $\mA=500$ (upper \cmsLeft plot), $600$ (upper \cmsRight plot), $800$ (lower \cmsLeft plot) and $1000\GeV$ (lower \cmsRight plot). The branching fraction of the $\Ph\to\PGt\PGt$ decay is set to the value predicted in the SM, ${\mathcal{B}}(\Ph\to\PGt\PGt)=0.062$~\cite{LHCHiggsCrossSectionWorkingGroup:2016ypw}. Computation of the best-fit point and determination of the observed and expected 68\% and 95\% \CL contours are described in the text.
  }
  \label{fig:2D_limits_2}
\end{figure*}

Results of the search are also provided in terms of two-dimensional constraints on the cross 
section times branching fraction, $\sigma{\mathcal{B}}(\PA\to\PZ\Ph)$, for the \ggA and \bbA production 
mechanisms. Constraints are derived assuming that the rates of \ggA and \bbA processes are nonnegative. 
Figures~\ref{fig:2D_limits_1} and~\ref{fig:2D_limits_2} present 68\% and 95\% \CL contours for eight 
representative \mA hypotheses. For each \mA, $\sigma{\mathcal{B}}(\PA\to\PZ\Ph)$ values are 
scanned in two dimensions, corresponding to the \ggA and \bbA production mechanisms. At each scanned 
point, \NLL is calculated, defined as the negative-log-likelihood (NLL) of the conditional fit to 
the background-only Asimov and observed datasets. The minimal value of the NLL in the scanned domain of 
nonnegative values of $\sigma(\ggA){\mathcal{B}}(\PA\to\PZ\Ph)$ and $\sigma(\bbA){\mathcal{B}}(\PA\to\PZ\Ph)$ 
defines the best fit point. For each scanned point, the difference between the NLL at a given point and 
the minimal value of NLL, \DNLL, is then computed. The  68\% and 95\% \CL boundaries are 
found at \DNLL values of 2.30 and 5.99, respectively. Maximum-likelihood fits to the background-only 
Asimov dataset are performed to extract 68\% and 95\% \CL expected contours in the absence of signal. 
Fits to data are performed to determine the observed 68\% and 95\% \CL contours.

The results of this analysis are also interpreted as constraints on the parameters  \tanbeta and \mA
within the \BSMEFT MSSM benchmark scenario~\cite{Bahl:2019ago}. For this scenario the Higgs boson 
masses, mixing angle $\alpha$, and effective Yukawa couplings have been calculated 
with the \textsc{FeynHiggs}~\cite{Heinemeyer:1998yj,Heinemeyer:1998np,Degrassi:2002fi,Frank:2006yh,Hahn:2013ria,Bahl:2016brp,Bahl:2017aev,Bahl:2018qog} program. 
Branching fractions for the $\PA\to\PZ\Ph$ and $\Ph\to\PGt\PGt$ decays have 
been obtained from a combination of the  \textsc{FeynHiggs} and
\textsc{hdecay}~\cite{Djouadi:2018xqq} programs, as described in Ref.~\cite{Bagnaschi:2021jaj} following the prescriptions given in Refs.~\cite{LHCHiggsCrossSectionWorkingGroup:2013rie,LHCHiggsCrossSectionWorkingGroup:2016ypw,Denner:2011mq}. For the \ggA process the cross sections are obtained with {\textsc{SusHi}}~1.7.0~\cite{Harlander:2012pb,Harlander:2016hcx}, which includes NLO corrections in \alpS for the \PQt- and \PQb-quark contributions to the cross section~\cite{Spira:1995rr,Harlander:2005rq}, NNLO corrections in \alphas 
in the heavy \PQt quark limit, for the \PQt quark contribution~\cite{Harlander:2002wh,Anastasiou:2002yz,Ravindran:2003um,Harlander:2002vv,Anastasiou:2002wq}, and next-to-NNLO contributions in \alpS for \Ph production~\cite{Anastasiou:2014lda,Anastasiou:2015yha,Anastasiou:2016cez}. Electroweak corrections mediated by light-flavour quarks are included at two-loop accuracy reweighting the SM results 
of Refs.~\cite{Aglietti:2004nj,Bonciani:2010ms}. Cross sections for the \bbA process rely on matched predictions~\cite{Bonvini:2015pxa,Bonvini:2016fgf,Forte:2015hba,Forte:2016sja}, which are based on the calculation at NNLO in \alphas in the five-flavour scheme~\cite{Harlander:2003ai} and the calculation at NLO in \alphas in the four-flavor scheme~\cite{Dittmaier:2003ej,Dawson:2003kb}.

At low \tanbeta values, the $\PA\to\PZ\Ph$ decay dominates 
the natural decay width of \PA  in the \mA range from 220 to 350\GeV. For higher values 
of \mA decays $\PA\to\ttbar$ become dominant. For $\tanbeta\lesssim 4$, the \PA  boson is produced mainly 
via gluon-gluon fusion process, but at higher \tanbeta values, associated production with bottom quarks takes over.

\begin{figure*}[hbtp]
\centering
   \includegraphics[width=0.7\textwidth]{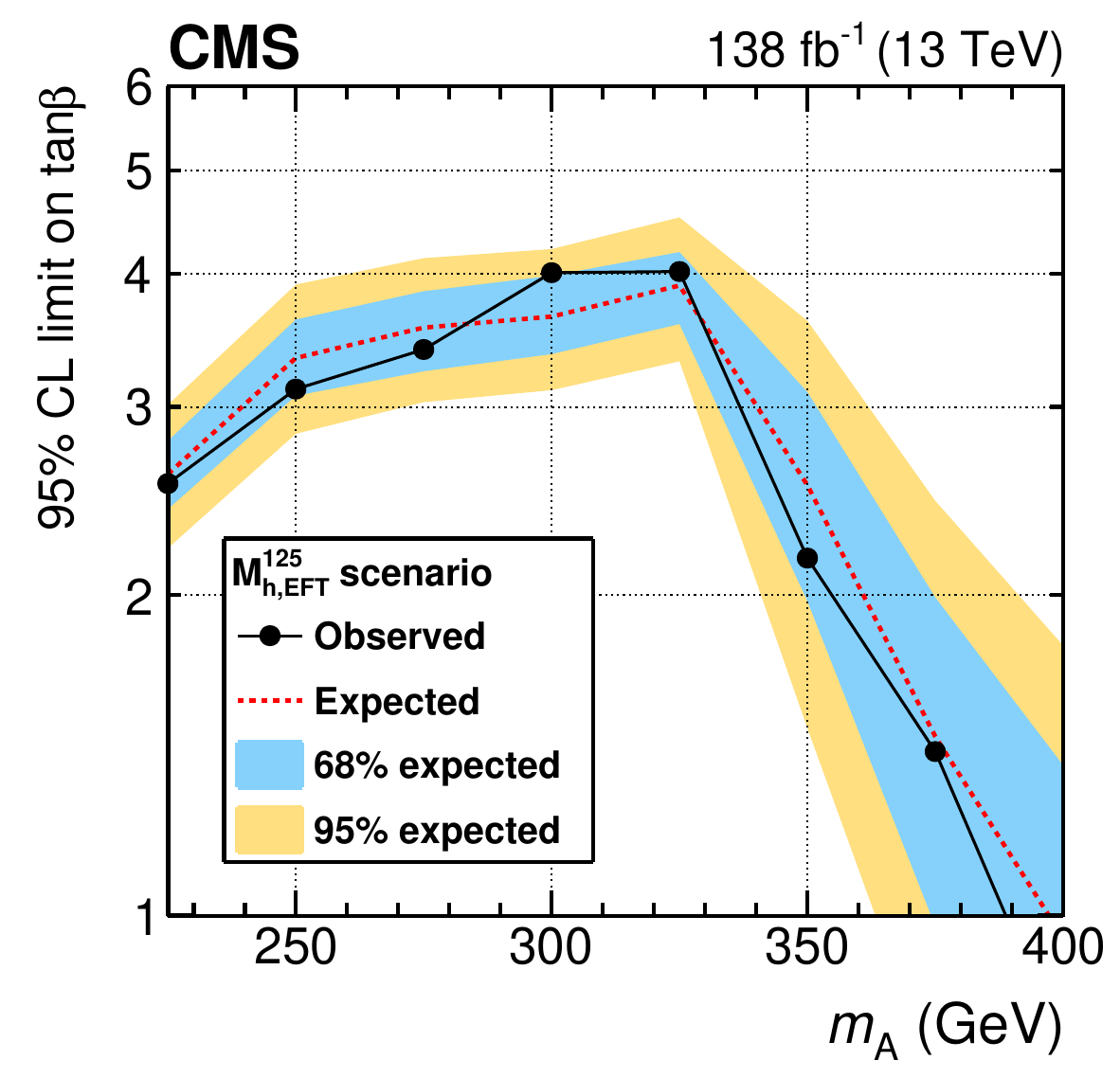}
  \caption{Lower 95\% \CL limit on \tanbeta as a function of \mA in the \BSMEFT MSSM scenario. Values below the black solid line are excluded at 95\% \CL.}
  \label{fig:MSSM_limits}
\end{figure*}

To derive exclusion contours at 95\% \CL, a scan in the $\mA$-$\tanbeta$ plane is performed. Signal yields 
are determined based on the theoretical predictions for $\sigma(\ggA)$, $\sigma(\bbA)$, ${\mathcal{B}}(\PA\to\PZ\Ph)$ and ${\mathcal{B}}(\Ph\to\PGt\PGt)$ at a given tested $\mA$-$\tanbeta$ point, and the \CLs value~\cite{Read:2002hq} is computed. Those points where \CLs falls below 5\% define the 95\% \CL exclusion contour for the benchmark scenario under consideration. Observed and expected lower limits at 95\% \CL on \tanbeta as functions of \mA are presented in Fig.~\ref{fig:MSSM_limits}. Observed (expected) limits on \tanbeta range from 1.4 (1.5) at $\mA=375\GeV$ to 4.0 (3.9) at $\mA=325\GeV$. This analysis excludes \tanbeta values below 2.2 at 95\% \CL in the mass range from 225 to 350\GeV. At higher probed values of \mA, $\PA\to\ttbar$ decay becomes kinematically allowed and suppresses $\PA\to\PZ\Ph$ decay, significantly reducing the sensitivity of this search. 

\section{Summary}
\label{sec:summary}

A search is presented for the decay of a heavy pseudoscalar boson \PA  to a \PZ boson and a 125\GeV Higgs boson, 
\Ph, in final states with two \PGt leptons and two light leptons ($\Pe\Pe$, $\PGm\PGm$). The study is based on 
proton-proton collision data collected by the CMS experiment at $\sqrt{s}=13\TeV$, corresponding to
an integrated luminosity of 138\fbinv. The analysis probes the 
gluon-gluon fusion process, \ggA, and bottom quark associated production, \bbA. No evidence for a signal 
is found in the data. Upper limits at 95\% confidence level are derived on the product of the cross section 
and branching fraction of the $\PA\to\PZ\Ph$ decay under the assumption that the scalar state \Ph has 
the properties of the 125\GeV SM Higgs boson. Observed limits range from 0.049 (0.053)\unit{pb} at 
$\mA=1\TeV$ to 1.02 (0.79)\unit{pb} at $\mA=250\GeV$ for the \ggA (\bbA) process.

The results of the search are also interpreted in terms of constraints on \tanbeta as a function of \mA 
within the \BSMEFT MSSM benchmark scenario. Values of \tanbeta below 2.2 are excluded at 95\% \CL in the 
mass range of $225<\mA<350\GeV$.

The present analysis supersedes the previous search for the $\PA\to\PZ\Ph$ decay carried 
out by the CMS Collaboration in the $(\PZ\to\PGn\PAGn/\Pell\Pell)(\Ph\to\bbbar)$ 
and $(\PZ\to\Pell\Pell)(\Ph\to\PGt\PGt)$ channels (where $\Pell=\Pe,\PGm$)~\cite{Sirunyan:2019xls,CMS:2019kca} 
on proton-proton collision data collected at $\sqrt{s}=13$\TeV and corresponding
to an integrated luminosity of 36\fbinv.

\begin{acknowledgments}
We congratulate our colleagues in the CERN accelerator departments for the excellent performance of the LHC and thank the technical and administrative staffs at CERN and at other CMS institutes for their contributions to the success of the CMS effort. In addition, we gratefully acknowledge the computing centres and personnel of the Worldwide LHC Computing Grid and other centres for delivering so effectively the computing infrastructure essential to our analyses. Finally, we acknowledge the enduring support for the construction and operation of the LHC, the CMS detector, and the supporting computing infrastructure provided by the following funding agencies: SC (Armenia), BMBWF and FWF (Austria); FNRS and FWO (Belgium); CNPq, CAPES, FAPERJ, FAPERGS, and FAPESP (Brazil); MES and BNSF (Bulgaria); CERN; CAS, MoST, and NSFC (China); MINCIENCIAS (Colombia); MSES and CSF (Croatia); RIF (Cyprus); SENESCYT (Ecuador); ERC PRG, RVTT3 and MoER TK202 (Estonia); Academy of Finland, MEC, and HIP (Finland); CEA and CNRS/IN2P3 (France); SRNSF (Georgia); BMBF, DFG, and HGF (Germany); GSRI (Greece); NKFIH (Hungary); DAE and DST (India); IPM (Iran); SFI (Ireland); INFN (Italy); MSIP and NRF (Republic of Korea); MES (Latvia); LMTLT (Lithuania); MOE and UM (Malaysia); BUAP, CINVESTAV, CONACYT, LNS, SEP, and UASLP-FAI (Mexico); MOS (Montenegro); MBIE (New Zealand); PAEC (Pakistan); MES and NSC (Poland); FCT (Portugal);  MESTD (Serbia); MCIN/AEI and PCTI (Spain); MOSTR (Sri Lanka); Swiss Funding Agencies (Switzerland); MST (Taipei); MHESI and NSTDA (Thailand); TUBITAK and TENMAK (Turkey); NASU (Ukraine); STFC (United Kingdom); DOE and NSF (USA).

\hyphenation{Rachada-pisek} Individuals have received support from the Marie-Curie programme and the European Research Council and Horizon 2020 Grant, contract Nos.\ 675440, 724704, 752730, 758316, 765710, 824093, 101115353, 101002207, and COST Action CA16108 (European Union); the Leventis Foundation; the Alfred P.\ Sloan Foundation; the Alexander von Humboldt Foundation; the Science Committee, project no. 22rl-037 (Armenia); the Fonds pour la Formation \`a la Recherche dans l'Industrie et dans l'Agriculture (FRIA-Belgium); the Beijing Municipal Science \& Technology Commission, No. Z191100007219010 and Fundamental Research Funds for the Central Universities (China); the Ministry of Education, Youth and Sports (MEYS) of the Czech Republic; the Shota Rustaveli National Science Foundation, grant FR-22-985 (Georgia); the Deutsche Forschungsgemeinschaft (DFG), among others, under Germany's Excellence Strategy -- EXC 2121 ``Quantum Universe" -- 390833306, and under project number 400140256 - GRK2497; the Hellenic Foundation for Research and Innovation (HFRI), Project Number 2288 (Greece); the Hungarian Academy of Sciences, the New National Excellence Program - \'UNKP, the NKFIH research grants K 131991, K 133046, K 138136, K 143460, K 143477, K 146913, K 146914, K 147048, 2020-2.2.1-ED-2021-00181, TKP2021-NKTA-64, and 2021-4.1.2-NEMZ\_KI-2024-00036 (Hungary); the Council of Science and Industrial Research, India; ICSC -- National Research Centre for High Performance Computing, Big Data and Quantum Computing and FAIR -- Future Artificial Intelligence Research, funded by the NextGenerationEU program (Italy); the Latvian Council of Science; the Ministry of Education and Science, project no. 2022/WK/14, and the National Science Center, contracts Opus 2021/41/B/ST2/01369 and 2021/43/B/ST2/01552 (Poland); the Funda\c{c}\~ao para a Ci\^encia e a Tecnologia, grant CEECIND/01334/2018 (Portugal); the National Priorities Research Program by Qatar National Research Fund;  MCIN/AEI/10.13039/501100011033, ERDF ``a way of making Europe", and the Programa Estatal de Fomento de la Investigaci{\'o}n Cient{\'i}fica y T{\'e}cnica de Excelencia Mar\'{\i}a de Maeztu, grant MDM-2017-0765 and Programa Severo Ochoa del Principado de Asturias (Spain); the Chulalongkorn Academic into Its 2nd Century Project Advancement Project, and the National Science, Research and Innovation Fund via the Program Management Unit for Human Resources \& Institutional Development, Research and Innovation, grant B39G670016 (Thailand); the Kavli Foundation; the Nvidia Corporation; the SuperMicro Corporation; the Welch Foundation, contract C-1845; and the Weston Havens Foundation (USA).
\end{acknowledgments}

\bibliography{auto_generated} 

\providecommand{\href}[2]{#2}\begingroup\raggedright\begin{thebibliography}{100}%
\makeatletter
\providecommand{\hrefCMSnoop }[0]{\@secondoftwo}%
\makeatother
\providecommand{\doi}{\texttt{doi:}\begingroup \urlstyle{tt}\Url}

\bibitem{Aad:2012tfa}
\hrefCMSnoop {}{{ATLAS Collaboration}, ``Observation of a new particle in the
  search for the standard model {Higgs} boson with the {ATLAS} detector at the
  {LHC}'',} \textit{ Phys. Lett. B} \textbf{ 716} (2012) 1,
  \href{http://dx.doi.org/10.1016/j.physletb.2012.08.020}{\doi{10.1016/j.physletb.2012.08.020}},
\href{http://www.arXiv.org/abs/1207.7214}{\texttt{arXiv:1207.7214}}.

\bibitem{Chatrchyan:2012xdj}
\hrefCMSnoop {}{{CMS Collaboration}, ``Observation of a new boson at a mass of
  125 {GeV} with the {CMS} experiment at the {LHC}'',} \textit{ Phys. Lett. B}
  \textbf{ 716} (2012) 30,
  \href{http://dx.doi.org/10.1016/j.physletb.2012.08.021}{\doi{10.1016/j.physletb.2012.08.021}},
\href{http://www.arXiv.org/abs/1207.7235}{\texttt{arXiv:1207.7235}}.

\bibitem{Chatrchyan:2013lba}
\hrefCMSnoop {}{{CMS Collaboration}, ``Observation of a new boson with mass
  near {$125\GeV$} in {$\Pp\Pp$} collisions at $\sqrt{s} = 7$ and {$8\TeV$}'',}
  \textit{ JHEP} \textbf{ 06} (2013) 081,
  \href{http://dx.doi.org/10.1007/JHEP06(2013)081}{\doi{10.1007/JHEP06(2013)081}},
\href{http://www.arXiv.org/abs/1303.4571}{\texttt{arXiv:1303.4571}}.

\bibitem{ATLAS:2022vkf}
\hrefCMSnoop {}{{ATLAS Collaboration}, ``{A detailed map of Higgs boson
  interactions by the ATLAS experiment ten years after the discovery}'',}
  \textit{ Nature} \textbf{ 607} (2022) 52,
  \href{http://dx.doi.org/10.1038/s41586-022-04893-w}{\doi{10.1038/s41586-022-04893-w}},
  \href{http://www.arXiv.org/abs/2207.00092}{\texttt{arXiv:2207.00092}}.
  [Erratum: \DOI{10.1038/s41586-022-05581-5}].

\bibitem{CMS:2022dwd}
\hrefCMSnoop {}{{CMS Collaboration}, ``{A portrait of the Higgs boson by the
  CMS experiment ten years after the discovery.}'',} \textit{ Nature} \textbf{
  607} (2022) 60,
  \href{http://dx.doi.org/10.1038/s41586-022-04892-x}{\doi{10.1038/s41586-022-04892-x}},
  \href{http://www.arXiv.org/abs/2207.00043}{\texttt{arXiv:2207.00043}}.
  [Author correction: \DOI{10.1038/s41586-023-06164-8}].

\bibitem{Khachatryan:2016vau}
\hrefCMSnoop {}{{ATLAS and CMS Collaborations}, ``Measurements of the {Higgs}
  boson production and decay rates and constraints on its couplings from a
  combined {ATLAS} and {CMS} analysis of the {LHC} pp collision data at
  {$\sqrt{s}=7$ and 8 TeV}'',} \textit{ JHEP} \textbf{ 08} (2016) 045,
  \href{http://dx.doi.org/10.1007/JHEP08(2016)045}{\doi{10.1007/JHEP08(2016)045}},
  \href{http://www.arXiv.org/abs/1606.02266}{\texttt{arXiv:1606.02266}}.

\bibitem{Sirunyan:2018koj}
\hrefCMSnoop {}{{CMS Collaboration}, ``Combined measurements of {Higgs} boson
  couplings in proton\textendash{}proton collisions at {$\sqrt{s}=13\,\text
  {Te}\text {V} $}'',} \textit{ Eur. Phys. J. C} \textbf{ 79} (2019) 421,
  \href{http://dx.doi.org/10.1140/epjc/s10052-019-6909-y}{\doi{10.1140/epjc/s10052-019-6909-y}},
  \href{http://www.arXiv.org/abs/1809.10733}{\texttt{arXiv:1809.10733}}.

\bibitem{Aad:2019mbh}
\hrefCMSnoop {}{{ATLAS Collaboration}, ``Combined measurements of {Higgs} boson
  production and decay using up to {$80$ fb$^{-1}$} of proton-proton collision
  data at {$\sqrt{s}=$ 13 TeV} collected with the {ATLAS experiment}'',}
  \textit{ Phys. Rev. D} \textbf{ 101} (2020) 012002,
  \href{http://dx.doi.org/10.1103/PhysRevD.101.012002}{\doi{10.1103/PhysRevD.101.012002}},
  \href{http://www.arXiv.org/abs/1909.02845}{\texttt{arXiv:1909.02845}}.

\bibitem{Sirunyan:2019twz}
\hrefCMSnoop {}{{CMS Collaboration}, ``Measurements of the {Higgs} boson width
  and anomalous {$\PH\PV\PV$} couplings from on-shell and off-shell production
  in the four-lepton final state'',} \textit{ Phys. Rev. D} \textbf{ 99} (2019)
  112003,
  \href{http://dx.doi.org/10.1103/PhysRevD.99.112003}{\doi{10.1103/PhysRevD.99.112003}},
  \href{http://www.arXiv.org/abs/1901.00174}{\texttt{arXiv:1901.00174}}.

\bibitem{CMS:2020xrn}
\hrefCMSnoop {}{{CMS Collaboration}, ``{A measurement of the Higgs boson mass
  in the diphoton decay channel}'',} \textit{ Phys. Lett. B} \textbf{ 805}
  (2020) 135425,
  \href{http://dx.doi.org/10.1016/j.physletb.2020.135425}{\doi{10.1016/j.physletb.2020.135425}},
  \href{http://www.arXiv.org/abs/2002.06398}{\texttt{arXiv:2002.06398}}.

\bibitem{PhysRevD.8.1226}
\hrefCMSnoop {}{T.~D. Lee, ``A theory of spontaneous {T} violation'',} \textit{
  Phys. Rev. D} \textbf{ 8} (1973) 1226,
  \href{http://dx.doi.org/10.1103/PhysRevD.8.1226}{\doi{10.1103/PhysRevD.8.1226}}.

\bibitem{Branco:2011iw}
G.~C. Branco\hrefCMSnoop {}{ { et~al.}, ``Theory and phenomenology of
  two-{H}iggs-doublet models'',} \textit{ Phys. Rept.} \textbf{ 516} (2012) 1,
  \href{http://dx.doi.org/10.1016/j.physrep.2012.02.002}{\doi{10.1016/j.physrep.2012.02.002}},
  \href{http://www.arXiv.org/abs/1106.0034}{\texttt{arXiv:1106.0034}}.

\bibitem{Fromme:2006cm}
\hrefCMSnoop {}{L.~Fromme, S.~J. Huber, and M.~Seniuch, ``{Baryogenesis in the
  two-Higgs doublet model}'',} \textit{ JHEP} \textbf{ 11} (2006) 038,
  \href{http://dx.doi.org/10.1088/1126-6708/2006/11/038}{\doi{10.1088/1126-6708/2006/11/038}},
\href{http://www.arXiv.org/abs/hep-ph/0605242}{\texttt{arXiv:hep-ph/0605242}}.

\bibitem{Fayet:1974pd}
\hrefCMSnoop {}{P.~Fayet, ``{Supergauge invariant extension of the Higgs
  mechanism and a model for the electron and its neutrino}'',} \textit{ Nucl.
  Phys. B} \textbf{ 90} (1975) 104,
  \href{http://dx.doi.org/10.1016/0550-3213(75)90636-7}{\doi{10.1016/0550-3213(75)90636-7}}.

\bibitem{Fayet:1977yc}
\hrefCMSnoop {}{P.~Fayet, ``Spontaneously broken supersymmetric theories of
  weak, electromagnetic and strong interactions'',} \textit{ Phys. Lett. B}
  \textbf{ 69} (1977) 489,
\href{http://dx.doi.org/10.1016/0370-2693(77)90852-8}{\doi{10.1016/0370-2693(77)90852-8}}.

\bibitem{Haber:1984rc}
\hrefCMSnoop {}{H.~E. Haber and G.~L. Kane, ``{The search for Supersymmetry:
  probing physics beyond the Standard Model}'',} \textit{ Phys. Rept.} \textbf{
  117} (1985) 75,
  \href{http://dx.doi.org/10.1016/0370-1573(85)90051-1}{\doi{10.1016/0370-1573(85)90051-1}}.

\bibitem{Bahl:2019ago}
\hrefCMSnoop {}{H.~Bahl, S.~Liebler, and T.~Stefaniak, ``{MSSM Higgs} benchmark
  scenarios for {Run 2} and beyond: the low {$\tan\beta$} region'',} \textit{
  Eur. Phys. J. C} \textbf{ 79} (2019) 279,
  \href{http://dx.doi.org/10.1140/epjc/s10052-019-6770-z}{\doi{10.1140/epjc/s10052-019-6770-z}},
  \href{http://www.arXiv.org/abs/1901.05933}{\texttt{arXiv:1901.05933}}.

\bibitem{Aad:2015wra}
\hrefCMSnoop {}{{ATLAS Collaboration}, ``{Search} for a {CP-odd} {Higgs} boson
  decaying to {Zh} in pp collisions at $\sqrt{s} = 8$ {TeV} with the {ATLAS}
  detector'',} \textit{ Phys. Lett. B} \textbf{ 744} (2015) 163,
  \href{http://dx.doi.org/10.1016/j.physletb.2015.03.054}{\doi{10.1016/j.physletb.2015.03.054}},
\href{http://www.arXiv.org/abs/1502.04478}{\texttt{arXiv:1502.04478}}.

\bibitem{Khachatryan:2015tha}
\hrefCMSnoop {}{{CMS Collaboration}, ``{Searches} for a heavy scalar boson {H}
  decaying to a pair of 125 {GeV} {Higgs} bosons hh or for a heavy pseudoscalar
  boson {A} decaying to {Zh}, in the final states with $h \to \tau \tau$'',}
  \textit{ Phys. Lett. B} \textbf{ 755} (2016) 217,
  \href{http://dx.doi.org/10.1016/j.physletb.2016.01.056}{\doi{10.1016/j.physletb.2016.01.056}},
\href{http://www.arXiv.org/abs/1510.01181}{\texttt{arXiv:1510.01181}}.

\bibitem{CMS:2019kca}
\hrefCMSnoop {}{{CMS Collaboration}, ``{Search for a heavy pseudoscalar Higgs
  boson decaying into a 125 GeV Higgs boson and a Z boson in final states with
  two tau and two light leptons at $\sqrt{s}=$ 13 TeV}'',} \textit{ JHEP}
  \textbf{ 03} (2020) 065,
  \href{http://dx.doi.org/10.1007/JHEP03(2020)065}{\doi{10.1007/JHEP03(2020)065}},
  \href{http://www.arXiv.org/abs/1910.11634}{\texttt{arXiv:1910.11634}}.

\bibitem{Aaboud:2017cxo}
\hrefCMSnoop {}{{ATLAS Collaboration}, ``{Search} for heavy resonances decaying
  into a {$W$} or {$Z$} boson and a {Higgs} boson in final states with leptons
  and $b$-jets in 36 fb$^{-1}$ of $\sqrt s = 13$ {TeV} $pp$ collisions with the
  {ATLAS} detector'',} \textit{ JHEP} \textbf{ 03} (2018) 174,
  \href{http://dx.doi.org/10.1007/JHEP03(2018)174}{\doi{10.1007/JHEP03(2018)174}},
  \href{http://www.arXiv.org/abs/1712.06518}{\texttt{arXiv:1712.06518}}.
[Erratum: \DOI{10.1007/JHEP11(2018)051}].

\bibitem{Sirunyan:2019xls}
\hrefCMSnoop {}{{CMS Collaboration}, ``{Search for a heavy pseudoscalar boson
  decaying to a Z and a Higgs boson at $\sqrt{s} =$ 13 TeV}'',} \textit{ Eur.
  Phys. J. C} \textbf{ 79} (2019) 564,
  \href{http://dx.doi.org/10.1140/epjc/s10052-019-7058-z}{\doi{10.1140/epjc/s10052-019-7058-z}},
\href{http://www.arXiv.org/abs/1903.00941}{\texttt{arXiv:1903.00941}}.

\bibitem{ATLAS:2022enb}
\hrefCMSnoop {}{{ATLAS Collaboration}, ``{Search for heavy resonances decaying
  into a $Z$ or $W$ boson and a Higgs boson in final states with leptons and
  $b$-jets in $139~$fb$^{-1}$ of $pp$ collisions at $\sqrt{s}=13~$TeV with the
  ATLAS detector}'',} \textit{ JHEP} \textbf{ 06} (2023) 016,
  \href{http://dx.doi.org/10.1007/JHEP06(2023)016}{\doi{10.1007/JHEP06(2023)016}},
  \href{http://www.arXiv.org/abs/2207.00230}{\texttt{arXiv:2207.00230}}.

\bibitem{Bianchini:2016yrt}
L.~Bianchini\hrefCMSnoop {}{ { et~al.}, ``{Reconstruction of the Higgs mass in
  events with Higgs bosons decaying into a pair of $\tau$ leptons using matrix
  element techniques}'',} \textit{ Nucl. Instrum. Meth. A} \textbf{ 862} (2017)
  54,
  \href{http://dx.doi.org/10.1016/j.nima.2017.05.001}{\doi{10.1016/j.nima.2017.05.001}},
\href{http://www.arXiv.org/abs/1603.05910}{\texttt{arXiv:1603.05910}}.

\bibitem{Matyszkiewicz:2025gal}
\hrefCMSnoop {}{W.~Matyszkiewicz and A.~Kalinowski, ``Tau-pair invariant mass
  estimation using maximum likelihood estimation and collinear
  approximation'',} \textit{ Acta Phys. Polon. Supp.} \textbf{ 18} (2025),
  no.~5, 5--A21,
  \href{http://dx.doi.org/10.5506/APhysPolBSupp.18.5-A21}{\doi{10.5506/APhysPolBSupp.18.5-A21}}.

\bibitem{CMS:2022prd}
\hrefCMSnoop {}{{CMS Collaboration}, ``{Identification of hadronic tau lepton
  decays using a deep neural network}'',} \textit{ JINST} \textbf{ 17} (2022)
  P07023,
  \href{http://dx.doi.org/10.1088/1748-0221/17/07/P07023}{\doi{10.1088/1748-0221/17/07/P07023}},
  \href{http://www.arXiv.org/abs/2201.08458}{\texttt{arXiv:2201.08458}}.

\bibitem{Bols:2020bkb}
E.~Bols\hrefCMSnoop {}{ { et~al.}, ``Jet flavour classification using
  {DeepJet}'',} \textit{ JINST} \textbf{ 15} (2020) P12012,
  \href{http://dx.doi.org/10.1088/1748-0221/15/12/P12012}{\doi{10.1088/1748-0221/15/12/P12012}},
  \href{http://www.arXiv.org/abs/2008.10519}{\texttt{arXiv:2008.10519}}.

\bibitem{hepdata}
\hrefCMSnoop {}{}{HEPD}ata record for this analysis, 2025.
\newblock
  \href{http://dx.doi.org/10.17182/hepdata.155628}{\doi{10.17182/hepdata.155628}}.

\bibitem{Sirunyan:2020zal}
\hrefCMSnoop {}{{CMS Collaboration}, ``Performance of the {CMS} {Level-1}
  trigger in proton-proton collisions at {$\sqrt{s} = 13\TeV$}'',} \textit{
  JINST} \textbf{ 15} (2020) P10017,
  \href{http://dx.doi.org/10.1088/1748-0221/15/10/P10017}{\doi{10.1088/1748-0221/15/10/P10017}},
  \href{http://www.arXiv.org/abs/2006.10165}{\texttt{arXiv:2006.10165}}.

\bibitem{Khachatryan:2016bia}
\hrefCMSnoop {}{{CMS Collaboration}, ``The {CMS} trigger system'',} \textit{
  JINST} \textbf{ 12} (2017) P01020,
  \href{http://dx.doi.org/10.1088/1748-0221/12/01/P01020}{\doi{10.1088/1748-0221/12/01/P01020}},
\href{http://www.arXiv.org/abs/1609.02366}{\texttt{arXiv:1609.02366}}.

\bibitem{CMS:2008xjf}
\hrefCMSnoop {}{{CMS Collaboration}, ``The {CMS} experiment at the {CERN
  LHC}'',} \textit{ JINST} \textbf{ 3} (2008) S08004,
  \href{http://dx.doi.org/10.1088/1748-0221/3/08/S08004}{\doi{10.1088/1748-0221/3/08/S08004}}.

\bibitem{Sirunyan:2017ulk}
\hrefCMSnoop {}{{CMS Collaboration}, ``Particle-flow reconstruction and global
  event description with the {CMS} detector'',} \textit{ JINST} \textbf{ 12}
  (2017) P10003,
  \href{http://dx.doi.org/10.1088/1748-0221/12/10/P10003}{\doi{10.1088/1748-0221/12/10/P10003}},
  \href{http://www.arXiv.org/abs/1706.04965}{\texttt{arXiv:1706.04965}}.

\bibitem{CMS-TDR-15-02}
\href {http://cds.cern.ch/record/2020886}{{CMS Collaboration}, ``Technical
  proposal for the {Phase-II} upgrade of the {Compact Muon Solenoid}'',} CMS
  Technical Proposal CERN-LHCC-2015-010, CMS-TDR-15-02, 2015.

\bibitem{CMS:2020uim}
\hrefCMSnoop {}{{CMS Collaboration}, ``Electron and photon reconstruction and
  identification with the {CMS} experiment at the {CERN LHC}'',} \textit{
  JINST} \textbf{ 16} (2021) P05014,
  \href{http://dx.doi.org/10.1088/1748-0221/16/05/P05014}{\doi{10.1088/1748-0221/16/05/P05014}},
  \href{http://www.arXiv.org/abs/2012.06888}{\texttt{arXiv:2012.06888}}.

\bibitem{Sirunyan:2018fpa}
\hrefCMSnoop {}{{CMS Collaboration}, ``Performance of the {CMS} muon detector
  and muon reconstruction with proton-proton collisions at {$\sqrt{s}=$ 13
  TeV}'',} \textit{ JINST} \textbf{ 13} (2018) P06015,
  \href{http://dx.doi.org/10.1088/1748-0221/13/06/P06015}{\doi{10.1088/1748-0221/13/06/P06015}},
  \href{http://www.arXiv.org/abs/1804.04528}{\texttt{arXiv:1804.04528}}.

\bibitem{Cacciari:2008gp}
\hrefCMSnoop {}{M.~Cacciari, G.~P. Salam, and G.~Soyez, ``The anti-\kt jet
  clustering algorithm'',} \textit{ JHEP} \textbf{ 04} (2008) 063,
  \href{http://dx.doi.org/10.1088/1126-6708/2008/04/063}{\doi{10.1088/1126-6708/2008/04/063}},
  \href{http://www.arXiv.org/abs/0802.1189}{\texttt{arXiv:0802.1189}}.

\bibitem{Cacciari:2011ma}
\hrefCMSnoop {}{M.~Cacciari, G.~P. Salam, and G.~Soyez, ``{FastJet} user
  manual'',} \textit{ Eur. Phys. J. C} \textbf{ 72} (2012) 1896,
  \href{http://dx.doi.org/10.1140/epjc/s10052-012-1896-2}{\doi{10.1140/epjc/s10052-012-1896-2}},
\href{http://www.arXiv.org/abs/1111.6097}{\texttt{arXiv:1111.6097}}.

\bibitem{CMS:2016lmd}
\hrefCMSnoop {}{{CMS Collaboration}, ``{Jet energy scale and resolution in the
  CMS experiment in pp collisions at 8 TeV}'',} \textit{ JINST} \textbf{ 12}
  (2017) P02014,
  \href{http://dx.doi.org/10.1088/1748-0221/12/02/P02014}{\doi{10.1088/1748-0221/12/02/P02014}},
\href{http://www.arXiv.org/abs/1607.03663}{\texttt{arXiv:1607.03663}}.

\bibitem{Sirunyan:2017ezt}
\hrefCMSnoop {}{{CMS Collaboration}, ``Identification of heavy-flavour jets
  with the {CMS} detector in pp collisions at {13 TeV}'',} \textit{ JINST}
  \textbf{ 13} (2018) P05011,
  \href{http://dx.doi.org/10.1088/1748-0221/13/05/P05011}{\doi{10.1088/1748-0221/13/05/P05011}},
  \href{http://www.arXiv.org/abs/1712.07158}{\texttt{arXiv:1712.07158}}.

\bibitem{CMS-DP-2018-058}
\href {https://cds.cern.ch/record/2646773}{{CMS Collaboration}, ``{Performance
  of the DeepJet b tagging algorithm using 41.9 $\mathrm{fb^{-1}}$ of data from
  proton-proton collisions at 13 TeV with Phase 1 CMS detector}'',} CMS
  Detector Performance Summary CMS-DP-2020-021, 2018.

\bibitem{Sirunyan:2018pgf}
\hrefCMSnoop {}{{CMS Collaboration}, ``Performance of reconstruction and
  identification of {$\tau$} leptons decaying to hadrons and {$\nu_\tau$} in pp
  collisions at {$\sqrt{s}=$ 13 TeV}'',} \textit{ JINST} \textbf{ 13} (2018)
  P10005,
  \href{http://dx.doi.org/10.1088/1748-0221/13/10/P10005}{\doi{10.1088/1748-0221/13/10/P10005}},
  \href{http://www.arXiv.org/abs/1809.02816}{\texttt{arXiv:1809.02816}}.

\bibitem{CMS:2019ctu}
\hrefCMSnoop {}{{CMS Collaboration}, ``{Performance of missing transverse
  momentum reconstruction in proton-proton collisions at $\sqrt{s} =$ 13 TeV
  using the CMS detector}'',} \textit{ JINST} \textbf{ 14} (2019) P07004,
  \href{http://dx.doi.org/10.1088/1748-0221/14/07/P07004}{\doi{10.1088/1748-0221/14/07/P07004}},
  \href{http://www.arXiv.org/abs/1903.06078}{\texttt{arXiv:1903.06078}}.

\bibitem{Alwall:2014hca}
J.~Alwall\hrefCMSnoop {}{ { et~al.}, ``The automated computation of tree-level
  and next-to-leading order differential cross sections, and their matching to
  parton shower simulations'',} \textit{ JHEP} \textbf{ 07} (2014) 079,
  \href{http://dx.doi.org/10.1007/JHEP07(2014)079}{\doi{10.1007/JHEP07(2014)079}},
\href{http://www.arXiv.org/abs/1405.0301}{\texttt{arXiv:1405.0301}}.

\bibitem{Artoisenet:2012st}
\hrefCMSnoop {}{P.~Artoisenet, R.~Frederix, O.~Mattelaer, and R.~Rietkerk,
  ``Automatic spin-entangled decays of heavy resonances in {Monte Carlo}
  simulations'',} \textit{ JHEP} \textbf{ 03} (2013) 015,
  \href{http://dx.doi.org/10.1007/JHEP03(2013)015}{\doi{10.1007/JHEP03(2013)015}},
  \href{http://www.arXiv.org/abs/1212.3460}{\texttt{arXiv:1212.3460}}.

\bibitem{Alwall:2007fs}
\hrefCMSnoop {}{J.~Alwall { et~al.}, ``{Comparative study of various algorithms
  for the merging of parton showers and matrix elements in hadronic
  collisions}'',} \textit{ Eur. Phys. J. C} \textbf{ 53} (2008) 473,
  \href{http://dx.doi.org/10.1140/epjc/s10052-007-0490-5}{\doi{10.1140/epjc/s10052-007-0490-5}},
  \href{http://www.arXiv.org/abs/0706.2569}{\texttt{arXiv:0706.2569}}.

\bibitem{Martin:1997ns}
\hrefCMSnoop {}{S.~P. Martin, ``{A Supersymmetry primer}'',} \textit{ Adv. Ser.
  Direct. HEP} \textbf{ 21} (2010) 1,
  \href{http://dx.doi.org/10.1142/9789814307505_0001}{\doi{10.1142/9789814307505_0001}},
\href{http://www.arXiv.org/abs/hep-ph/9709356}{\texttt{arXiv:hep-ph/9709356}}.

\bibitem{Nason:2004rx}
\hrefCMSnoop {}{P.~Nason, ``{A new method for combining NLO QCD with shower
  Monte Carlo algorithms}'',} \textit{ JHEP} \textbf{ 11} (2004) 040,
  \href{http://dx.doi.org/10.1088/1126-6708/2004/11/040}{\doi{10.1088/1126-6708/2004/11/040}},
  \href{http://www.arXiv.org/abs/hep-ph/0409146}{\texttt{arXiv:hep-ph/0409146}}.

\bibitem{Frixione:2007vw}
\hrefCMSnoop {}{S.~Frixione, P.~Nason, and C.~Oleari, ``{Matching NLO QCD
  computations with parton shower simulations: the POWHEG method}'',} \textit{
  JHEP} \textbf{ 11} (2007) 070,
  \href{http://dx.doi.org/10.1088/1126-6708/2007/11/070}{\doi{10.1088/1126-6708/2007/11/070}},
  \href{http://www.arXiv.org/abs/0709.2092}{\texttt{arXiv:0709.2092}}.

\bibitem{Alioli:2010xd}
\hrefCMSnoop {}{S.~Alioli, P.~Nason, C.~Oleari, and E.~Re, ``{A general
  framework for implementing NLO calculations in shower Monte Carlo programs:
  the POWHEG BOX}'',} \textit{ JHEP} \textbf{ 06} (2010) 043,
  \href{http://dx.doi.org/10.1007/JHEP06(2010)043}{\doi{10.1007/JHEP06(2010)043}},
  \href{http://www.arXiv.org/abs/1002.2581}{\texttt{arXiv:1002.2581}}.

\bibitem{Luisoni:2013cuh}
\hrefCMSnoop {}{G.~Luisoni, P.~Nason, C.~Oleari, and F.~Tramontano,
  ``{$\text{HW}^{\pm}$/HZ} + 0 and 1 jet at {NLO} with the {POWHEG BOX}
  interfaced to {GoSam} and their merging within {MiNLO}'',} \textit{ JHEP}
  \textbf{ 10} (2013) 083,
  \href{http://dx.doi.org/10.1007/JHEP10(2013)083}{\doi{10.1007/JHEP10(2013)083}},
  \href{http://www.arXiv.org/abs/1306.2542}{\texttt{arXiv:1306.2542}}.

\bibitem{Granata:2017iod}
\hrefCMSnoop {}{F.~Granata, J.~M. Lindert, C.~Oleari, and S.~Pozzorini, ``{NLO
  QCD+EW} predictions for {$\PH\PV$} and {$\PH\PV$+jet} production including
  parton-shower effects'',} \textit{ JHEP} \textbf{ 09} (2017) 012,
  \href{http://dx.doi.org/10.1007/JHEP09(2017)012}{\doi{10.1007/JHEP09(2017)012}},
  \href{http://www.arXiv.org/abs/1706.03522}{\texttt{arXiv:1706.03522}}.

\bibitem{Hartanto:2015uka}
\hrefCMSnoop {}{H.~B. Hartanto, B.~Jager, L.~Reina, and D.~Wackeroth, ``{Higgs
  boson production in association with top quarks in the POWHEG BOX}'',}
  \textit{ Phys. Rev. D} \textbf{ 91} (2015) 094003,
  \href{http://dx.doi.org/10.1103/PhysRevD.91.094003}{\doi{10.1103/PhysRevD.91.094003}},
  \href{http://www.arXiv.org/abs/1501.04498}{\texttt{arXiv:1501.04498}}.

\bibitem{Bolognesi:2012mm}
S.~Bolognesi\hrefCMSnoop {}{ { et~al.}, ``{On the spin and parity of a
  single-produced resonance at the LHC}'',} \textit{ Phys. Rev. D} \textbf{ 86}
  (2012) 095031,
  \href{http://dx.doi.org/10.1103/PhysRevD.86.095031}{\doi{10.1103/PhysRevD.86.095031}},
  \href{http://www.arXiv.org/abs/1208.4018}{\texttt{arXiv:1208.4018}}.

\bibitem{LHCHiggsCrossSectionWorkingGroup:2016ypw}
\hrefCMSnoop {}{{LHC Higgs Cross Section Working Group}, ``Handbook of {LHC
  Higgs} cross sections: 4. {D}eciphering the nature of the {Higgs} sector'',}
  CERN Report CERN-2017-002-M, 2016.
\newblock
  \href{http://dx.doi.org/10.23731/CYRM-2017-002}{\doi{10.23731/CYRM-2017-002}},
  \href{http://www.arXiv.org/abs/1610.07922}{\texttt{arXiv:1610.07922}}.

\bibitem{Campbell:2011bn}
\hrefCMSnoop {}{J.~M. Campbell, R.~K. Ellis, and C.~Williams, ``{Vector boson
  pair production at the LHC}'',} \textit{ JHEP} \textbf{ 07} (2011) 018,
  \href{http://dx.doi.org/10.1007/JHEP07(2011)018}{\doi{10.1007/JHEP07(2011)018}},
  \href{http://www.arXiv.org/abs/1105.0020}{\texttt{arXiv:1105.0020}}.

\bibitem{Nason:2013ydw}
\hrefCMSnoop {}{P.~Nason and G.~Zanderighi, ``{$W^+ W^-$ , $W Z$ and $Z Z$
  production in the POWHEG-BOX-V2}'',} \textit{ Eur. Phys. J. C} \textbf{ 74}
  (2014) 2702,
  \href{http://dx.doi.org/10.1140/epjc/s10052-013-2702-5}{\doi{10.1140/epjc/s10052-013-2702-5}},
  \href{http://www.arXiv.org/abs/1311.1365}{\texttt{arXiv:1311.1365}}.

\bibitem{Frederix:2012ps}
\hrefCMSnoop {}{R.~Frederix and S.~Frixione, ``{Merging meets matching in
  MC@NLO}'',} \textit{ JHEP} \textbf{ 12} (2012) 061,
  \href{http://dx.doi.org/10.1007/JHEP12(2012)061}{\doi{10.1007/JHEP12(2012)061}},
  \href{http://www.arXiv.org/abs/1209.6215}{\texttt{arXiv:1209.6215}}.

\bibitem{Alioli:2011as}
\hrefCMSnoop {}{S.~Alioli, S.-O. Moch, and P.~Uwer, ``{Hadronic top-quark
  pair-production with one jet and parton showering}'',} \textit{ JHEP}
  \textbf{ 01} (2012) 137,
  \href{http://dx.doi.org/10.1007/JHEP01(2012)137}{\doi{10.1007/JHEP01(2012)137}},
  \href{http://www.arXiv.org/abs/1110.5251}{\texttt{arXiv:1110.5251}}.

\bibitem{Caola:2016trd}
F.~Caola\hrefCMSnoop {}{ { et~al.}, ``{QCD corrections to vector boson pair
  production in gluon fusion including interference effects with off-shell
  Higgs at the LHC}'',} \textit{ JHEP} \textbf{ 07} (2016) 087,
  \href{http://dx.doi.org/10.1007/JHEP07(2016)087}{\doi{10.1007/JHEP07(2016)087}},
  \href{http://www.arXiv.org/abs/1605.04610}{\texttt{arXiv:1605.04610}}.

\bibitem{Gehrmann:2014fva}
T.~Gehrmann\hrefCMSnoop {}{ { et~al.}, ``{$W^+ W^-$} production at hadron
  colliders in next to next to leading order {QCD}'',} \textit{ Phys. Rev.
  Lett.} \textbf{ 113} (2014) 212001,
  \href{http://dx.doi.org/10.1103/PhysRevLett.113.212001}{\doi{10.1103/PhysRevLett.113.212001}},
  \href{http://www.arXiv.org/abs/1408.5243}{\texttt{arXiv:1408.5243}}.

\bibitem{Melnikov:2006kv}
\hrefCMSnoop {}{K.~Melnikov and F.~Petriello, ``Electroweak gauge boson
  production at hadron colliders through
  {$\mathcal{O}(\alpha_\text{s}^{2})$}'',} \textit{ Phys. Rev. D} \textbf{ 74}
  (2006) 114017,
  \href{http://dx.doi.org/10.1103/PhysRevD.74.114017}{\doi{10.1103/PhysRevD.74.114017}},
\href{http://www.arXiv.org/abs/hep-ph/0609070}{\texttt{arXiv:hep-ph/0609070}}.

\bibitem{Czakon:2011xx}
\hrefCMSnoop {}{M.~Czakon and A.~Mitov, ``{Top++}: A program for the
  calculation of the top-pair cross-section at hadron colliders'',} \textit{
  Comput. Phys. Commun.} \textbf{ 185} (2014) 2930,
  \href{http://dx.doi.org/10.1016/j.cpc.2014.06.021}{\doi{10.1016/j.cpc.2014.06.021}},
  \href{http://www.arXiv.org/abs/1112.5675}{\texttt{arXiv:1112.5675}}.

\bibitem{Campbell:2012dh}
\hrefCMSnoop {}{J.~M. Campbell and R.~K. Ellis, ``{$t\bar{t} W^{\pm}$
  production and decay at NLO}'',} \textit{ JHEP} \textbf{ 07} (2012) 052,
  \href{http://dx.doi.org/10.1007/JHEP07(2012)052}{\doi{10.1007/JHEP07(2012)052}},
  \href{http://www.arXiv.org/abs/1204.5678}{\texttt{arXiv:1204.5678}}.

\bibitem{Garzelli:2012bn}
\hrefCMSnoop {}{M.~V. Garzelli, A.~Kardos, C.~G. Papadopoulos, and
  Z.~Trocsanyi, ``{$t\bar{t} W^{\pm}$ and $t\bar{t} Z$} hadroproduction at
  {NLO} accuracy in {QCD} with parton shower and hadronization effects'',}
  \textit{ JHEP} \textbf{ 11} (2012) 056,
  \href{http://dx.doi.org/10.1007/JHEP11(2012)056}{\doi{10.1007/JHEP11(2012)056}},
  \href{http://www.arXiv.org/abs/1208.2665}{\texttt{arXiv:1208.2665}}.

\bibitem{Ball:2017nwa}
\hrefCMSnoop {}{{NNPDF} Collaboration, ``Parton distributions from
  high-precision collider data'',} \textit{ Eur. Phys. J. C} \textbf{ 77}
  (2017) 663,
  \href{http://dx.doi.org/10.1140/epjc/s10052-017-5199-5}{\doi{10.1140/epjc/s10052-017-5199-5}},
  \href{http://www.arXiv.org/abs/1706.00428}{\texttt{arXiv:1706.00428}}.

\bibitem{Sjostrand:2014zea}
T.~Sj{\"o}strand\hrefCMSnoop {}{ { et~al.}, ``An introduction to {PYTHIA
  8.2}'',} \textit{ Comput. Phys. Commun.} \textbf{ 191} (2015) 159,
  \href{http://dx.doi.org/10.1016/j.cpc.2015.01.024}{\doi{10.1016/j.cpc.2015.01.024}},
  \href{http://www.arXiv.org/abs/1410.3012}{\texttt{arXiv:1410.3012}}.

\bibitem{Sirunyan:2019dfx}
\hrefCMSnoop {}{{CMS Collaboration}, ``Extraction and validation of a new set
  of {CMS PYTHIA8} tunes from underlying-event measurements'',} \textit{ Eur.
  Phys. J. C} \textbf{ 80} (2020) 4,
  \href{http://dx.doi.org/10.1140/epjc/s10052-019-7499-4}{\doi{10.1140/epjc/s10052-019-7499-4}},
  \href{http://www.arXiv.org/abs/1903.12179}{\texttt{arXiv:1903.12179}}.

\bibitem{Agostinelli:2002hh}
\hrefCMSnoop {}{{GEANT4} Collaboration, ``{\GEANTfour}---a simulation
  toolkit'',} \textit{ Nucl. Instrum. Meth. A} \textbf{ 506} (2003) 250,
\href{http://dx.doi.org/10.1016/S0168-9002(03)01368-8}{\doi{10.1016/S0168-9002(03)01368-8}}.

\bibitem{Cousins:2018tiz}
\href {https://arxiv.org/abs/1807.05996}{R.~D. Cousins, ``Lectures on
  statistics in theory: Prelude to statistics in practice'',} 2018.
  \href{http://www.arXiv.org/abs/1807.05996}{\texttt{arXiv:1807.05996}}.

\bibitem{CMS:2024onh}
\hrefCMSnoop {}{{CMS Collaboration}, ``The {CMS} statistical analysis and
  combination tool: {\textsc{Combine}}'',} \textit{ Comput. Softw. Big Sci.}
  \textbf{ 8} (2024) 19,
  \href{http://dx.doi.org/10.1007/s41781-024-00121-4}{\doi{10.1007/s41781-024-00121-4}},
  \href{http://www.arXiv.org/abs/2404.06614}{\texttt{arXiv:2404.06614}}.

\bibitem{Butterworth:2015oua}
\hrefCMSnoop {}{J.~Butterworth { et~al.}, ``{PDF4LHC recommendations for LHC
  Run II}'',} \textit{ J. Phys. G} \textbf{ 43} (2016) 023001,
  \href{http://dx.doi.org/10.1088/0954-3899/43/2/023001}{\doi{10.1088/0954-3899/43/2/023001}},
  \href{http://www.arXiv.org/abs/1510.03865}{\texttt{arXiv:1510.03865}}.

\bibitem{Sirunyan:2017uzs}
\hrefCMSnoop {}{{CMS Collaboration}, ``Measurement of the cross section for top
  quark pair production in association with a {W} or {Z} boson in proton-proton
  collisions at $\sqrt{s} =$ 13 {TeV}'',} \textit{ JHEP} \textbf{ 08} (2018)
  011,
  \href{http://dx.doi.org/10.1007/JHEP08(2018)011}{\doi{10.1007/JHEP08(2018)011}},
\href{http://www.arXiv.org/abs/1711.02547}{\texttt{arXiv:1711.02547}}.

\bibitem{Bagnaschi:2021jaj}
E.~A. Bagnaschi\href {https://cds.cern.ch/record/2791954}{ { et~al.},
  ``Benchmark scenarios for {MSSM Higgs} boson searches at the {LHC}'',} CERN
  Report LHCHWG-2021-001, 2021.

\bibitem{CMS-LUM-17-003}
\hrefCMSnoop {}{{CMS Collaboration}, ``Precision luminosity measurement in
  proton-proton collisions at $\sqrt{s} =$ 13 {TeV} in 2015 and 2016 at
  {CMS}'',} \textit{ Eur. Phys. J. C} \textbf{ 81} (2021) 800,
  \href{http://dx.doi.org/10.1140/epjc/s10052-021-09538-2}{\doi{10.1140/epjc/s10052-021-09538-2}},
  \href{http://www.arXiv.org/abs/2104.01927}{\texttt{arXiv:2104.01927}}.

\bibitem{CMS-PAS-LUM-17-004}
\href {https://cds.cern.ch/record/2621960/}{{CMS Collaboration}, ``{CMS}
  luminosity measurement for the 2017 data-taking period at $\sqrt{s}$ = 13
  {TeV}'',} CMS Physics Analysis Summary CMS-PAS-LUM-17-004, 2018.

\bibitem{CMS-PAS-LUM-18-002}
\href {https://cds.cern.ch/record/2676164/}{{CMS Collaboration}, ``{CMS}
  luminosity measurement for the 2018 data-taking period at $\sqrt{s}$ = 13
  {TeV}'',} CMS Physics Analysis Summary CMS-PAS-LUM-18-002, 2019.

\bibitem{Barlow:1993dm}
\hrefCMSnoop {}{R.~J. Barlow and C.~Beeston, ``Fitting using finite {Monte}
  {Carlo} samples'',} \textit{ Comput. Phys. Commun.} \textbf{ 77} (1993) 219,
\href{http://dx.doi.org/10.1016/0010-4655(93)90005-W}{\doi{10.1016/0010-4655(93)90005-W}}.

\bibitem{Conway:2011in}
\hrefCMSnoop {}{J.~S. Conway, ``{Incorporating nuisance parameters in
  likelihoods for multisource spectra}'',} in \textit{ {PHYSTAT 2011}}, p.~115.
\newblock 2011.
\newblock \href{http://www.arXiv.org/abs/1103.0354}{\texttt{arXiv:1103.0354}}.
\newblock
  \href{http://dx.doi.org/10.5170/CERN-2011-006.115}{\doi{10.5170/CERN-2011-006.115}}.

\bibitem{Verkerke:2003ir}
\hrefCMSnoop {}{W.~Verkerke and D.~P. Kirkby, ``{The RooFit toolkit for data
  modeling}'',} in \textit{ {Proc. PHYSTAT 2003, Statistical Problems in
  Particle Physics, Astrophysics and Cosmology}}, L.~Lyons and M.~Karagoz,
  eds., p.~MOLT007.
\newblock 2003.
\newblock
  \href{http://www.arXiv.org/abs/physics/0306116}{\texttt{arXiv:physics/0306116}}.
\newblock {eConf C0303241}.

\bibitem{Moneta:2010pm}
L.~Moneta\hrefCMSnoop {}{ { et~al.}, ``The {RooStats} project'',} in \textit{
  {Proc. ACAT 2010, 13th International Workshop on Advanced Computing and
  Analysis Techniques in Physics Research}}, p.~057.
\newblock 2010.
\newblock \href{http://www.arXiv.org/abs/1009.1003}{\texttt{arXiv:1009.1003}}.
\newblock {PoS ACAT2010}.
  \href{http://dx.doi.org/10.22323/1.093.0057}{\doi{10.22323/1.093.0057}}.

\bibitem{CMS:2012zhx}
\hrefCMSnoop {}{{CMS Collaboration}, ``Combined results of searches for the
  standard model {Higgs} boson in {$pp$} collisions at {$\sqrt{s}=7$ TeV}'',}
  \textit{ Phys. Lett. B} \textbf{ 710} (2012) 26,
  \href{http://dx.doi.org/10.1016/j.physletb.2012.02.064}{\doi{10.1016/j.physletb.2012.02.064}},
  \href{http://www.arXiv.org/abs/1202.1488}{\texttt{arXiv:1202.1488}}.

\bibitem{Junk:1999kv}
\hrefCMSnoop {}{T.~Junk, ``{Confidence level computation for combining searches
  with small statistics}'',} \textit{ Nucl. Instrum. Meth. A} \textbf{ 434}
  (1999) 435,
  \href{http://dx.doi.org/10.1016/S0168-9002(99)00498-2}{\doi{10.1016/S0168-9002(99)00498-2}},
  \href{http://www.arXiv.org/abs/hep-ex/9902006}{\texttt{arXiv:hep-ex/9902006}}.

\bibitem{Read:2002hq}
\hrefCMSnoop {}{A.~L. Read, ``Presentation of search results: The {$CL_s$}
  technique'',} \textit{ J. Phys. G} \textbf{ 28} (2002) 2693,
  \href{http://dx.doi.org/10.1088/0954-3899/28/10/313}{\doi{10.1088/0954-3899/28/10/313}}.

\bibitem{Heinemeyer:1998yj}
\hrefCMSnoop {}{S.~Heinemeyer, W.~Hollik, and G.~Weiglein, ``{FeynHiggs: A}
  program for the calculation of the masses of the neutral {CP}-even {Higgs}
  bosons in the {MSSM}'',} \textit{ Comput. Phys. Commun.} \textbf{ 124} (2000)
  76,
  \href{http://dx.doi.org/10.1016/S0010-4655(99)00364-1}{\doi{10.1016/S0010-4655(99)00364-1}},
\href{http://www.arXiv.org/abs/hep-ph/9812320}{\texttt{arXiv:hep-ph/9812320}}.

\bibitem{Heinemeyer:1998np}
\hrefCMSnoop {}{S.~Heinemeyer, W.~Hollik, and G.~Weiglein, ``The masses of the
  neutral {CP}-even {Higgs} bosons in the {MSSM}: {A}ccurate analysis at the
  two-loop level'',} \textit{ Eur. Phys. J. C} \textbf{ 9} (1999) 343,
  \href{http://dx.doi.org/10.1007/s100529900006}{\doi{10.1007/s100529900006}},
\href{http://www.arXiv.org/abs/hep-ph/9812472}{\texttt{arXiv:hep-ph/9812472}}.

\bibitem{Degrassi:2002fi}
G.~Degrassi\hrefCMSnoop {}{ { et~al.}, ``Towards high-precision predictions for
  the {MSSM Higgs} sector'',} \textit{ Eur. Phys. J. C} \textbf{ 28} (2003)
  133,
  \href{http://dx.doi.org/10.1140/epjc/s2003-01152-2}{\doi{10.1140/epjc/s2003-01152-2}},
\href{http://www.arXiv.org/abs/hep-ph/0212020}{\texttt{arXiv:hep-ph/0212020}}.

\bibitem{Frank:2006yh}
M.~Frank\hrefCMSnoop {}{ { et~al.}, ``The {Higgs} boson masses and mixings of
  the complex {MSSM} in the {F}eynman-diagrammatic approach'',} \textit{ JHEP}
  \textbf{ 02} (2007) 047,
  \href{http://dx.doi.org/10.1088/1126-6708/2007/02/047}{\doi{10.1088/1126-6708/2007/02/047}},
\href{http://www.arXiv.org/abs/hep-ph/0611326}{\texttt{arXiv:hep-ph/0611326}}.

\bibitem{Hahn:2013ria}
T.~Hahn\hrefCMSnoop {}{ { et~al.}, ``High-precision predictions for the light
  {CP}-even {Higgs} boson mass of the minimal supersymmetric standard model'',}
  \textit{ Phys. Rev. Lett.} \textbf{ 112} (2014) 141801,
  \href{http://dx.doi.org/10.1103/PhysRevLett.112.141801}{\doi{10.1103/PhysRevLett.112.141801}},
\href{http://www.arXiv.org/abs/1312.4937}{\texttt{arXiv:1312.4937}}.

\bibitem{Bahl:2016brp}
\hrefCMSnoop {}{H.~Bahl and W.~Hollik, ``Precise prediction for the light {MSSM
  Higgs} boson mass combining effective field theory and fixed-order
  calculations'',} \textit{ Eur. Phys. J. C} \textbf{ 76} (2016) 499,
  \href{http://dx.doi.org/10.1140/epjc/s10052-016-4354-8}{\doi{10.1140/epjc/s10052-016-4354-8}},
  \href{http://www.arXiv.org/abs/1608.01880}{\texttt{arXiv:1608.01880}}.

\bibitem{Bahl:2017aev}
\hrefCMSnoop {}{H.~Bahl, S.~Heinemeyer, W.~Hollik, and G.~Weiglein,
  ``Reconciling {EFT} and hybrid calculations of the light {MSSM Higgs-boson}
  mass'',} \textit{ Eur. Phys. J. C} \textbf{ 78} (2018) 57,
  \href{http://dx.doi.org/10.1140/epjc/s10052-018-5544-3}{\doi{10.1140/epjc/s10052-018-5544-3}},
  \href{http://www.arXiv.org/abs/1706.00346}{\texttt{arXiv:1706.00346}}.

\bibitem{Bahl:2018qog}
H.~Bahl\hrefCMSnoop {}{ { et~al.}, ``Precision calculations in the {MSSM
  Higgs-boson} sector with {FeynHiggs 2.14}'',} \textit{ Comput. Phys. Commun.}
  \textbf{ 249} (2020) 107099,
  \href{http://dx.doi.org/10.1016/j.cpc.2019.107099}{\doi{10.1016/j.cpc.2019.107099}},
  \href{http://www.arXiv.org/abs/1811.09073}{\texttt{arXiv:1811.09073}}.

\bibitem{Djouadi:2018xqq}
\hrefCMSnoop {}{A.~Djouadi, J.~Kalinowski, M.~Muehlleitner, and M.~Spira,
  ``{HDECAY: Twenty${++}$ years after}'',} \textit{ Comput. Phys. Commun.}
  \textbf{ 238} (2019) 214,
  \href{http://dx.doi.org/10.1016/j.cpc.2018.12.010}{\doi{10.1016/j.cpc.2018.12.010}},
  \href{http://www.arXiv.org/abs/1801.09506}{\texttt{arXiv:1801.09506}}.

\bibitem{LHCHiggsCrossSectionWorkingGroup:2013rie}
\hrefCMSnoop {}{{LHC Higgs Cross Section Working Group}, ``{Handbook of LHC
  Higgs} cross sections: 3. {Higgs} properties'',} CERN Report CERN-2013-004,
  2013.
\newblock
  \href{http://dx.doi.org/10.5170/CERN-2013-004}{\doi{10.5170/CERN-2013-004}},
  \href{http://www.arXiv.org/abs/1307.1347}{\texttt{arXiv:1307.1347}}.

\bibitem{Denner:2011mq}
A.~Denner\hrefCMSnoop {}{ { et~al.}, ``Standard model {Higgs}-boson branching
  ratios with uncertainties'',} \textit{ Eur. Phys. J. C} \textbf{ 71} (2011)
  1753,
  \href{http://dx.doi.org/10.1140/epjc/s10052-011-1753-8}{\doi{10.1140/epjc/s10052-011-1753-8}},
  \href{http://www.arXiv.org/abs/1107.5909}{\texttt{arXiv:1107.5909}}.

\bibitem{Harlander:2012pb}
\hrefCMSnoop {}{R.~V. Harlander, S.~Liebler, and H.~Mantler, ``{SusHi: A}
  program for the calculation of {Higgs} production in gluon fusion and
  bottom-quark annihilation in the standard model and the {MSSM}'',} \textit{
  Comput. Phys. Commun.} \textbf{ 184} (2013) 1605,
  \href{http://dx.doi.org/10.1016/j.cpc.2013.02.006}{\doi{10.1016/j.cpc.2013.02.006}},
\href{http://www.arXiv.org/abs/1212.3249}{\texttt{arXiv:1212.3249}}.

\bibitem{Harlander:2016hcx}
\hrefCMSnoop {}{R.~V. Harlander, S.~Liebler, and H.~Mantler, ``{SusHi Bento}:
  Beyond {NNLO} and the heavy-top limit'',} \textit{ Comput. Phys. Commun.}
  \textbf{ 212} (2017) 239,
  \href{http://dx.doi.org/10.1016/j.cpc.2016.10.015}{\doi{10.1016/j.cpc.2016.10.015}},
  \href{http://www.arXiv.org/abs/1605.03190}{\texttt{arXiv:1605.03190}}.

\bibitem{Spira:1995rr}
\hrefCMSnoop {}{M.~Spira, A.~Djouadi, D.~Graudenz, and P.~M. Zerwas, ``{Higgs}
  boson production at the {LHC}'',} \textit{ Nucl. Phys. B} \textbf{ 453}
  (1995) 17,
  \href{http://dx.doi.org/10.1016/0550-3213(95)00379-7}{\doi{10.1016/0550-3213(95)00379-7}},
\href{http://www.arXiv.org/abs/hep-ph/9504378}{\texttt{arXiv:hep-ph/9504378}}.

\bibitem{Harlander:2005rq}
\hrefCMSnoop {}{R.~Harlander and P.~Kant, ``{Higgs} production and decay:
  analytic results at next-to-leading order {QCD}'',} \textit{ JHEP} \textbf{
  12} (2005) 015,
  \href{http://dx.doi.org/10.1088/1126-6708/2005/12/015}{\doi{10.1088/1126-6708/2005/12/015}},
\href{http://www.arXiv.org/abs/hep-ph/0509189}{\texttt{arXiv:hep-ph/0509189}}.

\bibitem{Harlander:2002wh}
\hrefCMSnoop {}{R.~V. Harlander and W.~B. Kilgore, ``Next-to-next-to-leading
  order {Higgs} production at hadron colliders'',} \textit{ Phys. Rev. Lett.}
  \textbf{ 88} (2002) 201801,
  \href{http://dx.doi.org/10.1103/PhysRevLett.88.201801}{\doi{10.1103/PhysRevLett.88.201801}},
\href{http://www.arXiv.org/abs/hep-ph/0201206}{\texttt{arXiv:hep-ph/0201206}}.

\bibitem{Anastasiou:2002yz}
\hrefCMSnoop {}{C.~Anastasiou and K.~Melnikov, ``{Higgs} boson production at
  hadron colliders in {NNLO QCD}'',} \textit{ Nucl. Phys. B} \textbf{ 646}
  (2002) 220,
  \href{http://dx.doi.org/10.1016/S0550-3213(02)00837-4}{\doi{10.1016/S0550-3213(02)00837-4}},
\href{http://www.arXiv.org/abs/hep-ph/0207004}{\texttt{arXiv:hep-ph/0207004}}.

\bibitem{Ravindran:2003um}
\hrefCMSnoop {}{V.~Ravindran, J.~Smith, and W.~L. van Neerven, ``{NNLO}
  corrections to the total cross-section for {Higgs} boson production in
  hadron-hadron collisions'',} \textit{ Nucl. Phys. B} \textbf{ 665} (2003)
  325,
  \href{http://dx.doi.org/10.1016/S0550-3213(03)00457-7}{\doi{10.1016/S0550-3213(03)00457-7}},
\href{http://www.arXiv.org/abs/hep-ph/0302135}{\texttt{arXiv:hep-ph/0302135}}.

\bibitem{Harlander:2002vv}
\hrefCMSnoop {}{R.~V. Harlander and W.~B. Kilgore, ``Production of a
  pseudo-scalar {Higgs} boson at hadron colliders at next-to-next-to leading
  order'',} \textit{ JHEP} \textbf{ 10} (2002) 017,
  \href{http://dx.doi.org/10.1088/1126-6708/2002/10/017}{\doi{10.1088/1126-6708/2002/10/017}},
\href{http://www.arXiv.org/abs/hep-ph/0208096}{\texttt{arXiv:hep-ph/0208096}}.

\bibitem{Anastasiou:2002wq}
\hrefCMSnoop {}{C.~Anastasiou and K.~Melnikov, ``Pseudoscalar {Higgs} boson
  production at hadron colliders in next-to-next-to-leading order {QCD}'',}
  \textit{ Phys. Rev. D} \textbf{ 67} (2003) 037501,
  \href{http://dx.doi.org/10.1103/PhysRevD.67.037501}{\doi{10.1103/PhysRevD.67.037501}},
\href{http://www.arXiv.org/abs/hep-ph/0208115}{\texttt{arXiv:hep-ph/0208115}}.

\bibitem{Anastasiou:2014lda}
C.~Anastasiou\hrefCMSnoop {}{ { et~al.}, ``{Higgs} boson gluon-fusion
  production beyond threshold in {N$^{3}$LO QCD}'',} \textit{ JHEP} \textbf{
  03} (2015) 091,
  \href{http://dx.doi.org/10.1007/JHEP03(2015)091}{\doi{10.1007/JHEP03(2015)091}},
  \href{http://www.arXiv.org/abs/1411.3584}{\texttt{arXiv:1411.3584}}.

\bibitem{Anastasiou:2015yha}
C.~Anastasiou\hrefCMSnoop {}{ { et~al.}, ``Soft expansion of
  double-real-virtual corrections to {Higgs} production at {N$^{3}$LO}'',}
  \textit{ JHEP} \textbf{ 08} (2015) 051,
  \href{http://dx.doi.org/10.1007/JHEP08(2015)051}{\doi{10.1007/JHEP08(2015)051}},
  \href{http://www.arXiv.org/abs/1505.04110}{\texttt{arXiv:1505.04110}}.

\bibitem{Anastasiou:2016cez}
C.~Anastasiou\hrefCMSnoop {}{ { et~al.}, ``High precision determination of the
  gluon fusion {Higgs} boson cross-section at the {LHC}'',} \textit{ JHEP}
  \textbf{ 05} (2016) 058,
  \href{http://dx.doi.org/10.1007/JHEP05(2016)058}{\doi{10.1007/JHEP05(2016)058}},
  \href{http://www.arXiv.org/abs/1602.00695}{\texttt{arXiv:1602.00695}}.

\bibitem{Aglietti:2004nj}
\hrefCMSnoop {}{U.~Aglietti, R.~Bonciani, G.~Degrassi, and A.~Vicini,
  ``Two-loop light fermion contribution to {Higgs} production and decays'',}
  \textit{ Phys. Lett. B} \textbf{ 595} (2004) 432,
  \href{http://dx.doi.org/10.1016/j.physletb.2004.06.063}{\doi{10.1016/j.physletb.2004.06.063}},
\href{http://www.arXiv.org/abs/hep-ph/0404071}{\texttt{arXiv:hep-ph/0404071}}.

\bibitem{Bonciani:2010ms}
\hrefCMSnoop {}{R.~Bonciani, G.~Degrassi, and A.~Vicini, ``On the generalized
  harmonic polylogarithms of one complex variable'',} \textit{ Comput. Phys.
  Commun.} \textbf{ 182} (2011) 1253,
  \href{http://dx.doi.org/10.1016/j.cpc.2011.02.011}{\doi{10.1016/j.cpc.2011.02.011}},
\href{http://www.arXiv.org/abs/1007.1891}{\texttt{arXiv:1007.1891}}.

\bibitem{Bonvini:2015pxa}
\hrefCMSnoop {}{M.~Bonvini, A.~S. Papanastasiou, and F.~J. Tackmann,
  ``Resummation and matching of b-quark mass effects in {$
  \PQb\overline{\PQb}\PH $} production'',} \textit{ JHEP} \textbf{ 11} (2015)
  196,
  \href{http://dx.doi.org/10.1007/JHEP11(2015)196}{\doi{10.1007/JHEP11(2015)196}},
  \href{http://www.arXiv.org/abs/1508.03288}{\texttt{arXiv:1508.03288}}.

\bibitem{Bonvini:2016fgf}
\hrefCMSnoop {}{M.~Bonvini, A.~S. Papanastasiou, and F.~J. Tackmann, ``Matched
  predictions for the {$ \PQb\overline{\PQb}\PH $} cross section at the 13 {TeV
  LHC}'',} \textit{ JHEP} \textbf{ 10} (2016) 053,
  \href{http://dx.doi.org/10.1007/JHEP10(2016)053}{\doi{10.1007/JHEP10(2016)053}},
  \href{http://www.arXiv.org/abs/1605.01733}{\texttt{arXiv:1605.01733}}.

\bibitem{Forte:2015hba}
\hrefCMSnoop {}{S.~Forte, D.~Napoletano, and M.~Ubiali, ``{Higgs} production in
  bottom-quark fusion in a matched scheme'',} \textit{ Phys. Lett. B} \textbf{
  751} (2015) 331,
  \href{http://dx.doi.org/10.1016/j.physletb.2015.10.051}{\doi{10.1016/j.physletb.2015.10.051}},
  \href{http://www.arXiv.org/abs/1508.01529}{\texttt{arXiv:1508.01529}}.

\bibitem{Forte:2016sja}
\hrefCMSnoop {}{S.~Forte, D.~Napoletano, and M.~Ubiali, ``{Higgs} production in
  bottom-quark fusion: matching beyond leading order'',} \textit{ Phys. Lett.
  B} \textbf{ 763} (2016) 190,
  \href{http://dx.doi.org/10.1016/j.physletb.2016.10.040}{\doi{10.1016/j.physletb.2016.10.040}},
  \href{http://www.arXiv.org/abs/1607.00389}{\texttt{arXiv:1607.00389}}.

\bibitem{Harlander:2003ai}
\hrefCMSnoop {}{R.~V. Harlander and W.~B. Kilgore, ``Higgs boson production in
  bottom quark fusion at next-to-next-to leading order'',} \textit{ Phys. Rev.
  D} \textbf{ 68} (2003) 013001,
  \href{http://dx.doi.org/10.1103/PhysRevD.68.013001}{\doi{10.1103/PhysRevD.68.013001}},
  \href{http://www.arXiv.org/abs/hep-ph/0304035}{\texttt{arXiv:hep-ph/0304035}}.

\bibitem{Dittmaier:2003ej}
\hrefCMSnoop {}{S.~Dittmaier, M.~Kr{\"a}mer, and M.~Spira, ``{Higgs} radiation
  off bottom quarks at the {Fermilab Tevatron} and the {CERN LHC}'',} \textit{
  Phys. Rev. D} \textbf{ 70} (2004) 074010,
  \href{http://dx.doi.org/10.1103/PhysRevD.70.074010}{\doi{10.1103/PhysRevD.70.074010}},
\href{http://www.arXiv.org/abs/hep-ph/0309204}{\texttt{arXiv:hep-ph/0309204}}.

\bibitem{Dawson:2003kb}
\hrefCMSnoop {}{S.~Dawson, C.~B. Jackson, L.~Reina, and D.~Wackeroth,
  ``Exclusive {Higgs} boson production with bottom quarks at hadron
  colliders'',} \textit{ Phys. Rev. D} \textbf{ 69} (2004) 074027,
  \href{http://dx.doi.org/10.1103/PhysRevD.69.074027}{\doi{10.1103/PhysRevD.69.074027}},
  \href{http://www.arXiv.org/abs/hep-ph/0311067}{\texttt{arXiv:hep-ph/0311067}}.

\end{thebibliography}\endgroup

\cleardoublepage \appendix\section{The CMS Collaboration \label{app:collab}}\begin{sloppypar}\hyphenpenalty=5000\widowpenalty=500\clubpenalty=5000\cmsinstitute{Yerevan Physics Institute, Yerevan, Armenia}
{\tolerance=6000
V.~Chekhovsky, A.~Hayrapetyan, V.~Makarenko\cmsorcid{0000-0002-8406-8605}, A.~Tumasyan\cmsAuthorMark{1}\cmsorcid{0009-0000-0684-6742}
\par}
\cmsinstitute{Institut f\"{u}r Hochenergiephysik, Vienna, Austria}
{\tolerance=6000
W.~Adam\cmsorcid{0000-0001-9099-4341}, J.W.~Andrejkovic, L.~Benato\cmsorcid{0000-0001-5135-7489}, T.~Bergauer\cmsorcid{0000-0002-5786-0293}, S.~Chatterjee\cmsorcid{0000-0003-2660-0349}, K.~Damanakis\cmsorcid{0000-0001-5389-2872}, M.~Dragicevic\cmsorcid{0000-0003-1967-6783}, P.S.~Hussain\cmsorcid{0000-0002-4825-5278}, M.~Jeitler\cmsAuthorMark{2}\cmsorcid{0000-0002-5141-9560}, N.~Krammer\cmsorcid{0000-0002-0548-0985}, A.~Li\cmsorcid{0000-0002-4547-116X}, D.~Liko\cmsorcid{0000-0002-3380-473X}, I.~Mikulec\cmsorcid{0000-0003-0385-2746}, J.~Schieck\cmsAuthorMark{2}\cmsorcid{0000-0002-1058-8093}, R.~Sch\"{o}fbeck\cmsAuthorMark{2}\cmsorcid{0000-0002-2332-8784}, D.~Schwarz\cmsorcid{0000-0002-3821-7331}, M.~Sonawane\cmsorcid{0000-0003-0510-7010}, W.~Waltenberger\cmsorcid{0000-0002-6215-7228}, C.-E.~Wulz\cmsAuthorMark{2}\cmsorcid{0000-0001-9226-5812}
% ...
\par}
\vskip\cmsinstskip
\dag:~Deceased\\
$^{1}$Also at Yerevan State University, Yerevan, Armenia\\
$^{2}$Also at TU Wien, Vienna, Austria\\
$^{3}$Also at Ghent University, Ghent, Belgium\\
$^{4}$Also at Universidade do Estado do Rio de Janeiro, Rio de Janeiro, Brazil\\
$^{5}$Also at FACAMP - Faculdades de Campinas, Sao Paulo, Brazil\\
$^{6}$Also at Universidade Estadual de Campinas, Campinas, Brazil\\
$^{7}$Also at Federal University of Rio Grande do Sul, Porto Alegre, Brazil\\
$^{8}$Also at University of Chinese Academy of Sciences, Beijing, China\\
$^{9}$Also at China Center of Advanced Science and Technology, Beijing, China\\
$^{10}$Also at University of Chinese Academy of Sciences, Beijing, China\\
$^{11}$Also at China Spallation Neutron Source, Guangdong, China\\
$^{12}$Now at Henan Normal University, Xinxiang, China\\
$^{13}$Also at University of Shanghai for Science and Technology, Shanghai, China\\
$^{14}$Now at The University of Iowa, Iowa City, Iowa, USA\\
$^{15}$Also at an institute formerly covered by a cooperation agreement with CERN\\
$^{16}$Also at Helwan University, Cairo, Egypt\\
$^{17}$Now at Zewail City of Science and Technology, Zewail, Egypt\\
$^{18}$Now at British University in Egypt, Cairo, Egypt\\
$^{19}$Now at Cairo University, Cairo, Egypt\\
$^{20}$Also at Purdue University, West Lafayette, Indiana, USA\\
$^{21}$Also at Universit\'{e} de Haute Alsace, Mulhouse, France\\
$^{22}$Also at Istinye University, Istanbul, Turkey\\
$^{23}$Also at an international laboratory covered by a cooperation agreement with CERN\\
$^{24}$Also at The University of the State of Amazonas, Manaus, Brazil\\
$^{25}$Also at University of Hamburg, Hamburg, Germany\\
$^{26}$Also at RWTH Aachen University, III. Physikalisches Institut A, Aachen, Germany\\
$^{27}$Also at Bergische University Wuppertal (BUW), Wuppertal, Germany\\
$^{28}$Also at Brandenburg University of Technology, Cottbus, Germany\\
$^{29}$Also at Forschungszentrum J\"{u}lich, Juelich, Germany\\
$^{30}$Also at CERN, European Organization for Nuclear Research, Geneva, Switzerland\\
$^{31}$Also at HUN-REN ATOMKI - Institute of Nuclear Research, Debrecen, Hungary\\
$^{32}$Now at Universitatea Babes-Bolyai - Facultatea de Fizica, Cluj-Napoca, Romania\\
$^{33}$Also at MTA-ELTE Lend\"{u}let CMS Particle and Nuclear Physics Group, E\"{o}tv\"{o}s Lor\'{a}nd University, Budapest, Hungary\\
$^{34}$Also at HUN-REN Wigner Research Centre for Physics, Budapest, Hungary\\
$^{35}$Also at Physics Department, Faculty of Science, Assiut University, Assiut, Egypt\\
$^{36}$Also at Punjab Agricultural University, Ludhiana, India\\
$^{37}$Also at University of Visva-Bharati, Santiniketan, India\\
$^{38}$Also at Indian Institute of Science (IISc), Bangalore, India\\
$^{39}$Also at Amity University Uttar Pradesh, Noida, India\\
$^{40}$Also at IIT Bhubaneswar, Bhubaneswar, India\\
$^{41}$Also at Institute of Physics, Bhubaneswar, India\\
$^{42}$Also at University of Hyderabad, Hyderabad, India\\
$^{43}$Also at Deutsches Elektronen-Synchrotron, Hamburg, Germany\\
$^{44}$Also at Isfahan University of Technology, Isfahan, Iran\\
$^{45}$Also at Sharif University of Technology, Tehran, Iran\\
$^{46}$Also at Department of Physics, University of Science and Technology of Mazandaran, Behshahr, Iran\\
$^{47}$Also at Department of Physics, Faculty of Science, Arak University, ARAK, Iran\\
$^{48}$Also at Italian National Agency for New Technologies, Energy and Sustainable Economic Development, Bologna, Italy\\
$^{49}$Also at Centro Siciliano di Fisica Nucleare e di Struttura Della Materia, Catania, Italy\\
$^{50}$Also at Universit\`{a} degli Studi Guglielmo Marconi, Roma, Italy\\
$^{51}$Also at Scuola Superiore Meridionale, Universit\`{a} di Napoli 'Federico II', Napoli, Italy\\
$^{52}$Also at Fermi National Accelerator Laboratory, Batavia, Illinois, USA\\
$^{53}$Also at Lulea University of Technology, Lulea, Sweden\\
$^{54}$Also at Consiglio Nazionale delle Ricerche - Istituto Officina dei Materiali, Perugia, Italy\\
$^{55}$Also at Institut de Physique des 2 Infinis de Lyon (IP2I ), Villeurbanne, France\\
$^{56}$Also at Department of Applied Physics, Faculty of Science and Technology, Universiti Kebangsaan Malaysia, Bangi, Malaysia\\
$^{57}$Also at Consejo Nacional de Ciencia y Tecnolog\'{i}a, Mexico City, Mexico\\
$^{58}$Also at Trincomalee Campus, Eastern University, Sri Lanka, Nilaveli, Sri Lanka\\
$^{59}$Also at Saegis Campus, Nugegoda, Sri Lanka\\
$^{60}$Also at National and Kapodistrian University of Athens, Athens, Greece\\
$^{61}$Also at Ecole Polytechnique F\'{e}d\'{e}rale Lausanne, Lausanne, Switzerland\\
$^{62}$Also at Universit\"{a}t Z\"{u}rich, Zurich, Switzerland\\
$^{63}$Also at Stefan Meyer Institute for Subatomic Physics, Vienna, Austria\\
$^{64}$Also at Laboratoire d'Annecy-le-Vieux de Physique des Particules, IN2P3-CNRS, Annecy-le-Vieux, France\\
$^{65}$Also at Near East University, Research Center of Experimental Health Science, Mersin, Turkey\\
$^{66}$Also at Konya Technical University, Konya, Turkey\\
$^{67}$Also at Izmir Bakircay University, Izmir, Turkey\\
$^{68}$Also at Adiyaman University, Adiyaman, Turkey\\
$^{69}$Also at Bozok Universitetesi Rekt\"{o}rl\"{u}g\"{u}, Yozgat, Turkey\\
$^{70}$Also at Marmara University, Istanbul, Turkey\\
$^{71}$Also at Milli Savunma University, Istanbul, Turkey\\
$^{72}$Also at Kafkas University, Kars, Turkey\\
$^{73}$Now at Istanbul Okan University, Istanbul, Turkey\\
$^{74}$Also at Hacettepe University, Ankara, Turkey\\
$^{75}$Also at Erzincan Binali Yildirim University, Erzincan, Turkey\\
$^{76}$Also at Istanbul University -  Cerrahpasa, Faculty of Engineering, Istanbul, Turkey\\
$^{77}$Also at Yildiz Technical University, Istanbul, Turkey\\
$^{78}$Also at School of Physics and Astronomy, University of Southampton, Southampton, United Kingdom\\
$^{79}$Also at IPPP Durham University, Durham, United Kingdom\\
$^{80}$Also at Monash University, Faculty of Science, Clayton, Australia\\
$^{81}$Also at Universit\`{a} di Torino, Torino, Italy\\
$^{82}$Also at Bethel University, St. Paul, Minnesota, USA\\
$^{83}$Also at Karamano\u {g}lu Mehmetbey University, Karaman, Turkey\\
$^{84}$Also at California Institute of Technology, Pasadena, California, USA\\
$^{85}$Also at United States Naval Academy, Annapolis, Maryland, USA\\
$^{86}$Also at Ain Shams University, Cairo, Egypt\\
$^{87}$Also at Bingol University, Bingol, Turkey\\
$^{88}$Also at Georgian Technical University, Tbilisi, Georgia\\
$^{89}$Also at Sinop University, Sinop, Turkey\\
$^{90}$Also at Erciyes University, Kayseri, Turkey\\
$^{91}$Also at Horia Hulubei National Institute of Physics and Nuclear Engineering (IFIN-HH), Bucharest, Romania\\
$^{92}$Now at another institute formerly covered by a cooperation agreement with CERN\\
$^{93}$Also at Texas A\&M University at Qatar, Doha, Qatar\\
$^{94}$Also at Kyungpook National University, Daegu, Korea\\
$^{95}$Also at Yerevan Physics Institute, Yerevan, Armenia\\
$^{96}$Also at another international laboratory covered by a cooperation agreement with CERN\\
$^{97}$Also at Imperial College, London, United Kingdom\\
$^{98}$Also at Institute of Nuclear Physics of the Uzbekistan Academy of Sciences, Tashkent, Uzbekistan\\
$^{99}$Also at another institute formerly covered by a cooperation agreement with CERN\\
\end{sloppypar}
\end{document}